\documentclass[longauth]{aa}  

\usepackage{graphicx}
\usepackage{txfonts}
\usepackage{lscape}
\usepackage{hyperref}
\begin{document}

\title{The CARMENES search for exoplanets around M dwarfs}
\subtitle{High-resolution optical and near-infrared spectroscopy of 324 survey stars}

\titlerunning{Optical and near-infrared spectroscopy of the CARMENES GTO sample}

    \author{A.~Reiners\inst{1}      
     \and M.~Zechmeister\inst{1}    
     \and J.A.~Caballero\inst{2,3}  
     \and I.~Ribas\inst{4}          
     \and J.C.~Morales\inst{4}
     \and S.V.~Jeffers\inst{1}
     \and P.~Sch\"ofer\inst{1}
     \and L.~Tal-Or\inst{1}
     \and A.~Quirrenbach\inst{3}  
     \and P.J.~Amado\inst{5}        
% ...
}
  \institute{ Institut f\"ur Astrophysik, Georg-August-Universit\"at, 
              Friedrich-Hund-Platz 1, D-37077 G\"ottingen, Germany\\
              \email{Ansgar.Reiners@phys.uni-goettingen.de}
         \and Centro de Astrobiolog\'ia (CSIC-INTA), Instituto Nacional de T\'ecnica Aeroespacial,
              Ctra. de Torrej\'on a Ajalvir, km 4, E-28850 Torrej\'on de Ardoz, Madrid, Spain
         \and Zentrum f\"ur Astronomie der Universt\"at Heidelberg, Landessternwarte,
              K\"onigstuhl 12, D-69117 Heidelberg, Germany
         \and Institut de Ci\`encies de l’Espai (CSIC-IEEC), Campus UAB, c/ de Can Magrans s/n, 
              E-08193 Bellaterra, Barcelona, Spain
         \and Instituto de Astrof\'isica de Andaluc\'ia (IAA-CSIC), Glorieta de la Astronom\'ia s/n, 
              E-18008 Granada, Spain
% ...
}


   \date{\today}


   \abstract{The CARMENES radial velocity (RV) survey is observing 324
     M dwarfs to search for any orbiting planets.  In this paper, we
     present the survey sample by publishing one CARMENES spectrum for
     each M dwarf. These spectra cover the wavelength range
     520--1710\,nm at a resolution of at least $R > 80,000$, and we
     measure its RV, H$\alpha$ emission, and projected rotation
     velocity. We present an atlas of high-resolution M-dwarf spectra
     and compare the spectra to atmospheric models. To quantify the RV
     precision that can be achieved in low-mass stars over the
     CARMENES wavelength range, we analyze our empirical information
     on the RV precision from more than 6500 observations.  We compare
     our high-resolution M-dwarf spectra to atmospheric models where
     we determine the spectroscopic RV information content, $Q$, and
     signal-to-noise ratio.  We find that for all M-type dwarfs, the
     highest RV precision can be reached in the wavelength range
     700--900\,nm. Observations at longer wavelengths are equally
     precise only at the very latest spectral types (M8 and M9). We
     demonstrate that in this spectroscopic range, the large amount of
     absorption features compensates for the intrinsic faintness of an
     M7 star. To reach an RV precision of 1\,m\,s$^{-1}$ in very low
     mass M dwarfs at longer wavelengths likely requires the use of a
     10\,m class telescope. For spectral types M6 and earlier, the
     combination of a red visual and a near-infrared spectrograph is
     ideal to search for low-mass planets and to distinguish between
     planets and stellar variability. At a 4\,m class telescope, an
     instrument like CARMENES has the potential to push the RV
     precision well below the typical jitter level of
     3--4\,m\,s$^{-1}$.}

 \keywords{Astronomical data bases -- Stars: rotation -- Stars: late-type -- Stars: low-mass -- Infrared: stars}

   \maketitle
%

\section{Introduction}
\label{sect:Introduction}

Spectroscopy of M dwarfs has become a very active research field
because potentially habitable planetary companions cause Doppler
variations that are more easily detectable around stars of lower mass
\citep{2005AN....326.1015M, 2007AsBio...7...85S,
  2007AsBio...7...30T}. These stars also constitute the vast majority
of potential planet hosts in our immediate vicinity, and a detailed
characterization of their planets is believed to be easier than in the
more massive, brighter, and more distant Sun-like stars \citep[see,
e.g.,][]{2016Natur.536..437A}.

The first radial velocity (RV) surveys for extrasolar planets focused
on objects in orbit around Sun-like stars \citep[see,
e.g.,][]{2007ARA&A..45..397U}. The lower end of the mass range of the
discovered planets was continuously extended until measurement
precision reached a level of about 1\,m\,s$^{-1}$
\citep{2009A&A...493..639M, 2016PASP..128f6001F}. At this precision, a
10\,M$_{\oplus}$ planet can be discovered on a 1 yr orbit in the
liquid-water habitable zone around a 1\,M$_{\odot}$ star. The shortcut
to potentially habitable planets similar to Earth is to look around
lighter stars; at the same RV precision, a 2\,M$_{\oplus}$ planet can
be found in the habitable zone around a 0.3\,M$_{\odot}$ star
\citep[see, e.g.,][]{2005ApJ...634..625R, 2009A&A...507..487M,
  2013A&A...549A.109B, 2013A&A...556A.126A, 2014MNRAS.443L..89A,
  2016Natur.536..437A}.

The CARMENES M-dwarf survey began operations on Jan 1, 2016. The
instrument is located at Calar Alto observatory in Almeria, Southern
Spain ($37\degr13'25''$N, $2\degr32'46''$W). It provides nearly
continuous wavelength coverage from 520 to 1710\,nm from its two
channels: the visual channel (VIS) with a spectral resolution of $R =
94,600$ covers the range $\lambda = 520$--960\,nm, and the
near-infrared channel (NIR) operates at $R = 80,400$ and $\lambda =
960$--1710\,nm \citep{2016SPIE.9908E..12Q}. For the M-dwarf survey, we
regularly observe about 300 M dwarfs across all M-spectral subtypes. A
total amount of 750 useful nights is reserved as Guaranteed Time
Observations (GTO) with the goal to collect approximately 70 spectra
for each target over the course of the program
\citep{2017A&A...604A..87G}.

The main motivation for building an optical and near-infrared spectrograph
with this large wavelength coverage is to measure RVs in very cool
stars \citep[e.g.,][]{2006ApJ...644L..75M} and to understand the
amount of RV information and stellar RV jitter as a function of
wavelength. It is well known that Sun-like stars provide most RV
information at blue optical wavelengths where astronomical
spectrographs already reached the 10\,m\,s$^{-1}$ level in the 1980s
\citep{1979PASP...91..540C, 1985srv..conf...87M}. However, cooler
low-mass stars provide more flux at near-infrared wavelengths, while
their spectrum is extremely rich in molecular features at optical
wavelengths, which makes detailed predictions about measurable RV
precisions difficult. There has not been a final answer so far to the
question in which spectral range the RV method is most sensitive for
low-mass stars (see Section\,\ref{sect:information}).

Radial velocity jitter can be caused by corotating active regions, magnetic
cycles, variations in stellar granulation, stellar oscillations, and
other mechanisms \citep[e.g.,][]{2010A&A...512A..38L,
  2010A&A...512A..39M, 2010A&A...519A..66M, 2014ApJ...780..104C,
  2016A&A...587A.103L}. Its amplitude is expected to depend on
wavelength \citep[e.g.,][]{2007A&A...473..983D, 2010ApJ...710..432R,
  2015ApJ...798...63M}. A spectrograph with large wavelength coverage
can help to distinguish between Keplerian signals from an orbiting
planet and RV variations caused by the star itself
\citep[e.g.,][]{1997ApJ...485..319S, 2011MNRAS.412.1599B,
  2017MNRAS.466.1733B, 2014MNRAS.438.2717J,
  2015MNRAS.448.3038K}. Furthermore, the pattern of RV variation as a
function of wavelength can itself provide important information about
the star, for example, about spot temperatures and Zeeman broadening
\citep{2013A&A...552A.103R}, or about a modal identification for
pulsating stars \citep{2007CoAst.150..311A, 2007CoAst.151...57A}.

While stellar atmosphere models have improved significantly over the
past decade and instruments are being designed for 10\,cm\,s$^{-1}$
precision \citep{2010SPIE.7735E..0FP}, there is a need for empirical
calibration of the possible RV precision across optical and
near-infrared wavelengths for M dwarfs. The growing amount of
transiting-planet candidates discovered by photometry missions such as
\emph{Kepler}, MEarth, APACHE, and future missions like \emph{TESS}
and \emph{PLATO}, requires a substantial infrastructure for
spectroscopic follow-up. A host of red optical and near-infrared
spectrographs are currently planned or under construction that will
provide the required data to determine down the mass of our nearest
transiting neighbors \citep[e.g.,][]{2014SPIE.9147E..1GM,
  2014SPIE.9147E..15A, 2014SPIE.9147E..14K, 2016SPIE.9908E..18S,
  2016SPIE.9908E..19C, 2016SPIE.9908E..1AC, 2016SPIE.9908E..6TJ,
  2016SPIE.9908E..70G}. For an efficient planning of RV follow-up in
low-mass stars, and for new or extended RV surveys of our closest
neighbors, empirical information on the RV performance across
different wavelengths is important. With our data from the CARMENES
program, we are in an excellent position to address this question.

In this paper, we introduce the CARMENES sample and provide detailed
information about the 324 target stars that we are surveying for
planets. After more than a year of observations, we have amassed
several thousand spectra with both CARMENES channels, enough to draw
statistically significant conclusions about the spectroscopic
properties and RV precision across the entire CARMENES wavelength
range. CARMENES is the first spectrograph that routinely delivers high-resolution spectra of low-mass stars at infrared wavelengths. Before
CARMENES, spectroscopic information of low-mass stars could only be
provided by a few instruments, most of them requiring multiple
settings and/or access to 10 m class telescopes
\citep[e.g.,][]{2012A&A...539A.109L}. As a service to the community,
with this paper we also publish one CARMENES spectrum for each survey
star.

We introduce the CARMENES GTO sample stars and the library of CARMENES
spectra in Section\,\ref{sect:Sample}, and we derive spectroscopic
information about rotation and radial velocities for each star. In
Section\,\ref{sect:atlas} we take a detailed look into our atlas of
high-resolution spectra for three example stars that represent
different M subtypes. The atlas itself is published in the online
version of the paper. We investigate the RV information
content of M dwarfs from our observations in
Section\,\ref{sect:information}. Finally, our results are summarized
in Section\,\ref{sect:Summary}.

\section{Library of M-type CARMENES spectra}
\label{sect:Sample}

\begin{figure}
  \centering
  \resizebox{\hsize}{!}{\includegraphics[bb = 25 5 625 470, clip=]{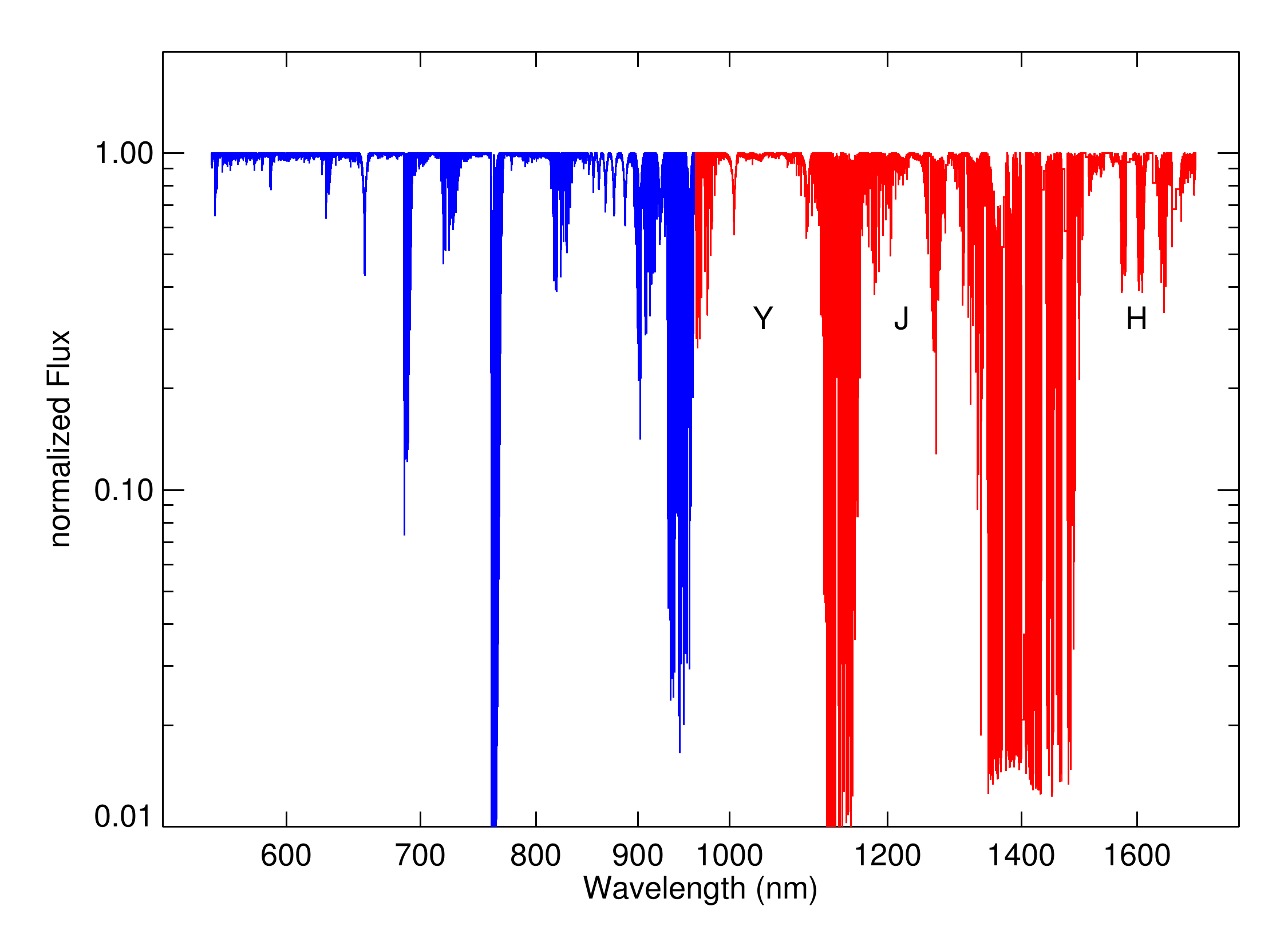}}
  \caption{\label{fig:SpectrumOverview}Overview of the CARMENES
    spectral range. The spectrum covered by the VIS channel is shown in
    blue, and the NIR channel in red. The target shown is the A2V star
    50\,Cas. The stellar spectrum only exihibits a few hydrogen lines,
    all strong features are from Earth's atmosphere.}
\end{figure}

The CARMENES spectral format covers the wavelength range
520--1710\,nm. In the telescope front-end, a beam splitter sends light
at wavelengths shorter than 960\,nm into the VIS channel and longer
wavelengths into the NIR channel. The two channels operate independently,
but see light from the same object. Data are reduced with our
automatic pipeline using the method of optimal extraction
\citep{2014A&A...561A..59Z}. Reduced spectra are stored at the Calar
Alto archive and analyzed for their RVs \citep{Zechmeister17}; see
also \citet{2016SPIE.9910E..0EC}.

The NIR detector is an array of two detectors that are separated by a small
gap. The spectral coverage is almost continuous, with additional small
gaps that grow toward long wavelengths; gaps are between zero and
15\,nm large from the very blue to the very red end. The spectral
format cannot be changed. An overview of the CARMENES spectral range
and telluric contamination is presented in
Fig.\,\ref{fig:SpectrumOverview}. The normalized spectrum is shown
logarithmically on both axes. It shows the telluric contamination of
the spectrum and the three main atmospheric windows, that is, the $z, J$,
and $H$\,bands, that are covered by the NIR channel.

As part of the GTO agreement, we provide early access to one CARMENES
spectrum for each of our sample targets (Table\,\ref{tab:Table}).
They can be downloaded from the CARMENES GTO Data Archive
\citep{2016SPIE.9910E..0EC}.\footnote{\url{http://carmenes.cab.inta-csic.es}}
Each spectrum is a single exposure obtained between Jan~1, 2016, and
Aug~31, 2017, and has a signal-to-noise ratio (S/N) typical for our
survey (see below). Details on how the S/N is calculated are given in
\citet{Zechmeister17}.

\subsection{Sample}

\begin{figure}
  \centering
  \resizebox{\hsize}{!}{\includegraphics[bb = 50 10 625 440, clip=]{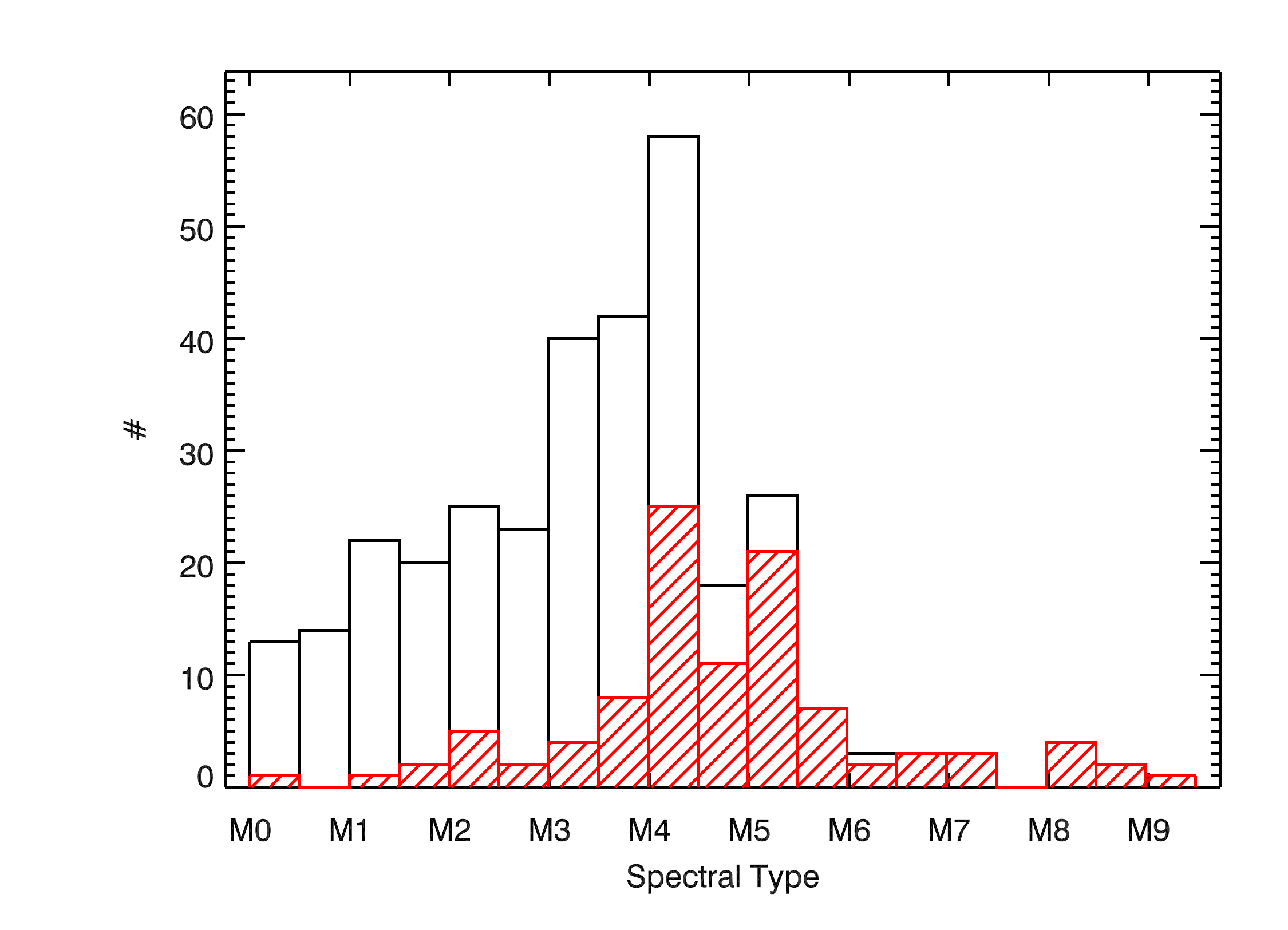}}
  \resizebox{\hsize}{!}{\includegraphics[bb = 50 10 625 440, clip=]{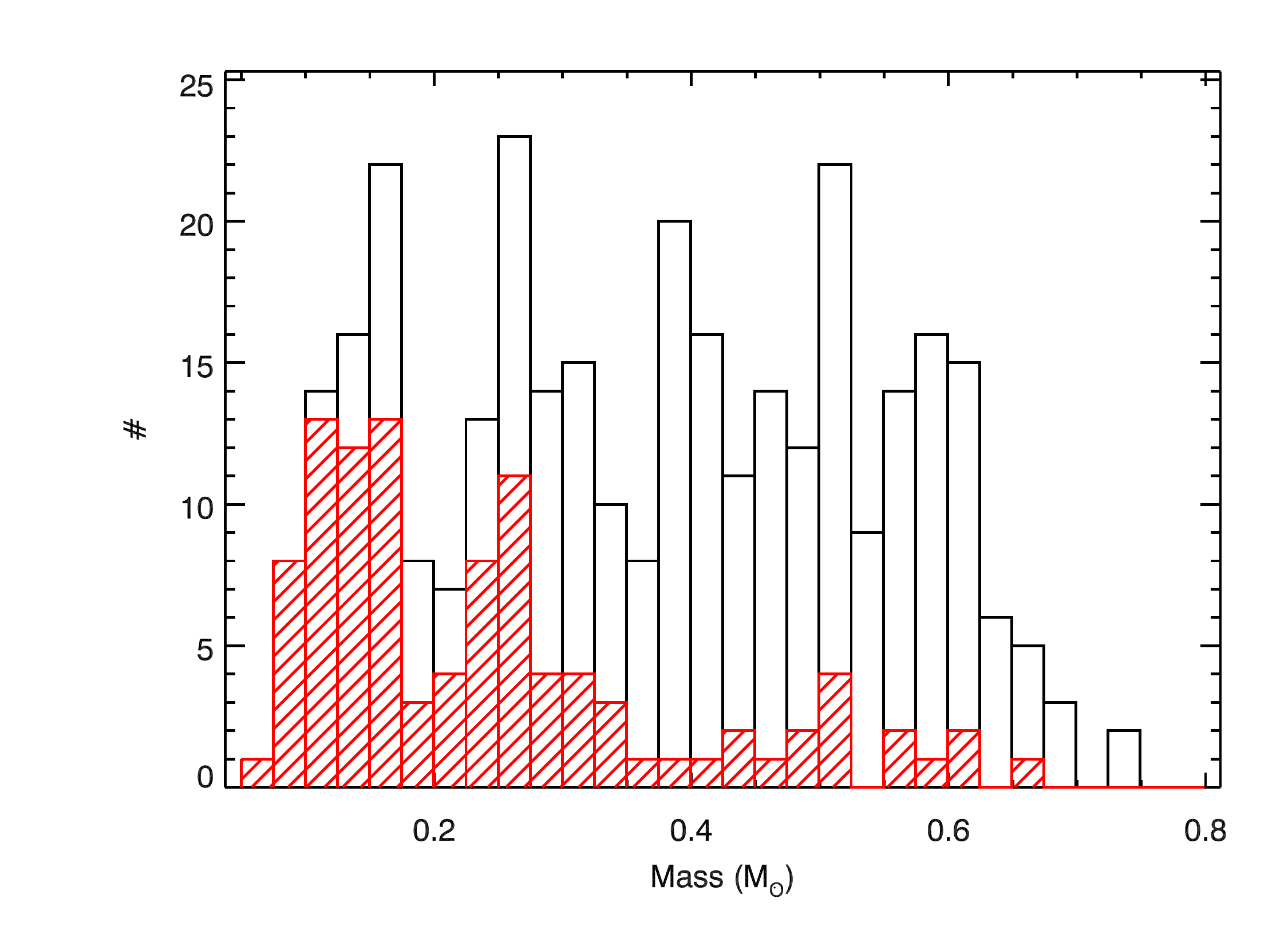}}
  \resizebox{\hsize}{!}{\includegraphics[bb = 50 10 625 440, clip=]{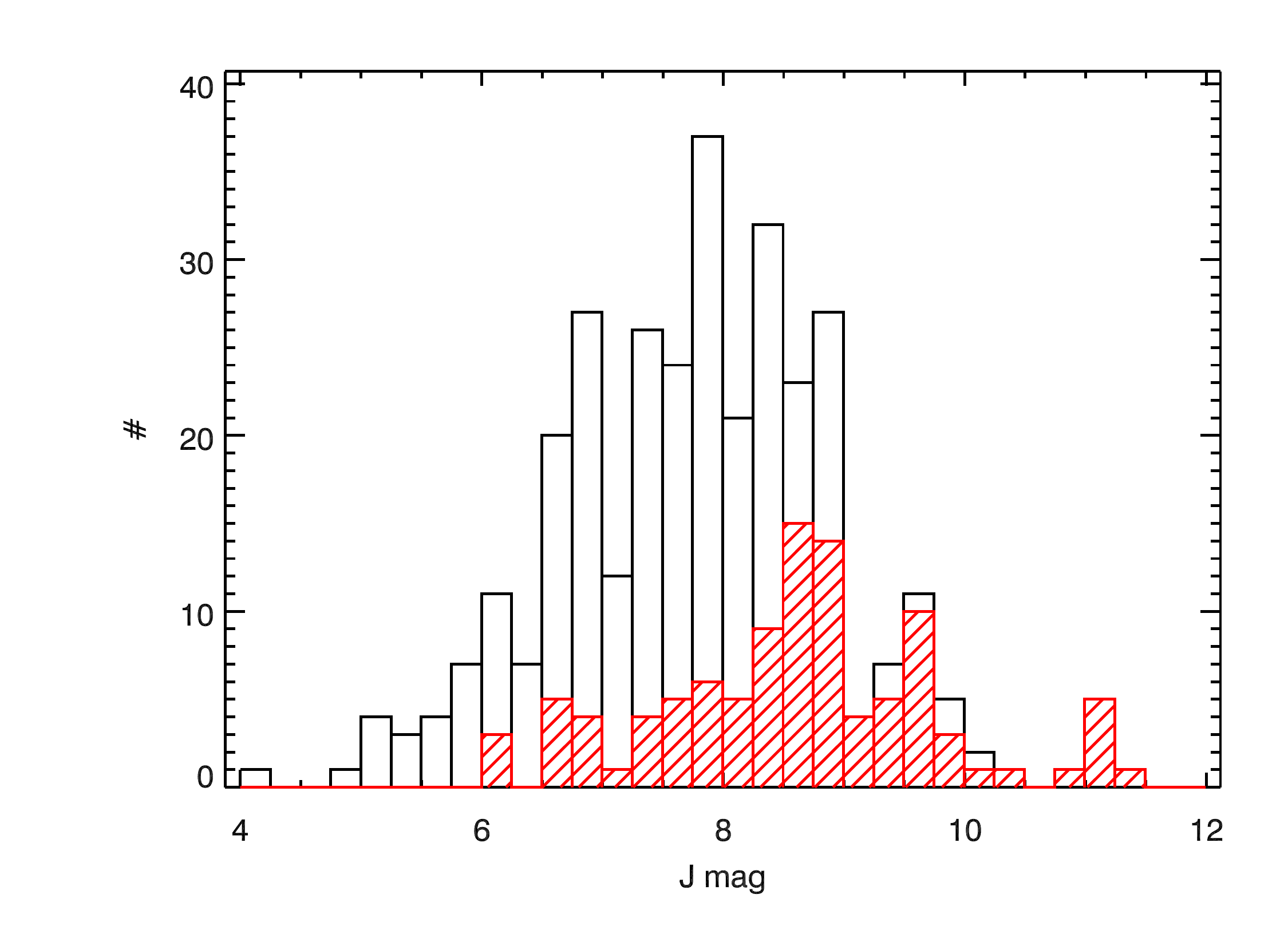}}
  \caption{\label{fig:Histo_Sample}From top to bottom: distribution of the CARMENES survey
    sample in spectral type, mass, and $J$-band magnitude. The subsample of
    stars with H$\alpha$ emission is shown in red.}
\end{figure}

To define our sample of M dwarfs, we selected the
brightest members of every spectral subtype that are visible from
Calar Alto ($\delta > -23^{\circ}$) and that are not known to be
members of multiple systems at separations closer than $5''$. We
carried out extensive preparatory observations and characterization to
define our survey sample. For more details on target preselection and
characterization, we refer to \citet{2015A&A...577A.128A},
\citet{2017A&A...597A..47C}, and \citet{Jeffers17}. In contrast to
other M-dwarf planet surveys, we explicitly did not bias our sample
with regard to age or chromospheric activity. One of the expected
advantages of the long-wavelength coverage of CARMENES is that RV
variations caused by stellar activity can to some extent be
distinguished from orbital motion. Learning about the RV signature
from starspots and stellar activity as a function of wavelength is one
of the science goals of the CARMENES M-dwarf survey.  We also did not
exclude stars with planets that were already known. Our sample
therefore has some overlap with other RV programs. Analyses of
CARMENES RVs for seven stars with known planets were presented in
\citet{Trifonov17}.

\begin{figure*}
  \centering
  \resizebox{.475\hsize}{!}{\includegraphics[bb = 25 5 625 470, clip=]{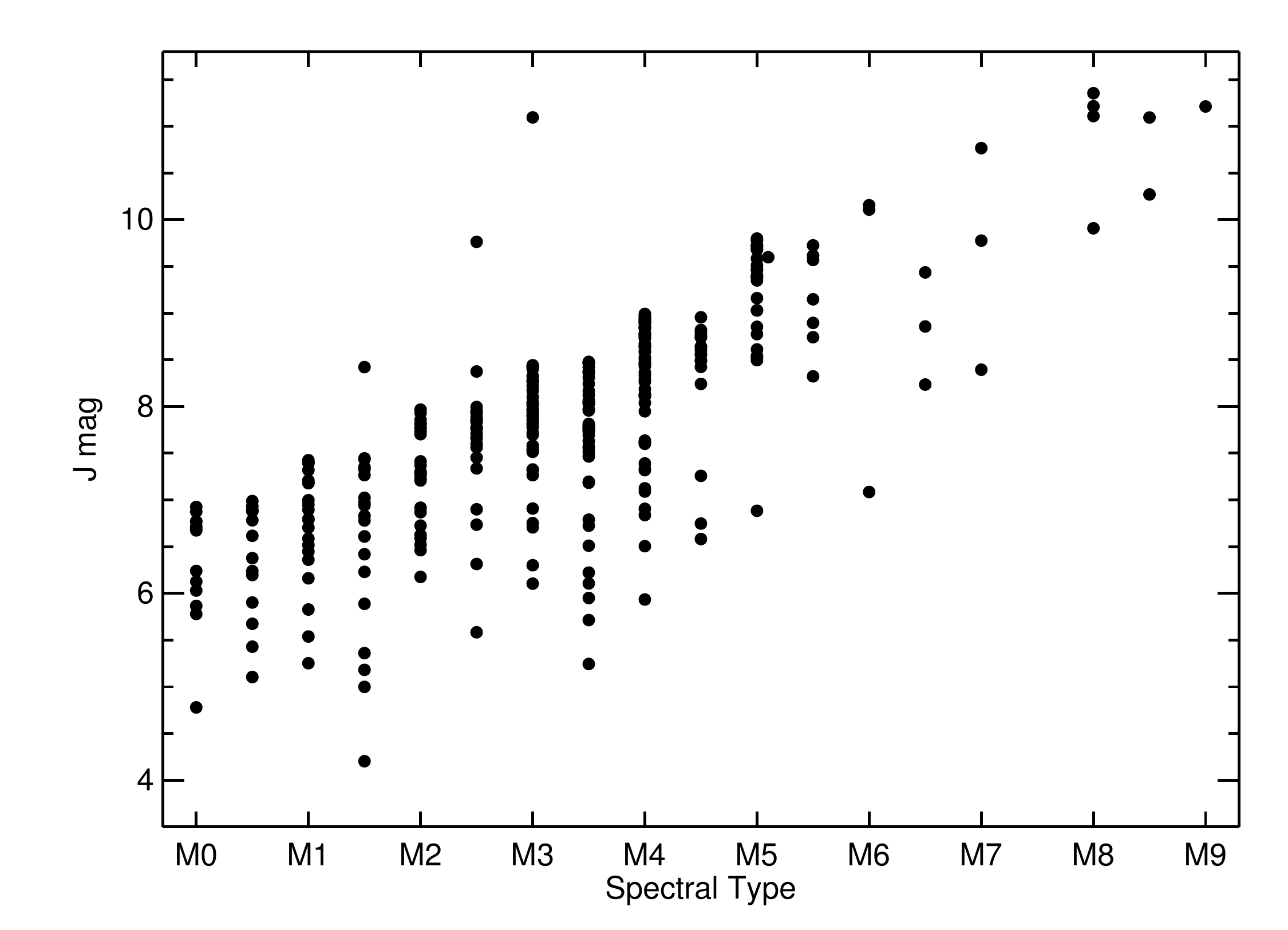}}\hspace{2mm}
  \resizebox{.475\hsize}{!}{\includegraphics[bb = 25 5 625 470, clip=]{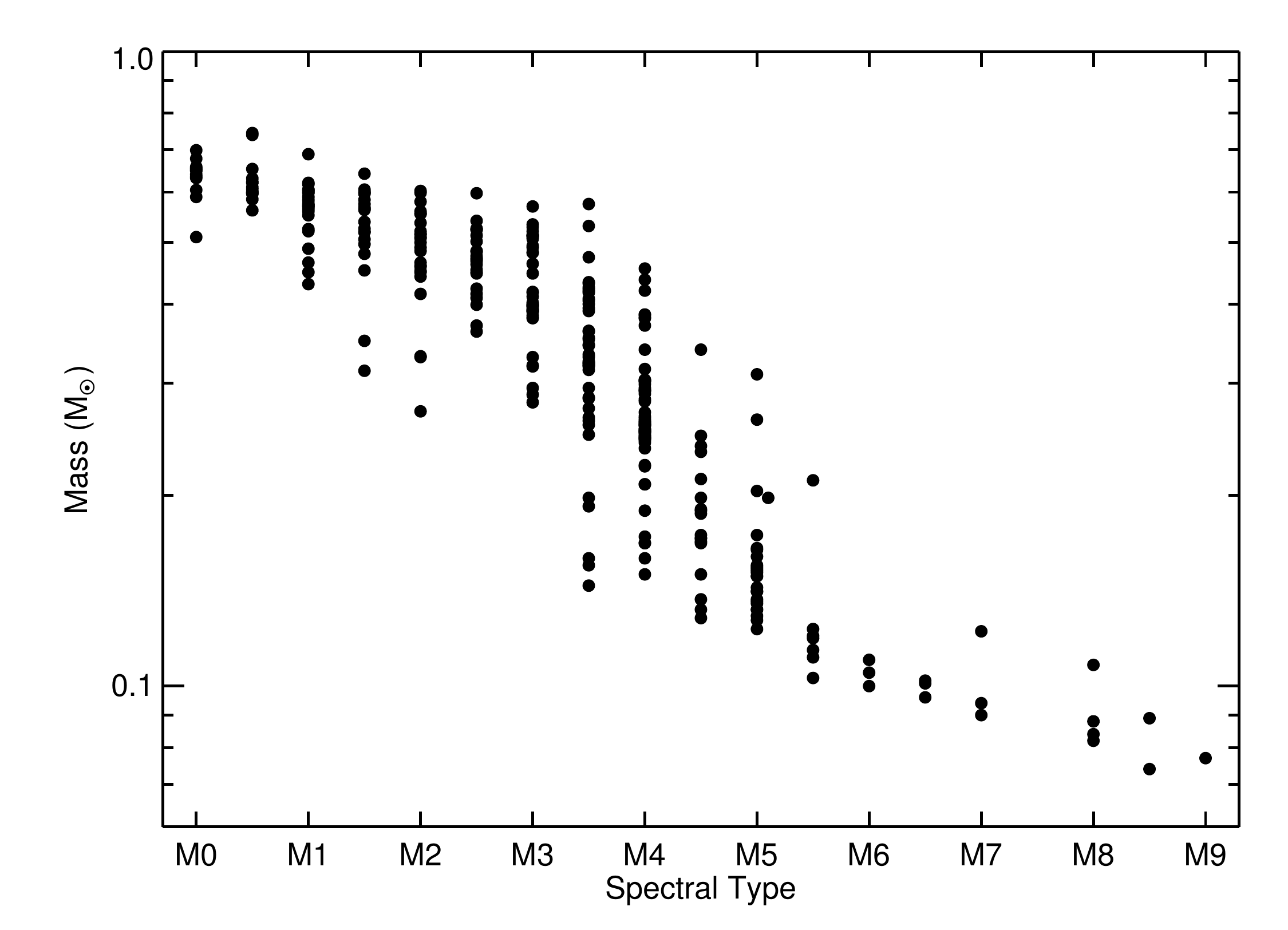}}
  \caption{\label{fig:JandMass}$J$-band magnitude (\emph{left panel})
    and mass (\emph{right panel}) for all stars of the CARMENES survey
    sample.}
\end{figure*}

After a few observations were taken for each star, we discovered
several double-lined spectroscopic binaries (SB2s) that will be
presented in a separate paper. After cleaning for SB2s, we ended up
with a survey sample of 324 stars; all stars are presented in
Table\,\ref{tab:Table}. With only a few exceptions, the spectral types
we used were adopted from the works of \citet{1995AJ....110.1838R,
  1996AJ....112.2799H, 2002AJ....123.3356G, 2013AJ....145..102L,
  2015A&A...577A.128A}. For details about spectral types we refer to \citet{2015A&A...577A.128A}. $J$-band magnitudes are taken
from
the Two Micron All Sky Survey \citep{2006AJ....131.1163S}. The typical
limit for the $J$-band magnitude for our survey is $J$\,=\,10\,mag, and
it is brighter for earlier spectral types. Some targets with known
transiting planets were added to the sample although they are fainter
than our typical survey targets. The distribution of the sample stars
in spectral type, mass, and $J$-band magnitude is shown in
Fig.\,\ref{fig:Histo_Sample}. As a consequence of the steep decrease
in luminosity toward late-type stars, only very few stars of our
sample are later than M5. On the other hand, the distribution of our
sample stars in mass is flatter and reaches down to the brown dwarf
limit. This is mainly because there is only little difference in mass
between dwarf stars in the spectral type range M6--M9. Individual
values of $J$-band magnitude and mass are shown versus spectral types
in Fig.\,\ref{fig:JandMass}. As a rule of thumb, for a mid-M $J =
9$\,mag star, CARMENES reaches a S/N of 150 per pixel in the $I$ band
after 25\,min exposure time.

We also report masses and activity level in terms of H$\alpha$ luminosity
relative to bolometric luminosity in Table\,\ref{tab:Table}. Masses are
computed from $K_{\rm s}$-band magnitudes according to the relations provided
in \citet{2000A&A...364..217D} and \citet{2016AJ....152..141B}. We caution
that masses below 0.1\,M$_{\odot}$ may be underestimated because these
relations lose predictive utility for $M_K > 10$\,mag
\citep{2016AJ....152..141B}.  H$\alpha$ luminosities are computed from
measuring equivalent widths and converting them into luminosities \cite[see,
e.g.,][]{2008ApJ...684.1390R}. More details on H$\alpha$ measurements and other
chromosperic lines in the CARMENES spectral range will be provided in a
forthcoming paper.

\subsection{Stellar rotation}
\label{sect:vsini}

\begin{table}
  \begin{centering}
    \caption{\label{tab:vsinirefs}Refererence stars used for the
      calculation of $\varv\sin{i}$. }
    \begin{tabular}{llccc}
      \hline
      \hline
      \noalign{\smallskip}
      SpT      & Reference & SpT & $P$ & $\varv_{\rm Eq} (P)$ \\
      interval & star & & (d) & (km\,s$^{-1}$) \\
      \noalign{\smallskip}
      \hline
      \noalign{\smallskip}
      M0.0 -- M0.5 & GJ 548A &                    M0.0 &   111  &   $  0.3 $ \\
      M1.0 -- M3.5 & GJ 849  &                    M3.5 &   39.2 &   $  0.6 $ \\
      M4.0 -- M9.5 & GJ 1256 &                    M4.5 &  105.4 &   $  0.1 $ \\
      \noalign{\smallskip}
      \hline
      \noalign{\smallskip}
    \end{tabular}
    \tablefoot{Radii and references for rotation
      periods are given in Table\,\ref{tab:Table}.}
  \end{centering}
\end{table}

We calculated the projected rotation velocities, $\varv\,\sin{i}$, from our
spectra taken with CARMENES-VIS with the cross-correlation method. We
computed the cross-correlation function (CCF) and calibrated the width
of the CCF against artificially broadened spectra of a reference star
\citep[see, e.g.,][]{2012AJ....143...93R, Jeffers17}. We used coadded
spectra from all observations for each star if more than five
exposures were available \citep[see][]{Zechmeister17}. For
cross-correlation reference, we selected stars that were observed
frequently (at least ten times), which guarantees a very high S/N
reference (coadded) spectrum, and for which information on the
rotation period from photometry is available. We used three different
reference stars to minimize systematic errors caused by spectral
mismatch (Table\,\ref{tab:vsinirefs}).  All reference stars are
relatively slow rotators, and their equatorial rotation velocity,
$\varv$, is estimated from the rotation period, $P$, and radius, $R$, to lie well
below our detection limit.

For each star, we computed a set of CCFs for individual orders of the
CARMENES spectral format. The adopted $\varv\sin{i}$ is the average of
values from orders that we found to provide reliable information about
stellar rotation. Our criteria for the selected orders are the absence
of significant telluric contamination and chromospheric emission
lines, high S/N, and small influence from strong spectroscopic
features, such as molecular band heads. The latter can introduce
substantial systematic errors for relatively small differences in the
spectral characteristics of our stars. The spectral regions that we
chose cover the wavelength ranges 592--610\,nm, 650.5--654\,nm, and
660--685\,nm\ in stars more massive than $M = 0.125$\,M$_{\odot}$, and
741--757\,nm, 774--810\,nm, 840--843\,nm, and 847--885\,nm\ in less
massive stars. As uncertainties, we report the standard deviations of
the set of values calculated in these spectral chunks. All values of
$\varv\sin{i}$ measured from the CARMENES spectra are given in
Table\,\ref{tab:Table}.

\subsubsection{Fast and slow rotation}
\label{sect:fastslow}

Many of the stars in our sample are relatively slow rotators. If
Doppler broadening from stellar rotation is too small compared to the
spectral resolution (and other broadening mechanisms), the effect
cannot be reliably detected. In the case of M stars, turbulence and
thermal broadening are on the order of 1--2\,km\,s$^{-1}$ in the lines
of heavy ions and molecules, so that instrumental resolution
determines the detection limit. For a criterion to reliably detect
stellar rotation, we estimate that its Doppler effect must broaden the
spectral lines by at least half a resolution element. In the case of
CARMENES-VIS with $R = 94,600$, this means that $\varv\sin{i} =
2$\,km\,s$^{-1}$ is a conservative lower detection limit. For stars
where we could not determine rotational broadening in excess of
2\,km\,s$^{-1}$, we report this value as an upper limit for
$\varv\sin{i}$ in Table\,\ref{tab:Table}.

We find that 75 of our 324 sample stars (23\,\%) show significant rotational
broadening. A detailed investigation of the fraction of active stars in our
sample and a comparison to volume-limited M-dwarf samples is carried out in
\citet{Jeffers17}. In that paper, a larger sample of stars was observed in preparation of the CARMENES survey. In the final CARMENES sample, we
included 40 stars for which no information on $\varv\sin{i}$ was available
before. We also compiled a list of rotation periods, $P$, measured from
photometry. These values are reported together with the expected equatorial
rotation velocity $\varv_{\rm Eq}$. To calculate $\varv_{\rm Eq}$ from $P$, we
determined the radius of each star according to the mass-radius relation for
4\,Gyr old stars of solar metallicity from \citet{1998A&A...337..403B}.

\begin{figure}
  \centering
  \resizebox{\hsize}{!}{\includegraphics[bb = 25 5 640 470, clip=]{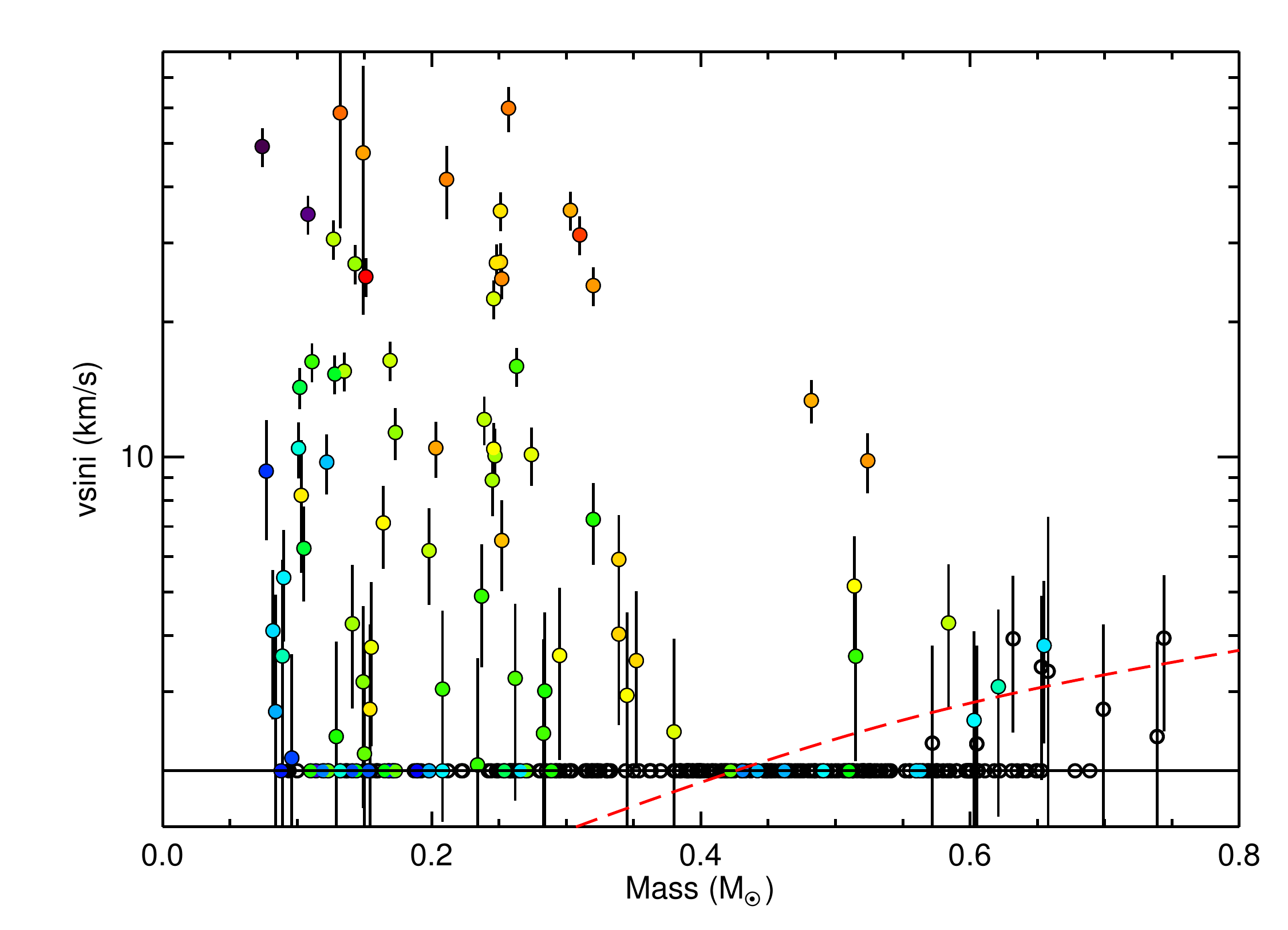}}

  \hspace{4.5cm}
  \resizebox{.35\hsize}{!}{\includegraphics[bb = 65 15 630 140, clip=]{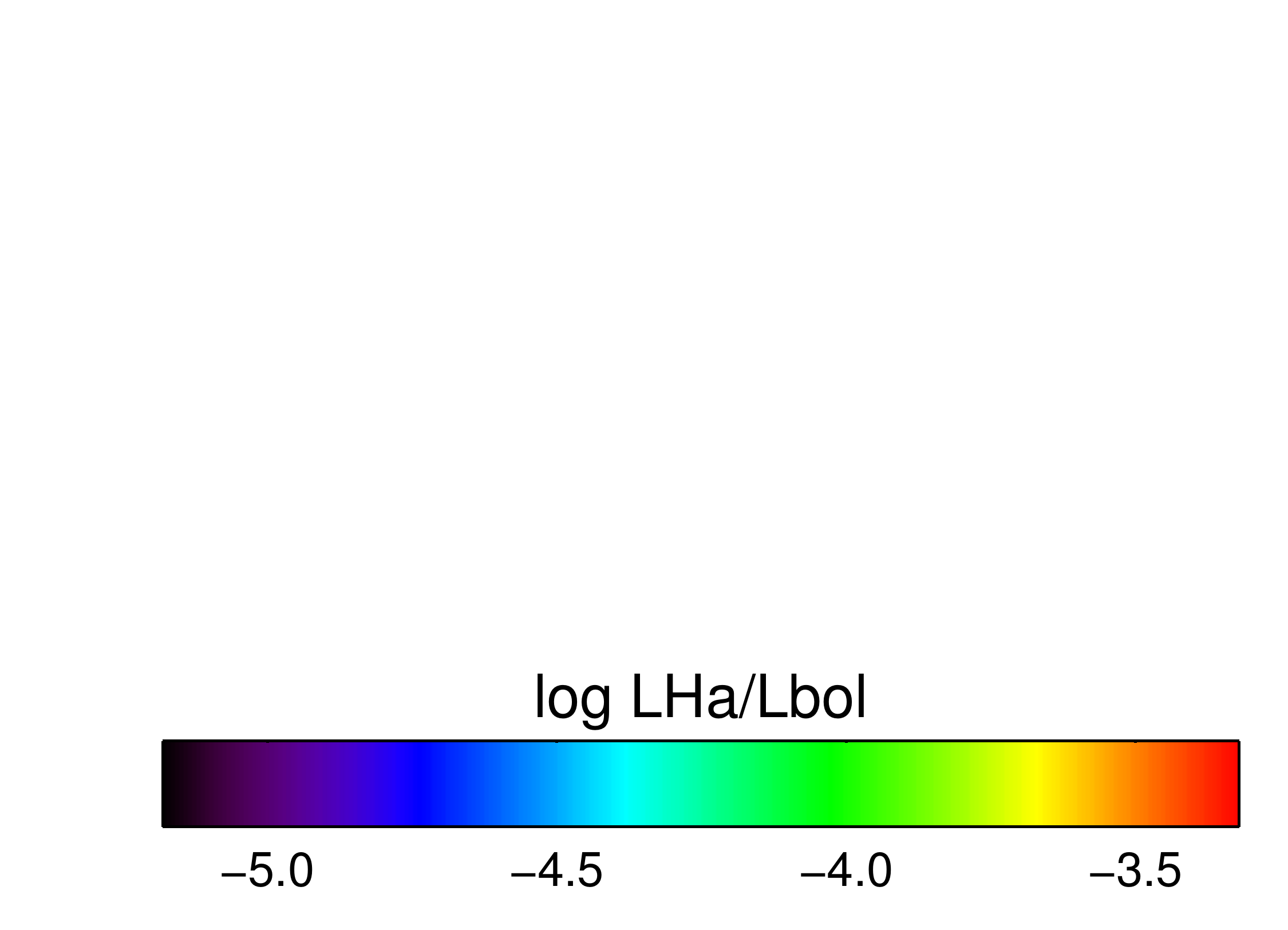}}
  \caption{\label{fig:vsini_mass} Projected rotation velocity
    $\varv\sin{i}$ as a function of stellar mass. Stars with detected
    H$\alpha$ emission are shown as filled circles, and stars with no
    H$\alpha$ emission are shown as open circles. Colors indicate the
    level of H$\alpha$ emission as shown in the legend. Slow rotators
    are plotted at our detection limit of $\varv\sin{i} =
    2$\,km\,s$^{-1}$. The red dashed line indicates values of
    $\varv_{\rm Eq}$ for stars with rotation periods $P = 10$\,d. Above
    this line, H$\alpha$ is always expected in emission (see text).}
\end{figure}

The CARMENES results for $\varv\sin{i}$ are shown as a function of
mass in Fig.\,\ref{fig:vsini_mass}. In this figure, we plot all stars
with H$\alpha$ emission as filled circles and those without
H$\alpha$ emission as open circles. All stars with significant
rotational broadening and masses below $M=0.55$\,M$_{\odot}$ are also
H$\alpha$-emitters; there is no inactive fast rotator below that mass
limit. The fraction of active stars is higher at lower mass, a result
that was found previously by other investigations \cite[cf.][]{Jeffers17}. 

A handful of inactive stars with high masses ($M>0.55$\,M$_{\odot}$)
also show significant line broadening. The absence of H$\alpha$
emission in these stars deserves some deeper discussion. One possible
explanation is that the spectra of these stars, in particular the
wings of their atomic lines, are intrinsically different from
``normal'' M0 star spectra. The masses of these targets are relatively
high for spectral type M0, which may point to peculiarities in the
composition or age of these stars. This could lead to systematic
differences between the spectra of our template M0 star and these
apparently massive M0 stars, which cause a systematic offset in our
determination of $\varv\sin{i}$. A second plausible explanation is
that these stars are in fact rotating significantly faster than the
inactive stars that are less massive, but here rotation is still not
fast enough to produce H$\alpha$ emission. The left panel of Figure\,8
in \citet{Jeffers17} shows that normalized H$\alpha$ emission is
proportional to rotation period in all M stars of spectral type
M0--M4.5. H$\alpha$ emission, however, becomes undetectable at a level
of $\log{L_{\rm H\alpha}/L_{\rm bol}} = -4.5$ in these stars. While
all stars rotating faster than $P = 10$\,d show H$\alpha$ in emission,
activity in a growing fraction of the slower rotators falls below that
threshold and cannot be detected anymore. In other words, at rotation
periods of $P = 10$\,d and slower, not all stars exhibit H$\alpha$ in
emission. For our measurements, this means that the critical value of
$\varv\sin{i}$ above which we always expect to find H$\alpha$
emission is a function of radius (or mass). We plot the critical
value of $\varv\sin{i}$ that corresponds to $P = 10$\,d as a red
dashed line in Fig.\,\ref{fig:vsini_mass}. The critical value is below
$\varv\sin{i} = 2$\,km\,s$^{-1}$ in stars less massive than
$M=0.45$\,M$_{\odot}$ but rises above our detection limit towards
higher masses. The five stars with detected surface rotation but no
H$\alpha$ emission are rotating at rates around the critical rate of
$P = 10$\,d or slower. We therefore conclude that our measurements of
$\varv\sin{i} > 2$\,km\,s$^{-1}$ are consistent with the lack of
H$\alpha$ emission in these stars.

\subsubsection{Comparing photometric period to surface rotation}

\begin{figure}
  \centering
  \resizebox{\hsize}{!}{\includegraphics[bb = 25 5 645 470, clip=]{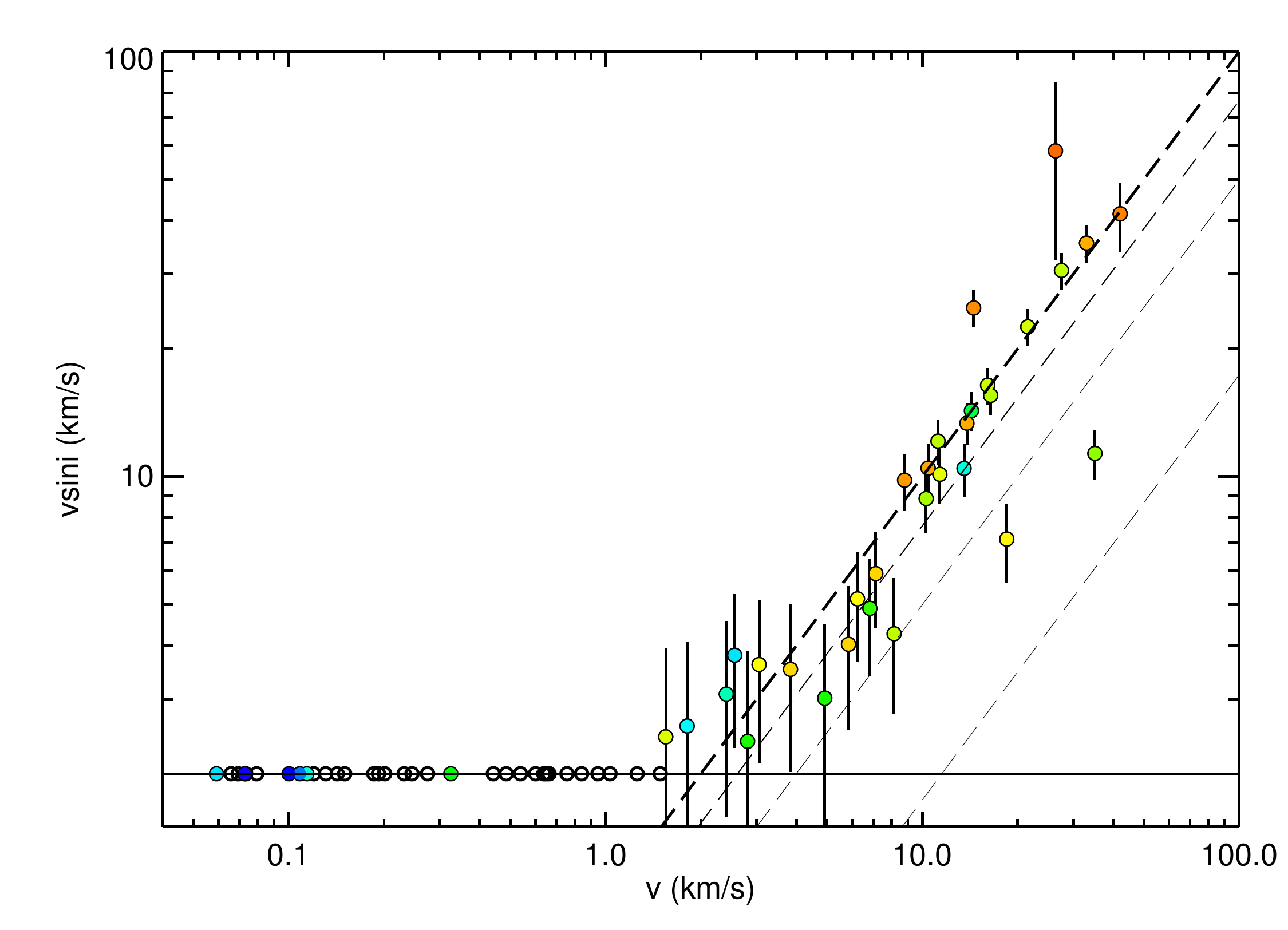}}
  
  \hspace{-3.5cm}
  \resizebox{.35\hsize}{!}{\includegraphics[bb = 65 15 630 140, clip=]{Hacollegend.pdf}}
  \caption{\label{fig:veq_vsini}Comparison between $\varv\sin{i}$ and
    equatorial velocity $\varv$ estimated from photometric
    period and radius. Active stars are indicated as filled circles as in
    Fig.\,\ref{fig:vsini_mass}. Dashed lines show expected relations
    between $\varv\sin{i}$ and $\varv$ for inclination angles
    $i = 90\degr, 50\degr, 30\degr$, and $10\degr$.}
\end{figure}

For many of our target stars we have information on rotational periods
that we can combine with our measurements of projected surface
rotation, $\varv\sin{i}$ \citep[see][]{2012AJ....143...93R}. We plot
$\varv\sin{i}$ against the expected equatorial rotation velocity
according to photometric period and stellar radius in
Fig.\,\ref{fig:veq_vsini}. We also show lines indicating
$\varv\sin{i}$ as expected for a given $\varv_{\rm Eq}$ observed under
inclination angles of $i = 90\degr, 50\degr, 30\degr$, and
$10\degr$. We include our estimate on the inclination values from this
comparison in Table\,\ref{tab:Table}. 

The majority of the stars with available $P$ and $\varv\sin{i}$ follow the
relation for $i = 90\degr$ or a little below. Useful estimates of the
inclination angle $i$ can be given in cases where a photometric period and a
measurement of surface rotation above the detection limit are available. Of
these stars, about one half (15) have values of $\varv\sin{i}$ \emph{higher}
than $\varv_{\rm Eq}$. Most of them are consistent with inclination angles $i
= 90$\,deg within the measurement uncertainties. We find only one star with a very
small uncertainty in $\varv\sin{i,}$ but with a photometric period indicating
much slower rotation (RX~J0506.2); its rotation period of 0.89\,d is
inconsistent with the line broadening seen in our spectra. Nevertheless, we do
not expect such high a fraction of stars observed nearly equator-on in a
sample of stars in which inclination angles should be randomly distributed
(uniform distribution in $\cos{i}$). Possible reasons for a bias toward large
inclination angles include $i$) underestimated stellar radii, perhaps caused
by a systematic bias in metallicity; $ii$) overestimated rotational periods,
perhaps caused by misidentifying harmonics of $P$ as the true rotational
period or by differential rotation; $iii$) a systematically higher detection
efficiency for photometric periods in stars observed under high inclination
angles; and $iv$) overestimated values of $\varv\sin{i}$. For the latter,
spectral mismatch between the targes and their reference stars is one obvious
candidate. This may be particularly important for the three stars with
$\varv\sin{i} > \varv_{\rm Eq}$ that have surface rotation velocities between
2 and 4\,km\,s$^{-1}$ and belong to the group of relatively massive stars
discussed above (see Section\,\ref{sect:fastslow}).

One way to adjust the distribution of inclination angles is to select a
different set of reference stars. We have experimented with other reference
stars, and as expected, found other choices that can produce systematically
lower values for $\varv\sin{i}$. Our principal requirement for the set of
reference stars, however, was that spectral line broadening should be
negligible and that external information from photometric measurements should
be available. We conclude that the absolute values of $\varv\sin{i}$ need to
be interpreted with great care, in particular when they are compared to photometric
periods.

\subsection{Absolute radial velocities}
\label{sect:vrad}

For our sample stars, we computed absolute radial velocities from the
same data as were used for the calculation of $\varv\,\sin{i}$. From the CCFs
that were calculated for a set of spectral orders, we derived RV offsets between each star and the reference. Because we here
use  coadded spectra, our results represent the mean
radial velocities averaged over all observations for each
star. An analysis of the RV variability of each target will be performed
in detail for our planet search, but this is beyond the scope of this
paper.

From our cross-correlation analysis, we obtained RVs for
each star relative to its reference according to spectral type (see
Table\,\ref{tab:vsinirefs}). Before coadding, all observed spectra
were corrected for barycentric motion. Thus, the radial velocities are
the true differences between the motion of the stars with respect to
the solar system. The absolute values of the radial velocities,
however, are unknown. We shifted our RVs to the absolute scale of the
\emph{Gaia} pre-launch catalog of RV standard stars
provided by \citet{2013A&A...552A..64S}. Our samples have three
targets in common: Gl\,450, Gl\,514, and Gl\,526. We computed the
relative radial velocity shift between Gl\,514 and our three reference
stars, and shifted all our RVs such that the RVs of Gl\,514 match the
value from \citet[][$\varv_{\rm
  rad}=14.386$\,km\,s$^{-1}$]{2013A&A...552A..64S}. After calibration,
the RVs of the two other stars, Gl\,450 and Gl\,526, also agree with
the literature values within approximately 100\,m\,s$^{-1}$
\citep[$\varv_{\rm rad} = 0.221$\,km\,s$^{-1}$ and
15.570\,km\,s$^{-1}$, respectively;][]{2013A&A...552A..64S}.

Absolute RVs for all our stars are provided in
Table\,\ref{tab:Table}. The typical relative uncertainties for our
measurements are on the order of 10\,m\,s$^{-1}$; they are dominated by
spectral differences between the stars (note that we compute the
radial velocity from the CCF relative to observed spectra of our
reference stars). Faster rotators typically have larger uncertainties
for the same reasons as discussed in Section\,\ref{sect:vsini}.

\section{Spectral atlas}
\label{sect:atlas}

\begin{table*}
  \caption{\label{table:atlasexposures}Basic information about spectra shown in the spectral atlas. }
  \centering                   
  \begin{tabular}{l c c c c c c }     
    \hline\hline               
    \noalign{\smallskip}
    Target & SpT & Date & Exp.Time & S/N & Air mass & Doppler shift$^{a}$ \\
           &     & (UT) &  (s)     & 874\,nm / 1120\,nm  &    & (km\,s$^{-1}$)\\
    \noalign{\smallskip}
    \hline
    \noalign{\smallskip}
    GX~And           & M1   & 2016, Nov 10 20:57 & 300\,s & 290 / 420 & 1.01 &  22  \\
    Luyten's star    & M3.5 & 2016, Dec 31 00:58 & 230\,s & 150 / 230 & 1.18 &  10 \\
    Teegarden's star & M7   & 2016, Nov 16 22:32 & 1500\,s & \phantom{1}72 / 130 & 1.08 & 73 \\
    \noalign{\smallskip}
    \hline
  \end{tabular}
\tablefoot{$^a$ Doppler shift is applied to the model spectra.}
\end{table*}

We present a spectral atlas of three representative M dwarfs of
spectral types M1 (GX~And), M3.5 (Luyten's star), and M7 (Teegarden's
star). The three objects are among the brightest targets of their
spectral type and provide high-quality M-dwarf data at very different
effective temperatures. One spectrum of each star is shown in the
spectral atlas in Figs.\,\ref{fig:atlas00}--\ref{fig:atlas42}. We show
the entire wavelength range from H$\alpha$ up to the red end of our
spectral format (645--1710\,nm). Information on the individual
observations is summarized in Table\,\ref{table:atlasexposures}. Each
figure of the atlas covers an increase of approximately 2.3\,\% in
wavelength, that is, $\sim 2100$ resolution elements. The top panels show a
spectrum of the telluric standard star 50\,Cas as in
Fig.\,\ref{fig:SpectrumOverview}. In the other three panels, we
present the three M-dwarf spectra in black, together with a PHOENIX
model spectrum from \citet{2013A&A...553A...6H} in red, calculated for
approximately the atmospheric parameters we expect for the stars'
spectral types, that is, 3700\,K (M1), 3400\,K (M3.5), and 2600\,K
(M7). We chose $\log{g} = 5.0$ and solar metallicity for this
comparison. Model spectra are artificially broadened to match the
spectral resolution of our observations. In the top panel of the atlas
figures, the spectrum of the telluric standard is shown with
annotations of the most prominent absorption features. The hydrogen
\ion{H}{i} lines are stellar features, positions are taken from The
Atomic Line List
v2.04.\footnote{\url{http://www.pa.uky.edu/~peter/atomic/}} In
the second panel, we show the spectrum of GX~And together with
annotations of the most prominent atomic and molecular absorption
features seen in M dwarfs. Line positions of atomic lines are taken
from VALD \citep{2015PhyS...90e4005R}, information on molecular bands
was compiled from \citet{2005ApJ...623.1115C}. We chose to show our
observed spectra without applying any Doppler shift so that telluric
lines appear at the same position in all panels of our atlas. Because
of barycentric motion and the stars' radial velocities, the spectral
features are therefore slightly shifted. In order to match the
features of the model spectra to our observations, we Doppler-shifted
the model spectra accordingly. The values that we applied are given in
Table\,\ref{table:atlasexposures}.

Our wavelength range covers parts of the range shown in the atlas from
\citet{1998MNRAS.301.1031T}, which contains the spectral range up to
920\,nm. In the following, we discuss the spectroscopic features
observed at wavelengths redward of 920\,nm\ only. For a discussion
of M-dwarf spectra at shorter wavelengths, we refer to
\citet{1998MNRAS.301.1031T}, for instance.

\subsection{$Y$ band}

The spectral range around 970\,nm contains a number of strong Ti lines
that are very useful for the study of M-dwarf magnetic fields
\citep[e.g.,][]{2017ApJ...835L...4K, 2017NatAs...1E.184S}. These lines
are embedded in telluric water absorption, which needs to be corrected
for. The strong water absorption ends at around 980\,nm, just before
the well-known Wing-Ford band of molecular FeH sets in \citep[see,
e.g.,][]{2006ApJ...644..497R}. The wavelength range 985--1100\,nm is
virtually free of telluric absorption. It contains mostly FeH lines,
several strong atomic Ti lines, a prominent Ca line, and a handful
of weaker lines from other atoms.  Absorption from the FeH band has
intensively been used to measure rotation and magnetic fields in M
dwarfs.  The great advantage of the Wing-Ford band is that it contains
a large number of relatively isolated absorption lines that are not as
densely packed as the TiO or VO lines in the visual wavelength range,
which allows a line profile analysis of individual lines at least in
early- and mid-M stars \citep[e.g.,][]{2007A&A...467..259R}. FeH lines
have very different Land\'e g-factors that are useful for differential
investigation of Zeeman broadening or stellar metallicity
\citep{2011MNRAS.418.2548S}.

Some visual spectrographs based on CCDs can reach out to this
wavelength range, such as UVES/VLT, HIRES/Keck, or ESPADONS/CFHT. They
have been used to provide spectral quality that so far was superior to
most of the spectra taken with infrared detectors. CARMENES NIR also
covers this range with an infrared detector, but here the efficiency
was optimized to reach high throughput at these
wavelengths. Therefore, CARMENES has become one of the most efficient
instruments specifically in this wavelength range.

Our comparison between observations and model spectra at all three
spectral types shows that many molecular and atomic lines are well
reproduced by the models. However, the intensity of the FeH molecular
lines is often overestimated, and a set of strong absorption features
appears in the M7 model at wavelengths around 1050\,nm that are not detected in the observed spectrum. We did not make an attempt to find
the models that are most similar to our observed spectra, but the physical setup of these particular models is
clearly either lacking adequate input data (e.g., better molecular line data)
or the models inappropriately simulate the physical situation at very low
effective temperatures (at least for this wavelength region), or
both.\footnote{We note that observations and models also disagree in
  several strong alkali lines where the models predict too wide
  damping wings \citep[see, e.g.,][]{2007A&A...473..245R}, and the
  models overpredict the strength of the lithium line at 670.8\,nm
  because lithium depletion was not taken into account.}

Heavy water absorption between 1100\,nm and 1200\,nm separates the
$Y$ band from the $J$ band. At high spectral resolution, we can
identify stellar absorption lines from Na, K, Fe, Cr, and Mg embedded
in the forest of water lines. Some of them are strong and isolated
enough, so that they might be useful for a line profile analysis.

\subsection{$J$ band}

The $J$ band starts around 1200\,nm, where the water absorption band
becomes much weaker. The transparent atmospheric window extends until
water is again detected redward of 1300\,nm. The region in between is
not as free of telluric lines as the $Y$ band because there is an
additional molecular band from O$_2$ around 1270\,nm. The M-dwarf
spectra in the $J$ band are very poor in absorption lines compared to
the other spectral regions in the visual and near-infrared
range. There are no substantial molecular bands similar to the TiO,
VO, or FeH bands at shorter wavelengths. The two \ion{K}{i} lines near
1250\,nm stand out prominently, and there are a few lines from Ca, Ti,
Fe, and Mn around 1286\,nm. A \ion{Na}{i} line is buried in telluric
O$_2$ absorption, and two \ion{Al}{i} lines appear at the red end of
the $J$ band, where water absorption is relevant again (the two
\ion{Al}{i} lines are unfortunately lost in a gap of the CARMENES
spectral format). Another line of \ion{Mn}{i} is visible around
1330\,nm before very strong water absorption separates the $J$ from
the $H$ band. We do not show the wavelength range between 13,850 and
1480\,nm because almost no starlight passes through the dense water
features.

The spectral models show good agreement with the observed spectra. In
the $J$ band, no prominent features seem to be missing in the models,
and none of the features predicted by the models are missing in the
spectra. The long exposure of the M7 star shows a number of sky
emission lines from OH airglow \citep{2015A&A...581A..47O}.

\subsection{$H$ band}

At the blue end of the $H$ band, water absorption is significantly
reduced at around 1500\,nm until the end of our spectral format at
1710\,nm. Additional telluric absorption bands from CO$_2$ are visible
at 1540\,nm, 1570\,nm, 1600\,nm, and 1640\,nm
\citep{1950msms.book.....H}.

Around 1480\,nm, a series of stellar molecular OH absorption lines
appears among the water lines. The set of OH lines covers the entire
$H$ band and adds a significant number of stellar absorption
features. In terms of atomic lines, the situation is similar to the
$J$ band, with only a few scattered weak lines throughout the entire
band. These lines are \ion{Mg}{i} at 1505\,nm (partially lost in
spectral format gaps), very weak lines of \ion{K}{i} 1517\,nm, a
handful of lines from \ion{Fe}{i} and \ion{Ti}{i} throughout the band,
and a set of \ion{Ca}{i} lines at 1615\,nm. Overall, these
lines are all relatively weak, which implies that the amount of
information available for line profile analysis using a
high-resolution spectrograph is limited.

Another component of stellar absorption that becomes stronger toward
later spectral types appears throughout the entire $H$ band. This
component is significantly above the noise level and is also predicted
by the models. The absorption is probably due to FeH, but the lines are
too dense for an individual line analysis.

\section{Radial velocity information content}
\label{sect:information}

The RV precision that can be achieved in a spectroscopic RV
measurement depends on the number of photons that can be collected at
some particular wavelength, that is, on the S/N and also on the amount of
spectroscopic information that is available to measure the Doppler
shift, that is, the RV information content \citep{1985Ap&SS.110..211C,
  1996PASP..108..500B, 2001A&A...374..733B}. At stellar photospheric
temperatures, electronic transitions from atoms and molecules
typically generate many spectral lines at optical wavelengths, but
fewer at longer wavelengths. In low-mass stars, ro-vibrational
transitions add molecular bands at red optical and near-infrared
wavelengths. Thus, the RV information content is typically higher at short
optical wavelengths, but toward later spectral types, the
photon-dominated S/N grows dramatically from blue optical toward
infrared wavelengths.

Detailed simulations of the photon-limited RV precision in M dwarfs
were carried out by \citet{2010ApJ...710..432R, 2011A&A...532A..31R,
  2013PASP..125..240B, 2015arXiv150301770P, 2016A&A...586A.101F}. With
only very few exceptions, these studies relied on synthetic spectra
from model atmosphere simulations. These works agree with the general
picture that more RV information is available at short wavelengths,
but about the details of which spectral range is better than another,
after taking into account also the available number of photons, there
are discrepancies by a factor of two or more. One reason for
the discrepancies is that different models were used; model predictions
for the occurrence of molecular bands depend on input parameters about
the stellar atmosphere and molecular physics. Another reason is the
treatment of telluric lines, which becomes a very significant factor
at near-infrared wavelengths.

The spectra from our survey provide empirical information on the RV
precision that is achievable in M dwarfs across the wavelength range
520--1710\,nm. A major motivation to observe M dwarfs at longer
wavelengths is that the measurement uncertainty in any spectral bin,
(S/N)$^{-1}$, decreases dramatically from $V$ band to $J$
band. Nevertheless, there are far fewer spectroscopic features, that
is, spectroscopic RV information, at longer wavelengths. The ratio of
the two is the RV precision that can be achieved in an observation. In
the following, we provide the first empirical information on the
wavelength dependence of the RV precision in M dwarfs from a
significant number of observations.

We follow the definition of \citet{2001A&A...374..733B}; the RV
precision that can be reached in an observation of a star in a given
wavelength range can be written as

\begin{equation}
  \label{eq:q-factor}
  \delta \varv_{\rm rms} = \frac{c}{Q \cdot {\rm S/N}},
\end{equation}

where $c$ is the speed of light and $Q$ is the quality factor. The latter is
essentially the cumulative spectral gradient across the wavelength
range that is used for analysis. It depends on the intrinsic spectral
features of the star, but also on additional line broadening, in
particular, spectral resolution and stellar surface rotation,
$\varv\,\sin{i}$.

\subsection{Empirical radial velocity information}

\begin{figure*}
  \centering
  \resizebox{.89\hsize}{!}{\includegraphics{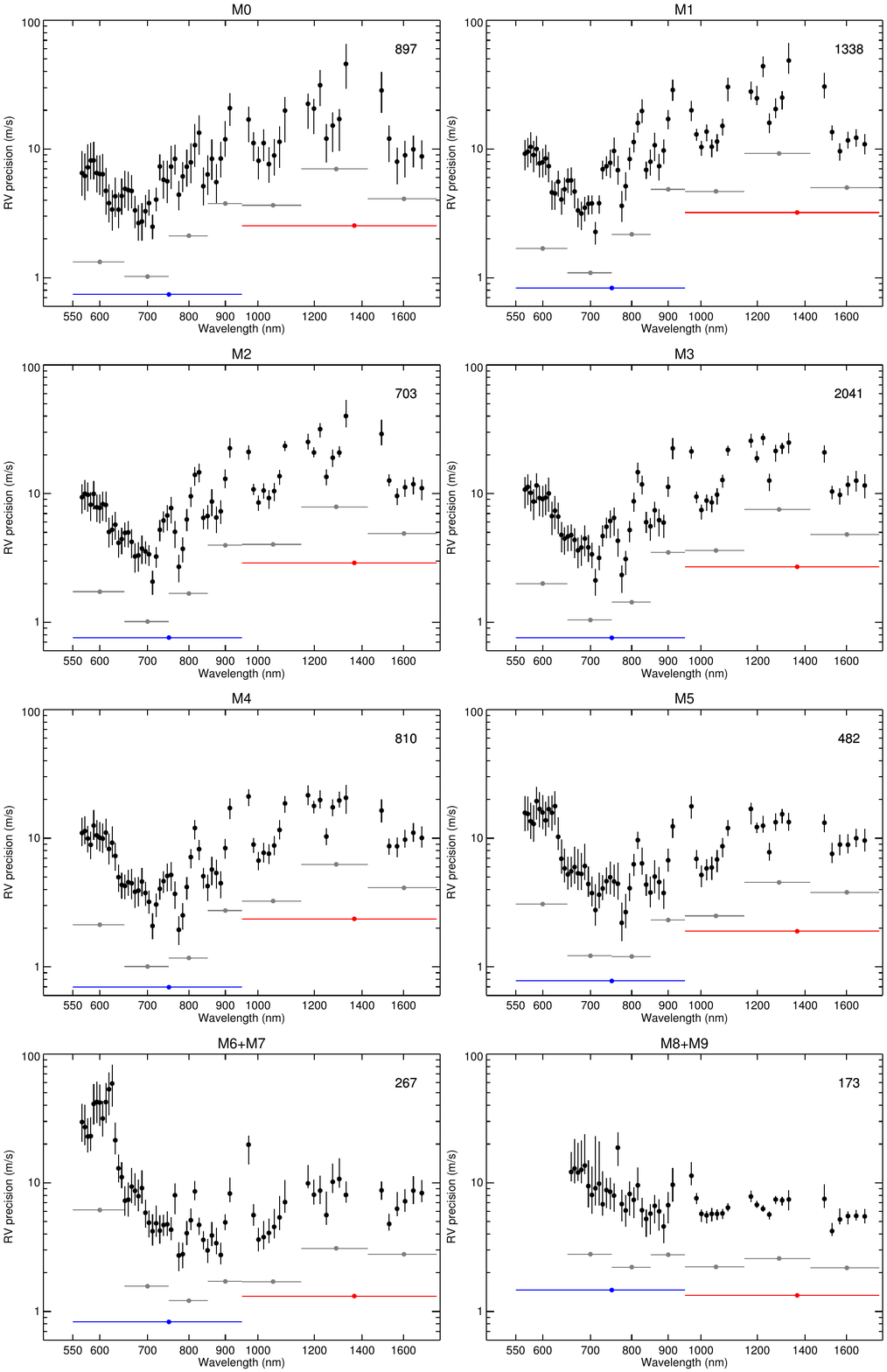}}
  \caption{\label{fig:accuracy}Empirical RV precision ($\delta
    \varv_{\rm rms}$) for individual spectral orders of CARMENES
    observed in the sample stars. Each panel shows for each order the
    median value for all observations taken of stars with spectral
    types M0 and M0.5 (upper left panel), M1 and M1.5 (upper right
    panel), and so forth. Error bars mark the 25th and 75th percentile
    of all observations. Gray points show quadratically added $\delta
    \varv_{\rm rms}$ for 100 or 200\,nm wide spectral windows, as
    indicated by the horizontal bars. Blue and red symbols show the
    CARMENES VIS and NIR spectral ranges. The number of observations
    per spectral type is given in the upper right corner of each
    panel.}
\end{figure*}

For all CARMENES observations, we have empirical values of $\delta
\varv_{\rm rms}$ for each individual spectral order. For details on
the calculation of $\delta \varv_{\rm rms}$ , we refer to
\citet{Zechmeister17}, where it is called $\epsilon_{\varv}$. In short,
our pipeline fits each individual spectrum to a coadded template
calculated from all our observations of that star. The value of
$\delta \varv_{\rm rms}$ is the uncertainty of the optimal radial
velocity in this fit. We begin our analysis of the RV precision by looking
into its dependence on wavelength for different M-star spectral
types. For this analysis, we used 6625 observations that were taken
with both CARMENES channels simultaneously. Observations were only
considered for the 261 stars that were observed more than five times
with each of the channels. Of these data, the values of $\delta
\varv_{\rm rms}$ differ even for stars of the same spectral types
because the individual stars are not equally bright and they are
observed under varying sky conditions, which results in different
S/Ns. Furthermore, the quality factor can differ between stars of the
same spectral type because of rotation. In order to make our
observations comparable, we rescaled the RV precisions $\delta
\varv_{\rm rms}$ for each individual spectrum according to S/N and
$\varv\,\sin{i}$. Before doing so, we checked in the individual values
that $\delta \varv_{\rm rms}$ could be described as a function of both
S/N and $\varv\,\sin{i}$. In both cases, we found clear relations for
all stars of a given spectral type by comparing $\delta \varv_{\rm
  rms}$ in one order. The relations are identical between the
different spectral types. As expected, we find that $\delta \varv_{\rm
  rms}$ is directly proportional to S/N$^{-1}$. For the dependence on
$\varv\,\sin{i}$, we find a relation of $\delta \varv_{\rm rms}
\propto (\varv\,\sin{i})^{0.6}$. We scaled the values of $\delta
\varv_{\rm rms}$ between different stars and observations to a
fiducial observation of a slowly rotating star observed at an S/N of
150 per resolution element in the $J$ band.

We show in Fig.\,\ref{fig:accuracy} the rescaled RV precision $\delta
\varv_{\rm rms}$ for all individual spectral orders for which RVs were
calculated by our analysis procedure (excluding orders with very heavy
telluric contamination or too low S/N). The wavelength range we
considered for this analysis is 550--1700\,nm. From the rescaled values,
we computed for each order the median value from all observations of
stars with similar spectral types. These are the black points plotted
in the individual panels of Fig.\,\ref{fig:accuracy}. The error bars
shown are the 25th and the 75th percentile of all observations, that is,
half of all observations for any given spectral type and spectral
order fall within the range of the error bars. The number of
observations used for each spectral type plot is given in upper right
corner of each panel.

The information from individual orders is very useful to assess the
amount of information in small spectral regions. To gain a better idea
about the RV precision calculated from larger wavelength areas, we
quadratically added the RV information from different spectral orders in
100\,nm (VIS) or 200\,nm (NIR) bands. This choice is rather arbitrary,
but it partially reflects the fact that at constant spectral
resolution, and at wavelengths that are longer by a factor of two, the
same number of resolution elements are contained in a range that is a
factor of two longer in units of wavelength. In the NIR channel, our
200\,nm chunks are also similar to the $Y$, $J$, and $H$ bands. Their
values are shown as gray circles with horizontal lines that mark the
wavelength ranges covered in each band. Finally, we quadratically
added RV information across the wavelength ranges that are covered by the CARMENES VIS
and NIR channels in blue and red, respectively.

From our general considerations about S/N and the distribution of
spectroscopic features (Section\,\ref{sect:Introduction}), we expect
that in early-M type stars the larger amount of spectroscopic
information at shorter wavelengths leads to better RV precision at
visual wavelengths than at infrared wavelengths. In late-M type stars,
however, the general lack of photons at visual wavelengths must lead
to a severe loss of information so that the RV precision at infrared
wavelengths improves relative to the performance at visual
wavelengths. This picture is consistent with the empirical results we
obtain from Fig.\,\ref{fig:accuracy}. In all stars of spectral type
M0--M5, the RV precision improves from the shortest wavelengths at 550\,nm
toward 700\,nm, where much  RV information is contained in the TiO
band that starts at 706\,nm. The next spectral orders up to about
900\,nm perform somewhat worse, but still relatively well. We find
another region of excellent RV precision around 770\,nm, where the TiO
system that sets in around 760\,nm is no longer contaminated by the
oxygen A band. Spectral orders at wavelengths longer than 900\,nm
carry significantly less RV information than those at shorter
wavelengths in stars of spectral types M0--M5. Specifically, when we
combined all RV information in the wavelength ranges covered by
CARMENES VIS and NIR (550--960\,nm and 960--1700\,nm, respectively),
we found that in stars of spectral type M5, the RV information content
is about a factor of 2.5 higher in the VIS channel than in the NIR
channel.

For stars of spectral type M6 and later, we can observe in
Fig.\,\ref{fig:accuracy} the influence of the spectral energy distribution on
RV precision. In our M6 and M7 stars, we can still compute RVs at wavelengths
down to 550\,nm, but the poor S/N leads to typical RV precisions exceeding
20\,m\,s$^{-1}$ (the spectral orders at wavelengths shorter than 650\,nm are
no longer computed in M8 and M9 stars). Nevertheless, the RV precision we
observe in stars of spectral types M6--M9 is still relatively high (the values
are low) at red visual wavelengths. In M6 and M7 stars, most RV information
is available at wavelengths 700--900\,nm, and the combined VIS channel
information outperforms the NIR channel information by approximately
50\,\%. In our combined sample of M8 and M9 stars, the VIS and the NIR channel
are roughly equal in terms of RV information. In particular, these stars still
show a high RV signal at 800--900\,nm. Similar performance is reached in the
$Y$ and $H$ bands; the $J$ band provides slightly less information. 

\subsection{Spectroscopic information and signal-to-noise ratio}

\begin{figure}
  \centering
  \resizebox{\hsize}{!}{\includegraphics{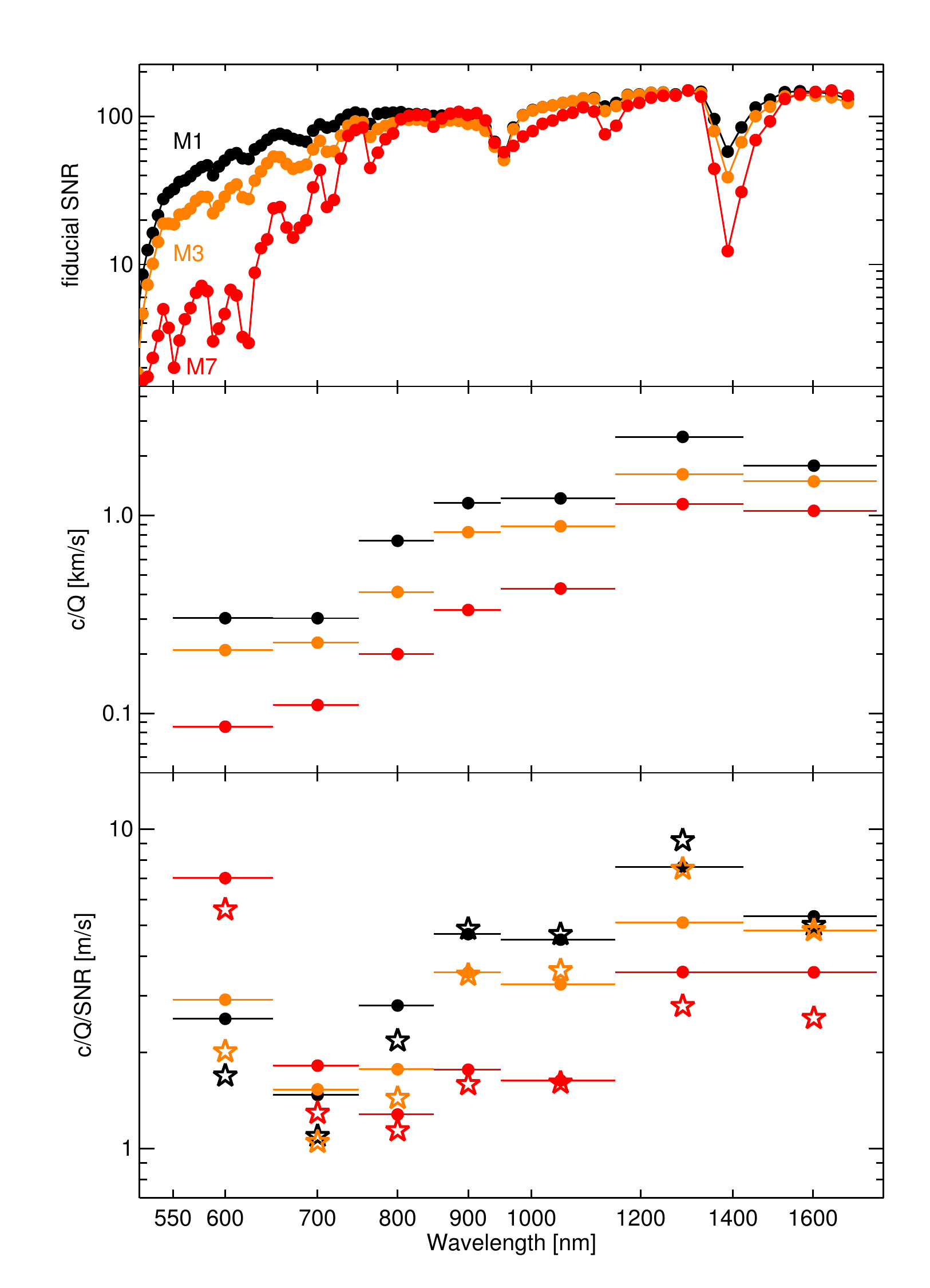}}
  \caption{\label{fig:acc_comp}Comparison between empirical values for
    RV precision $\delta \varv_{\rm rms}$ and its constituents S/N and
    $c/Q$. \emph{Top panel}: normalized S/N from individual spectra of
    GX~And (M1), Luyten's star (M3.5), and Teegarden's star (M7).
    \emph{Middle panel}: cumulative spectroscopic information $c/Q$
    calculated from coadded spectra of the three example stars.
    \emph{Bottom panel}: ratio between $c/Q$ and S/N (filled circles)
    together with empirical values of $\delta \varv_{\rm rms}$ (open
    stars, see Fig.\,\ref{fig:accuracy}).}
\end{figure}

As next step, we aim to distinguish the factors of $\delta \varv_{\rm
  rms}$, $Q,$ and S/N, and individually show them as a function of
wavelength. Our goal here is to see in which way the influence of S/N
and the amount of spectral information, $Q$, determines the RV
precision in M dwarfs. As an example of the S/N as a function of
wavelength, we show normalized S/N curves for three stars in the top
panel of Fig.\,\ref{fig:acc_comp} as observed with CARMENES. For our
example, we again selected the three stars shown in our spectroscopic
atlas: M1: GX~And (black), M3.5: Luyten's star (orange), and M7:
Teegarden's star (red). The S/N curves are normalized to S/N\,=\,150
at $J$ band. We clearly see the lack of signal at short wavelengths in
the very cool M7 star. For the S/N curve, we used all available
spectral orders, including those with strong telluric absorption. Thus,
the S/N here is a combination of the stellar flux and telluric
absorption.

In the center panel of Fig.\,\ref{fig:acc_comp}, we visualize the
amount of spectroscopic information ($c/Q$) that we calculated from
observed spectra of the three example stars. We computed this value as
the cumulative spectral gradient across the wavelength ranges, see
Eq.\,(6) in \citet{1996PASP..108..500B}. The wavelength regions are
the same 100\,nm and 200\,nm wide windows as in
Fig.\,\ref{fig:accuracy}.  For this computation, we used the coadded
spectra from all observations of our three example stars; see
\citet{Zechmeister17} for details. During coadding, wavelength
regions contaminated by telluric lines were masked. As discussed below,
this is a potential source of uncertainty particularly at long
wavelengths, where telluric contamination is significant. For the
computation of $c/Q$, the photon noise in the coadded spectra is
negligible. The number of observations used per star at VIS/NIR are
GX~And: 181/163; Luyten's star: 691/677;  and Teegarden's star: 48/43. In
all three example stars, there is significantly more RV information at
shorter wavelengths. Furthermore, the spectra of cooler stars carry
more RV information than the spectra from early-M dwarfs because they
exhibit more spectral features from molecular absorption.

In the bottom panel of Fig.\,\ref{fig:acc_comp}, we compare RV
uncertainties for CARMENES spectra following two different approaches:
a) assuming a signal-to-noise ratio of $S/N = 150$ at the $J$
band, we
calculate the amount of spectroscopic RV information, $(c/Q)/S/N$,
using a coadded high-quality spectrum for each of our example stars
(filled circles; see Eq.\,\ref{eq:q-factor}); and b) from a large
number of observations (e.g., 2041 CARMENES observations of M3 and
M3.5 stars), we calculated the median of their internal RV
uncertainties, $\delta\varv_{\rm rms}$, (open stars) from fitting
every individual spectrum to a coadded template of the observed star
as shown in Fig.\,\ref{fig:accuracy} (scaled to $S/N = 150$ at the
$J$ band and slow rotation, see above). With this comparison, we aim
to reveal any systematic problems in our RV determination. In general,
the values calculated following the two different approaches compare
very well; for all wavelength bands and spectral types, they agree
within a factor of two; most of the values agree even
better. Consistent with our calculations from spectral fitting, our
analysis of the S/N and the amount of spectrosopic
features shows that in all M dwarfs down to M7, the highest RV
precision can be reached at around 700\,nm or 800\,nm. At 600\,nm,
observations of M dwarfs show considerably lower RV precision because
of their very low S/N. At 700 and 800\,nm, however, the signal is
still high enough to provide higher RV precision than regions at
longer wavelengths (it levels out only in M8/9 stars, see
Fig.\,\ref{fig:accuracy}). The reason for this is the substantial loss
in spectroscopic features from VIS to NIR wavelengths.

One important source of uncertainty in our calculation of $Q$ is the
occurrence of telluric features, in particular, at infrared
wavelengths. In the bottom panel of Fig.\,\ref{fig:acc_comp}, we
observe that the values of $(c/Q)/S/N$ are lower than our measured RV
precision, in particular in the $J$ band. A possible explanation is
that our coadded spectra are contaminated by telluric features. These
features can contribute to $Q$ because they introduce additional
spectral gradients. They also increase the value of our empirically
measured RV uncertainty because the individual observations and the
coadded spectra do not resemble each other at these features. The
result of poorly treated telluric lines is therefore a bias toward
higher values in our empirical RV uncertainty and
toward lower values in the uncertainty estimate
$(c/Q)/S/N$. We identify the treatment of telluric lines as one of the
main areas where infrared (and also red visual) RV precision can be
improved, but we also note that in the presence of a sufficient amount
of spectroscopic features like in our M7 star, this appears to be a
negligible problem.

There are a number of instrumental effects that can affect our
analysis of precision vs.\ $\lambda$ from the CARMENES data. These
effects include global efficiency offsets between the VIS and the NIR,
varying spectrograph throughput and detector efficiency, or
$\lambda$-dependent read noise. The difference in spectral resolution
between the VIS and the NIR channel slightly penalizes the NIR channel
RV precision, but the expected effect is on the order of only $10\,\%$
and not significant for our conclusions. Although our analysis was
carried out using observed data and S/N values that were determined
empirically, we note that instrumental effects may be
hidden. However, the two CARMENES channels are very similar in their
optical design and performance, and our results can probably be
taken as fairly typical for other high-resolution spectrographs.

\section{Summary}
\label{sect:Summary}

After all targets of the CARMENES survey were observed at least once,
we derived basic spectroscopic information for each target about
radial velocity, H$\alpha$ emission, and projected surface
rotation. We provide this information together with the list of the
324 targets of our CARMENES M-dwarf survey in this work. For each
star, we provide one spectrum observed with the two CARMENES channels, which
cover the wavelength range 550--1700\,nm. This is the first large
library of high-resolution near-infrared spectra of low-mass stars. We
also show a spectroscopic atlas of three example M dwarfs (M1, M3.5,
M7) and compare their spectra to a telluric standard and synthetic
models at very high spectral resolution. We find that the
synthetic spectra in general succeed in predicting the main features in M-dwarf
spectra, but the amount of spectroscopic features is sometimes higher
than in the models.

Our analysis of stellar rotation and activity adds precise
$\varv\,\sin{i}$ measurements to the available catalogues of M-dwarf
rotation. Several of our stars rotate as fast as $\varv\,\sin{i} =
20$\,km\,s$^{-1}$ and more. Our values are included in the more
detailed analysis of rotation and activity by \citet{Jeffers17}. They
are consistent with the general picture of M-dwarf activity. We find
some very early M-dwarfs that exhibit significant rotation but no
H$\alpha$ emission. These stars have rotation periods close to $P =
10$\,d, and they are probably rotating slowly enough to explain
the lack of H$\alpha$ emission \citep[see also][]{Jeffers17}.

We employed several thousand observations of M dwarfs taken at visual
and near-infrared wavelengths to calculate the RV uncertainty as a
function of wavelength for individual spectral subtypes. We conclude
that the wavelength range 700--900\,nm
provides an excellent source of RV information for all M dwarfs. At shorter
wavelengths, the RV precision is lower and deteriorates toward later
spectral types. At longer wavelengths, the RV precision is significantly
lower up to spectral types M6/M7. The turnover point where RV
information content at NIR wavelengths becomes comparable to the one
at VIS wavelengths is located at spectral types as late as M8/M9.

Our results answer the question at which wavelength the best RV
precision can be reached in observations of M dwarfs with a stabilized
spectrograph: the optimal range for M-dwarf RV spectroscopy is the
spectral range 700--900\,nm. Calculating the spectroscopic quality
factor $Q$ and the S/N for three example stars, we showed that our
conclusions about the RV uncertainties are consistent with the
spectroscopic information we find in the spectra. At wavelengths
shorter than 700\,nm, the lack of photons limits the RV precision; at
wavelengths longer than 900\,nm, the amount of spectroscopic
information is so much lower than at shorter wavelengths that the
advantage in S/N cannot compensate for the loss in RV precision. Only
in the latest spectral types (M8 and M9) did we find the RV precision
at near-infrared wavelengths to match the amount of information at
shorter wavelengths.

Our result is especially interesting for other planned M-dwarf surveys
and the spectroscopic follow-up of transiting M dwarfs. Instruments
used for this purpose are often mounted at 4m class telescopes that have a
similar performance as the 3.5m telescope at Calar Alto that
is used for the
CARMENES survey. For these, very late-M dwarfs are difficult to
observe because they are extremely faint (at all wavelengths), and the
limited RV precision is further affected by their often high rotation
rates. Thus, instruments at 4m class telescopes will typically reach
an RV precision on the order of 1\,m\,s$^{-1}$ only in early- and
mid-M dwarfs, where the spectral range 700-900\,nm is the most
efficient. However, an RV precision of 2\,m\,s$^{-1}$ can be reached
at 4m class telescopes within reasonable times in many stars of
spectral type M6 and earlier. Here, NIR RV data cannot outperform red
visual RV observations, but NIR RVs can reach below the typical limit
of radial velocity jitter in M dwarfs that is on the order of
3-4\,m\,s$^{-1}$ \citep{2013A&A...549A.109B}.  The main reasons for
this jitter are corotating active regions (including their effect on
convective blueshift) and granulation, and their RV signal is expected
to be wavelength dependent. The combination of RV observations at
visual and near-infrared wavelengths is ideal to distinguish between
Keplerian signals and stellar variability. CARMENES delivers these data
across a very large wavelength range. It is therefore optimally suited
to search for low-mass planets around mid-type M dwarfs where
variability is a serious concern.

\begin{acknowledgements}
  We thank an anonymous referee for prompt attention and helpful
  comments that helped to improve the quality of this paper. CARMENES
  is an instrument for the Centro Astron\'omico Hispano-Alem\'an de
  Calar Alto (CAHA, Almer\'{\i}a, Spain).  CARMENES is funded by the
  German Max-Planck-Gesellschaft (MPG), the Spanish Consejo Superior
  de Investigaciones Cient\'{\i}ficas (CSIC), the European Union
  through FEDER/ERF FICTS-2011-02 funds, and the members of the
  CARMENES Consortium (Max-Planck-Institut f\"ur Astronomie, Instituto
  de Astrof\'{\i}sica de Andaluc\'{\i}a, Landessternwarte
  K\"onigstuhl, Institut de Ci\`encies de l'Espai, Insitut f\"ur
  Astrophysik G\"ottingen, Universidad Complutense de Madrid,
  Th\"uringer Landessternwarte Tautenburg, Instituto de
  Astrof\'{\i}sica de Canarias, Hamburger Sternwarte, Centro de
  Astrobiolog\'{\i}a and Centro Astron\'omico Hispano-Alem\'an), with
  additional contributions by the Spanish Ministry of Economy, the
  German Science Foundation through the Major Research Instrumentation
  Programme and DFG Research Unit FOR2544 ``Blue Planets around Red
  Stars'', the Klaus Tschira Stiftung, the states of
  Baden-W\"urttemberg and Niedersachsen, and by the Junta de
  Andaluc\'{\i}a.  This work has made use of the VALD database,
  operated at Uppsala University, the Institute of Astronomy RAS in
  Moscow, and the University of Vienna. We acknowledge the following
  funding programs: European Research Council (ERC-279347), Deutsche
  Forschungsgemeinschaft (RE 1664/12-1, RE 2694/4-1),
  Bundesministerium f\"ur Bildung und Forschung (BMBF-05A14MG3,
  BMBF-05A17MG3), Spanish Ministry of Economy and Competitiveness
  (MINECO, grants AYA2015-68012-C2-2-P, AYA2016-79425-C3-1,2,3-P,
  AYA2015-69350-C3-2-P, AYA2014-54348-C03-01, AYA2014-56359-P,
  AYA2014-54348-C3-2-R, AYA2016-79425-C3-3-P and 2013 Ram\`on y Cajal
  program RYC-2013-14875), Fondo Europeo de Desarrollo Regional
  (FEDER, grant ESP2016-80435-C2-1-R, ESP2015-65712- C5-5-R),
  Generalitat de Catalunya/CERCA programme, Spanish Ministerio de
  Educaci\'on, Cultura y Deporte, programa de Formaci\'on de
  Profesorado Universitario (grant FPU15/01476), Deutsches Zentrum
  f\"ur Luft- und Raumfahrt (grants 50OW0204 and 50OO1501), Office of
  Naval Research Global (award no. N62909-15-1-2011), Mexican CONACyT
  grant CB-2012-183007.
\end{acknowledgements}

\bibliographystyle{aa}
\bibliography{refs}

\begin{appendix}

\section{Atlas of near-infrared spectra}

\begin{figure*}
  \resizebox{.97\hsize}{!}{\includegraphics[angle=90]{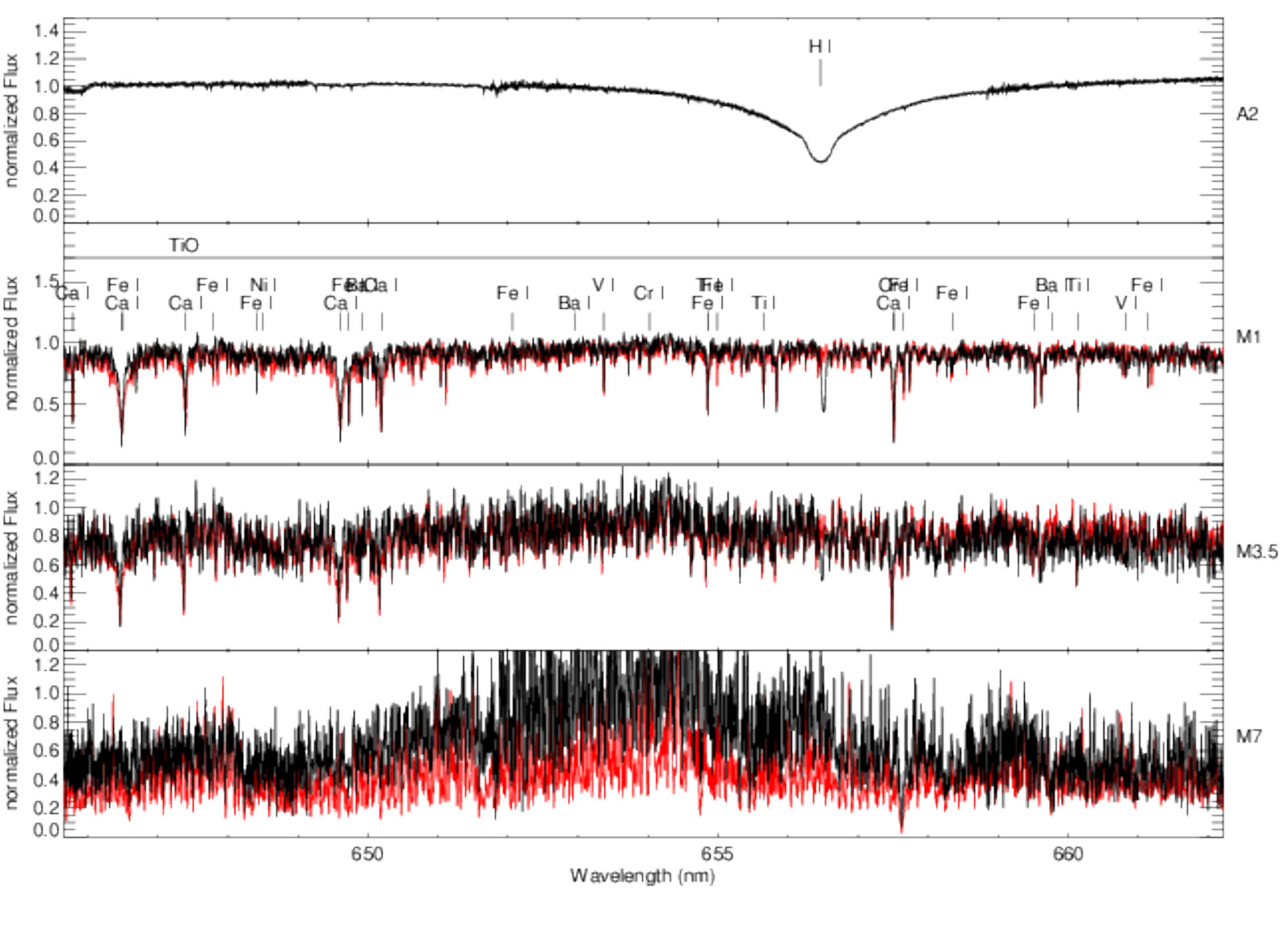}}
  \caption{\label{fig:atlas00}CARMENES spectral atlas.}
\end{figure*}

\begin{figure*}
  \resizebox{.97\hsize}{!}{\includegraphics[angle=90]{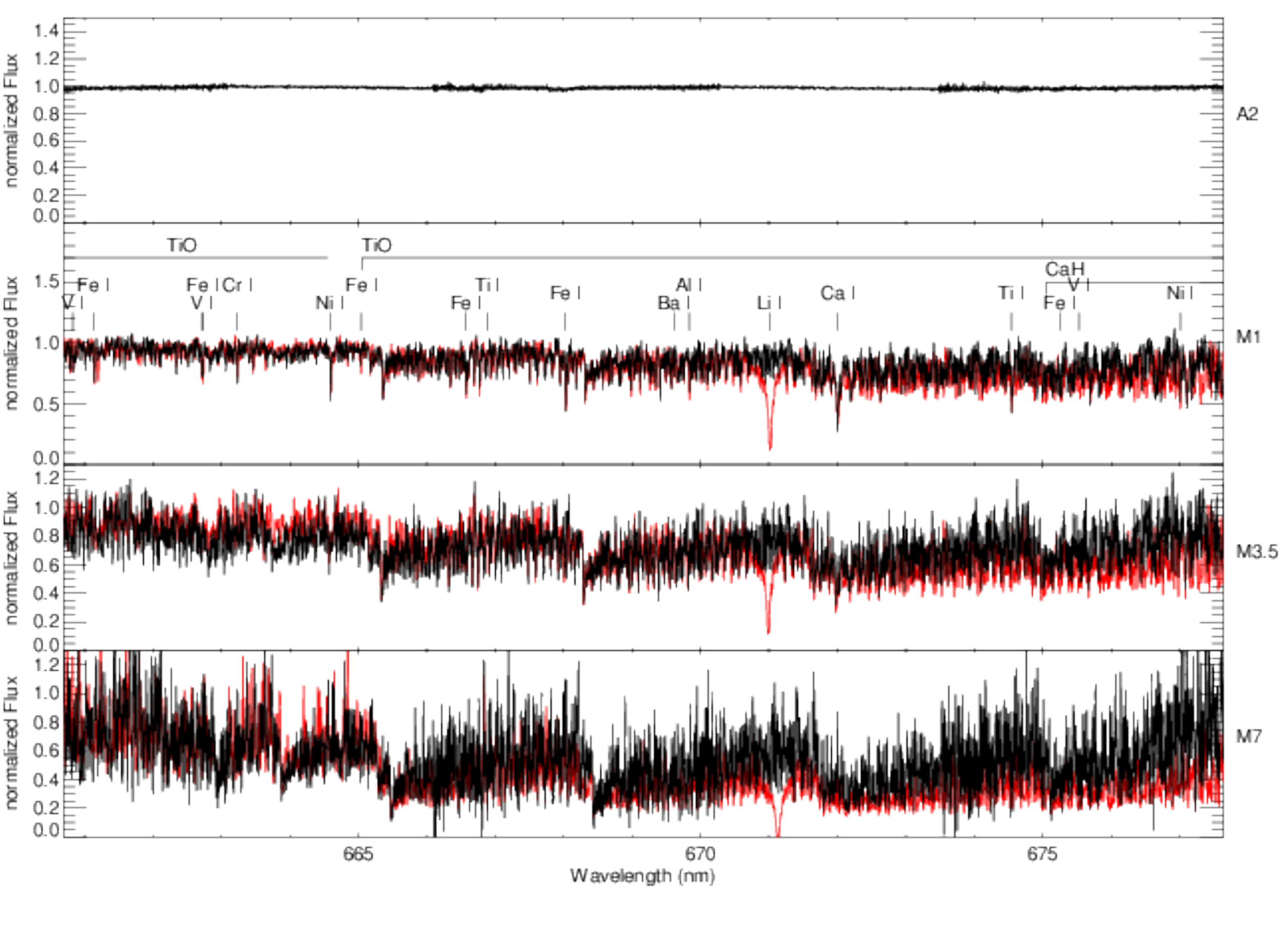}}
  \caption{\label{fig:atlas01}CARMENES spectral atlas.}
\end{figure*}

\begin{figure*}
  \resizebox{.97\hsize}{!}{\includegraphics[angle=90]{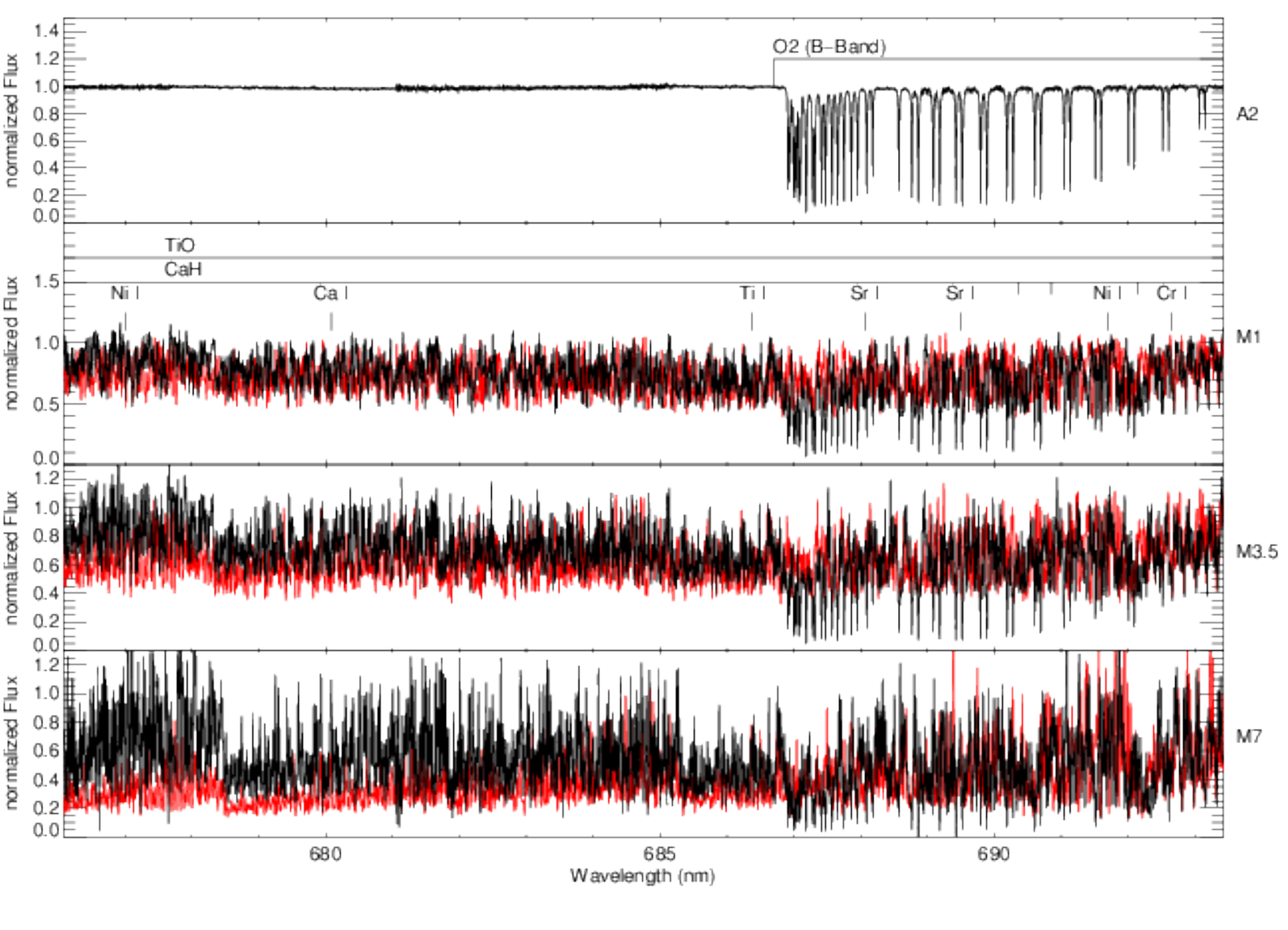}}
  \caption{\label{fig:atlas02}CARMENES spectral atlas.}
\end{figure*}

\begin{figure*}
  \resizebox{.97\hsize}{!}{\includegraphics[angle=90]{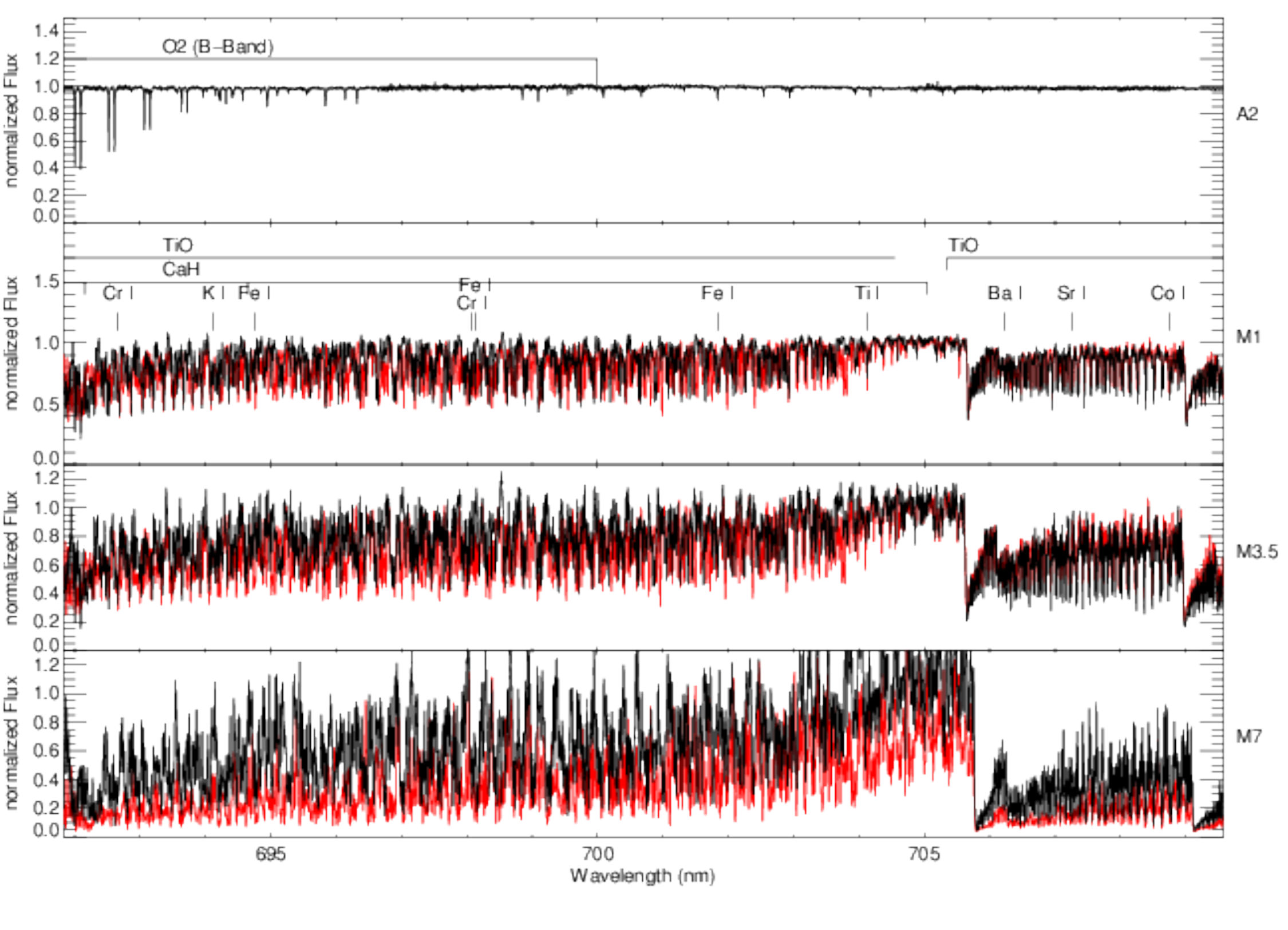}}
  \caption{\label{fig:atlas03}CARMENES spectral atlas.}
\end{figure*}

\begin{figure*}
  \resizebox{.97\hsize}{!}{\includegraphics[angle=90]{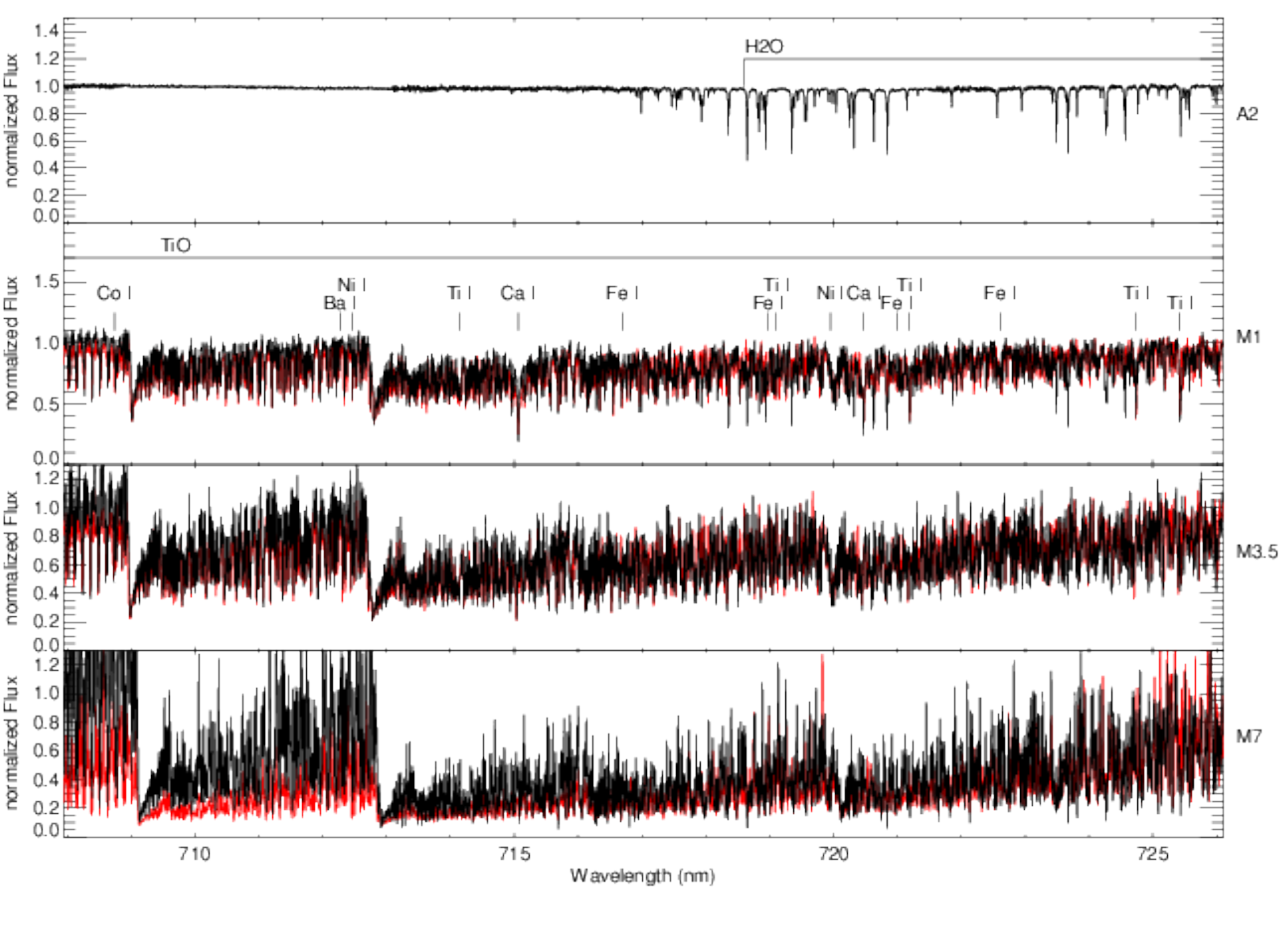}}
  \caption{\label{fig:atlas04}CARMENES spectral atlas.}
\end{figure*}

\begin{figure*}
  \resizebox{.97\hsize}{!}{\includegraphics[angle=90]{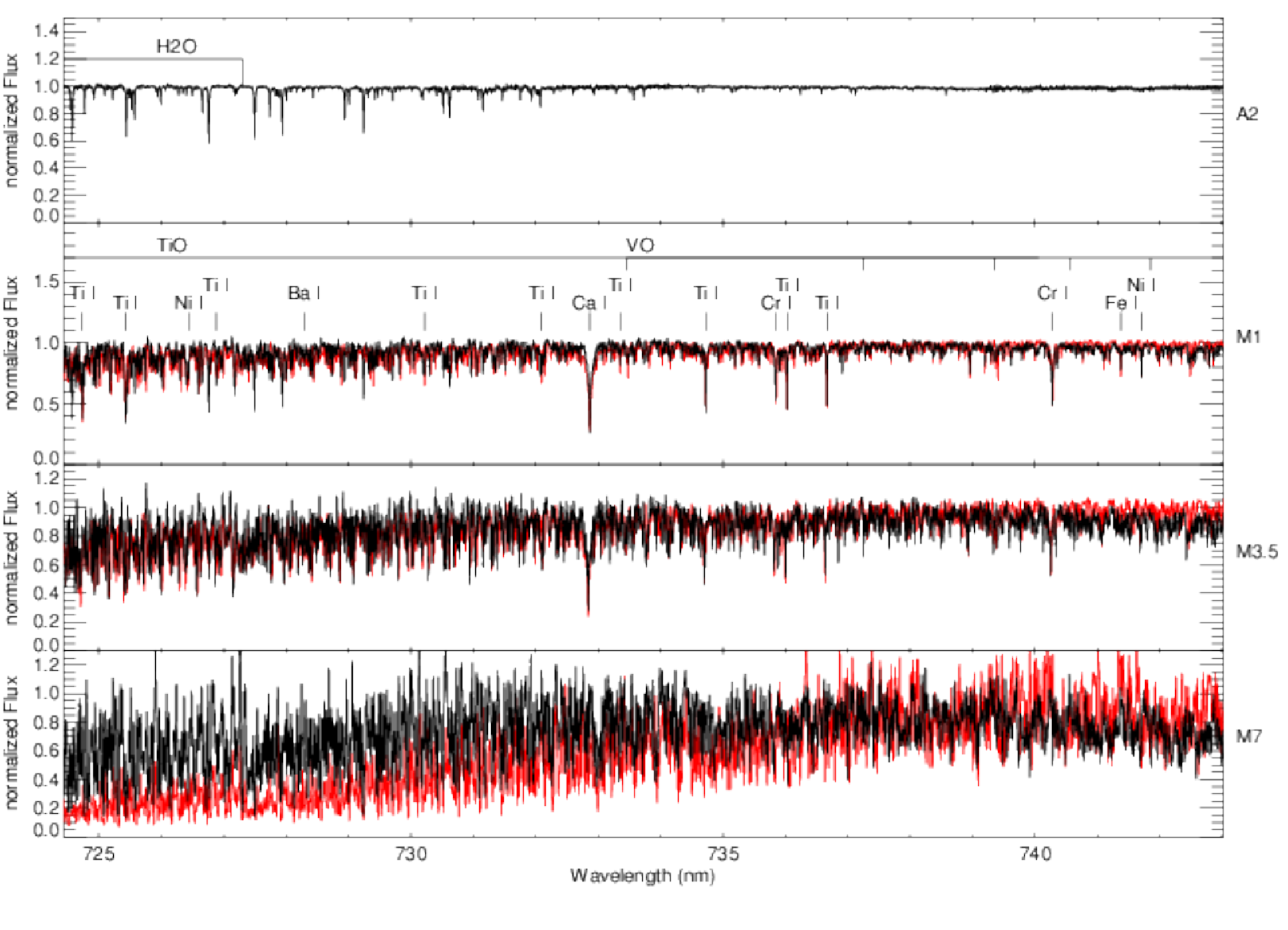}}
  \caption{\label{fig:atlas05}CARMENES spectral atlas.}
\end{figure*}

\begin{figure*}
  \resizebox{.97\hsize}{!}{\includegraphics[angle=90]{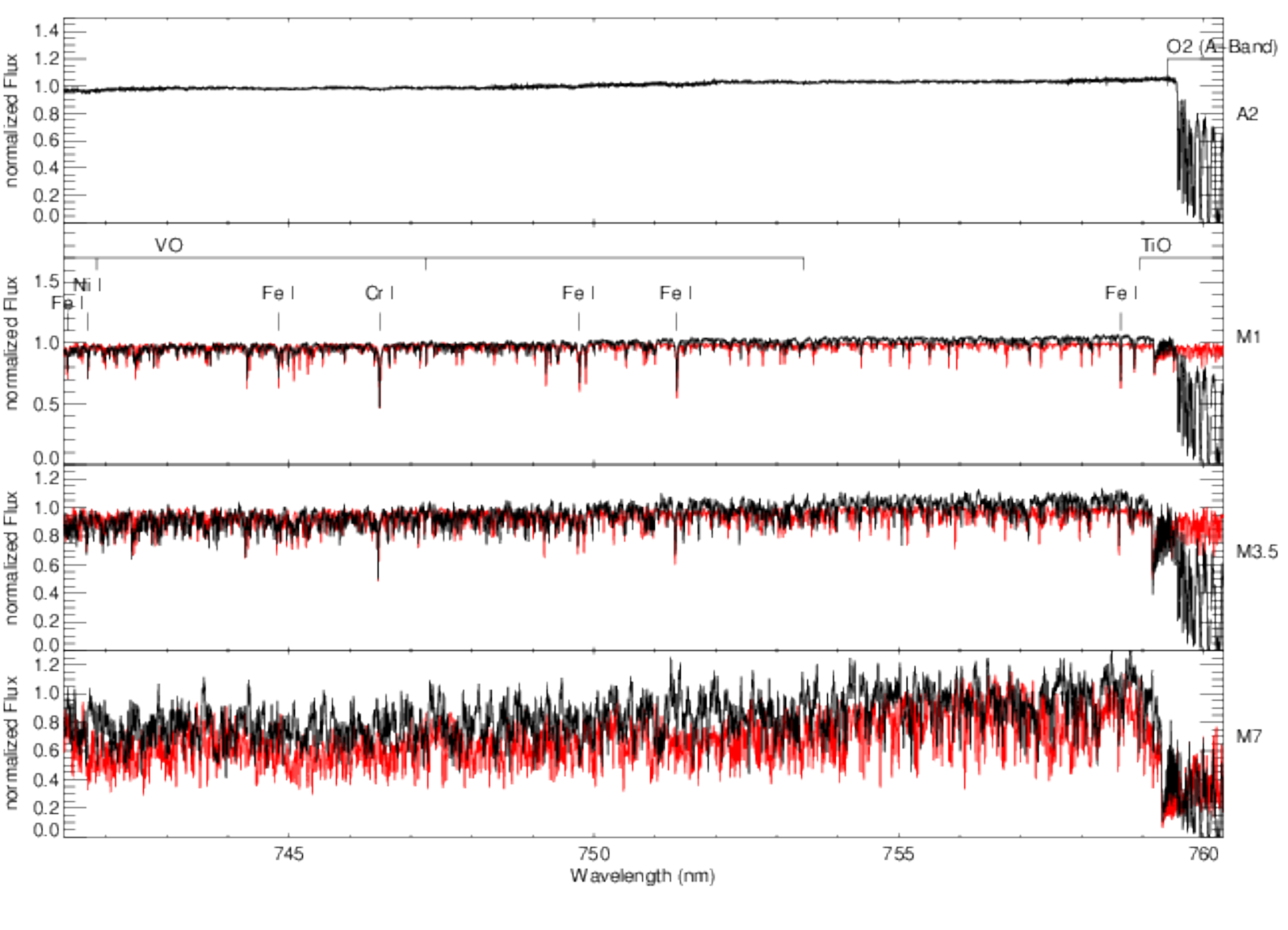}}
  \caption{\label{fig:atlas06}CARMENES spectral atlas.}
\end{figure*}

\begin{figure*}
  \resizebox{.97\hsize}{!}{\includegraphics[angle=90]{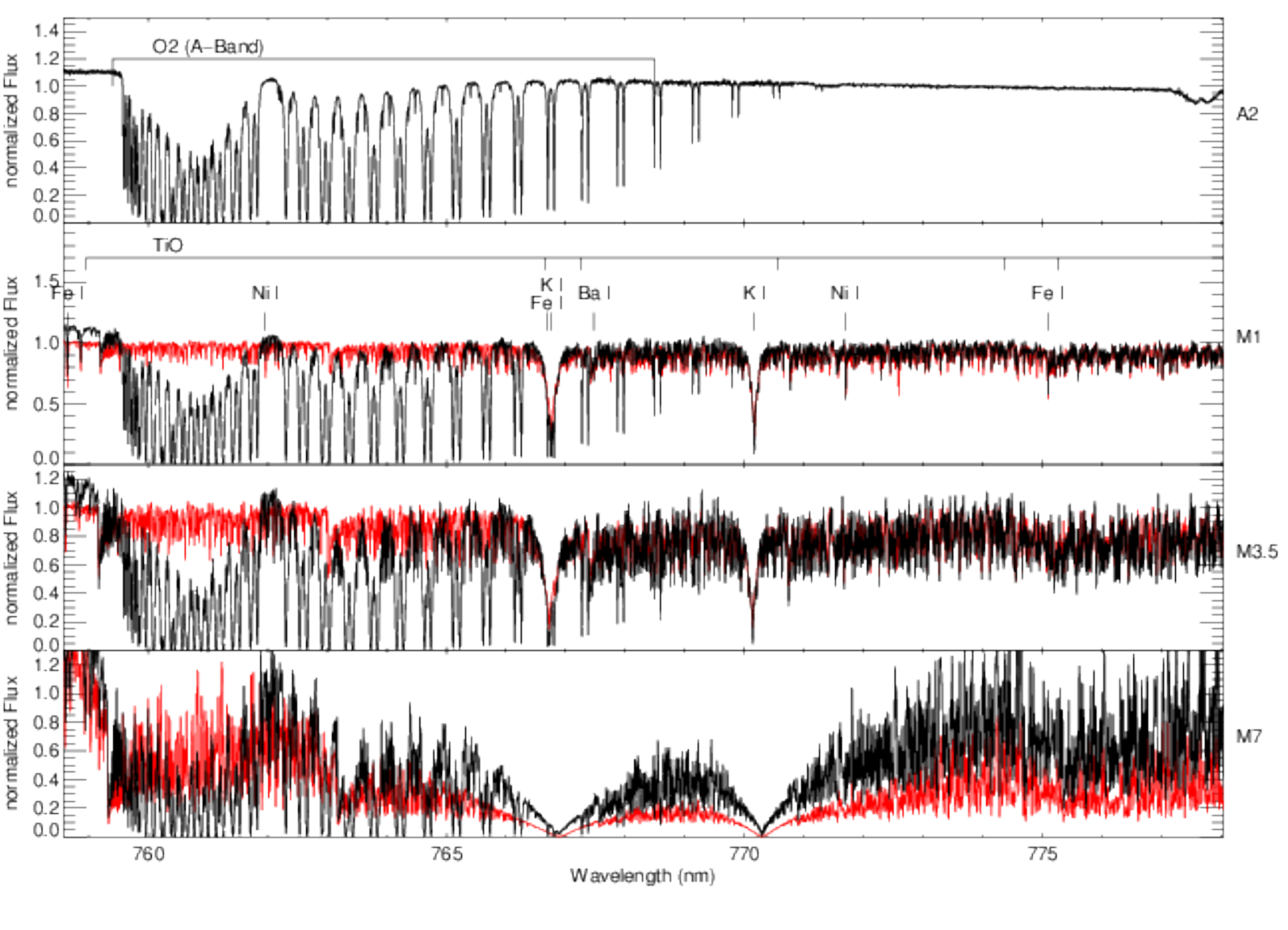}}
  \caption{\label{fig:atlas07}CARMENES spectral atlas.}
\end{figure*}

\begin{figure*}
  \resizebox{.97\hsize}{!}{\includegraphics[angle=90]{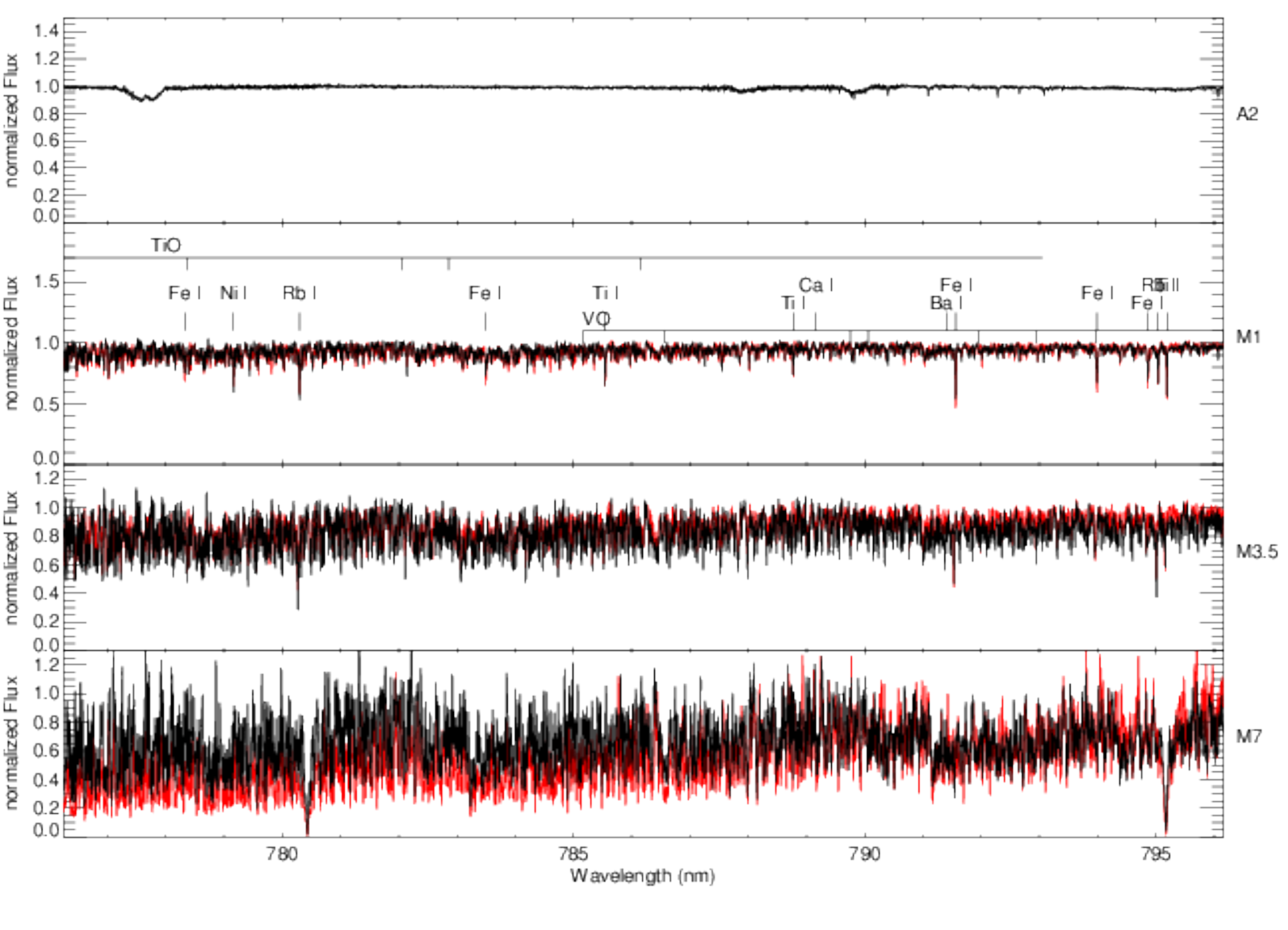}}
  \caption{\label{fig:atlas08}CARMENES spectral atlas.}
\end{figure*}

\begin{figure*}
  \resizebox{.97\hsize}{!}{\includegraphics[angle=90]{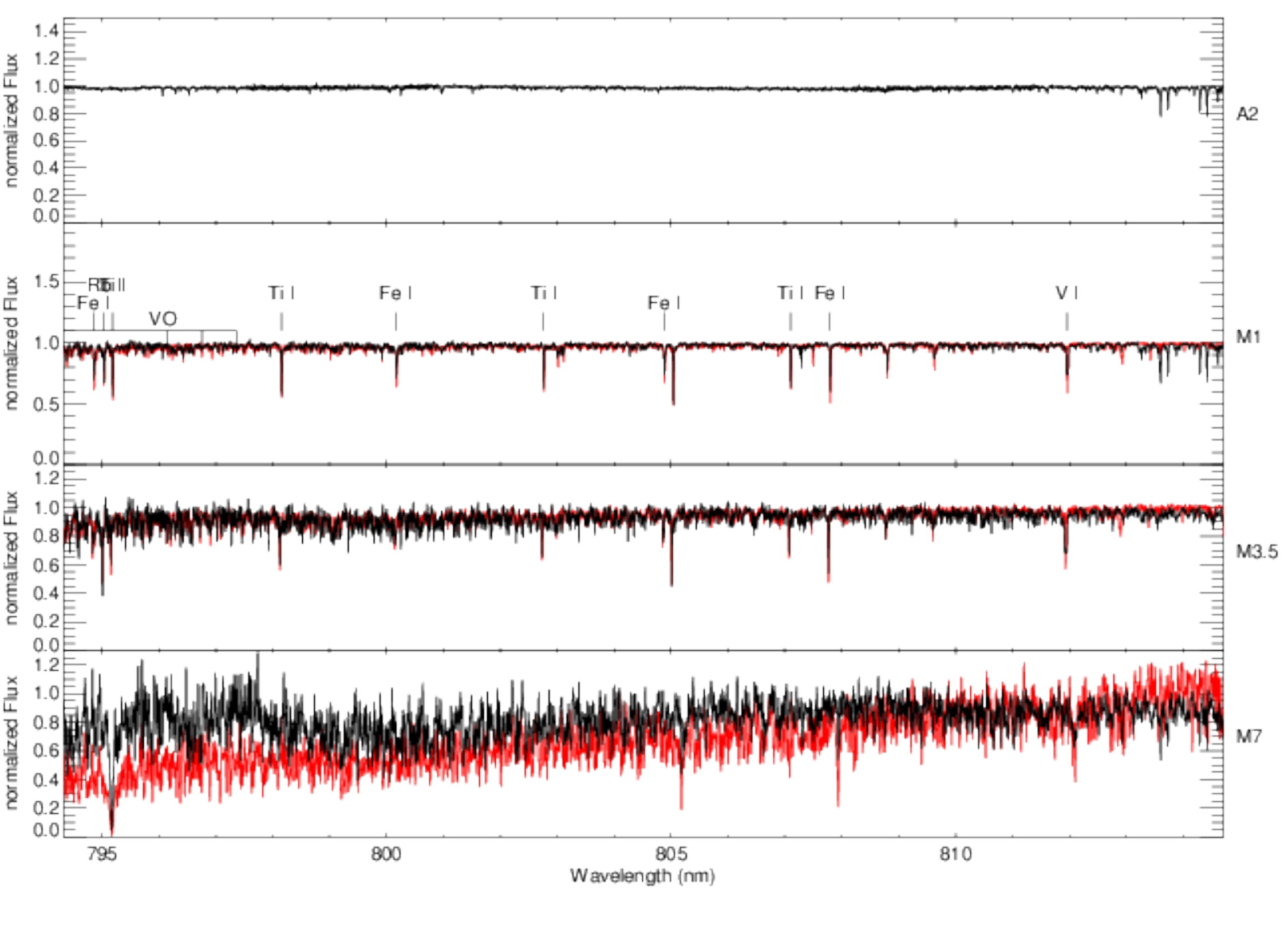}}
  \caption{\label{fig:atlas09}CARMENES spectral atlas.}
\end{figure*}

\begin{figure*}
  \resizebox{.97\hsize}{!}{\includegraphics[angle=90]{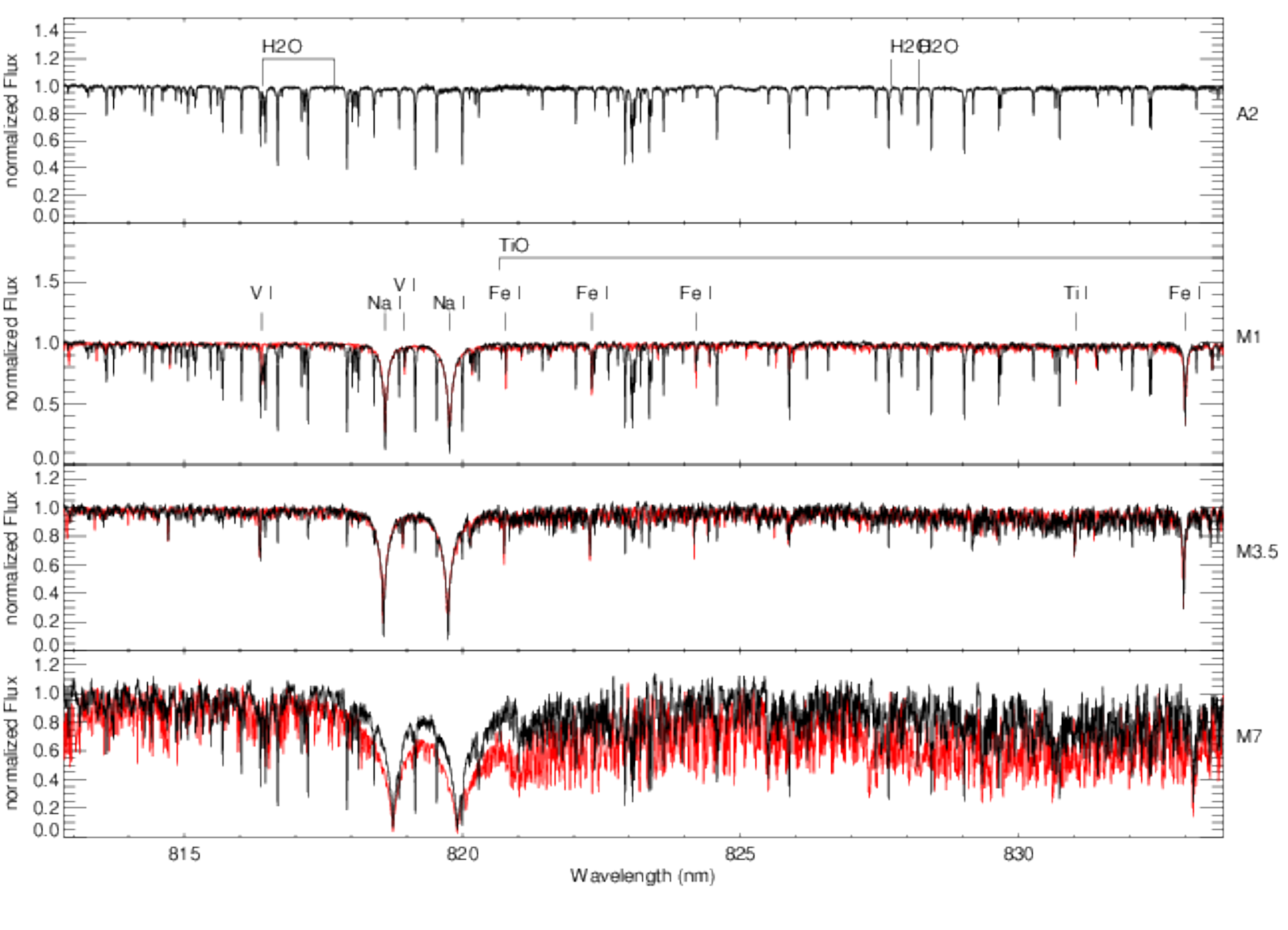}}
  \caption{\label{fig:atlas10}CARMENES spectral atlas.}
\end{figure*}

\begin{figure*}
  \resizebox{.97\hsize}{!}{\includegraphics[angle=90]{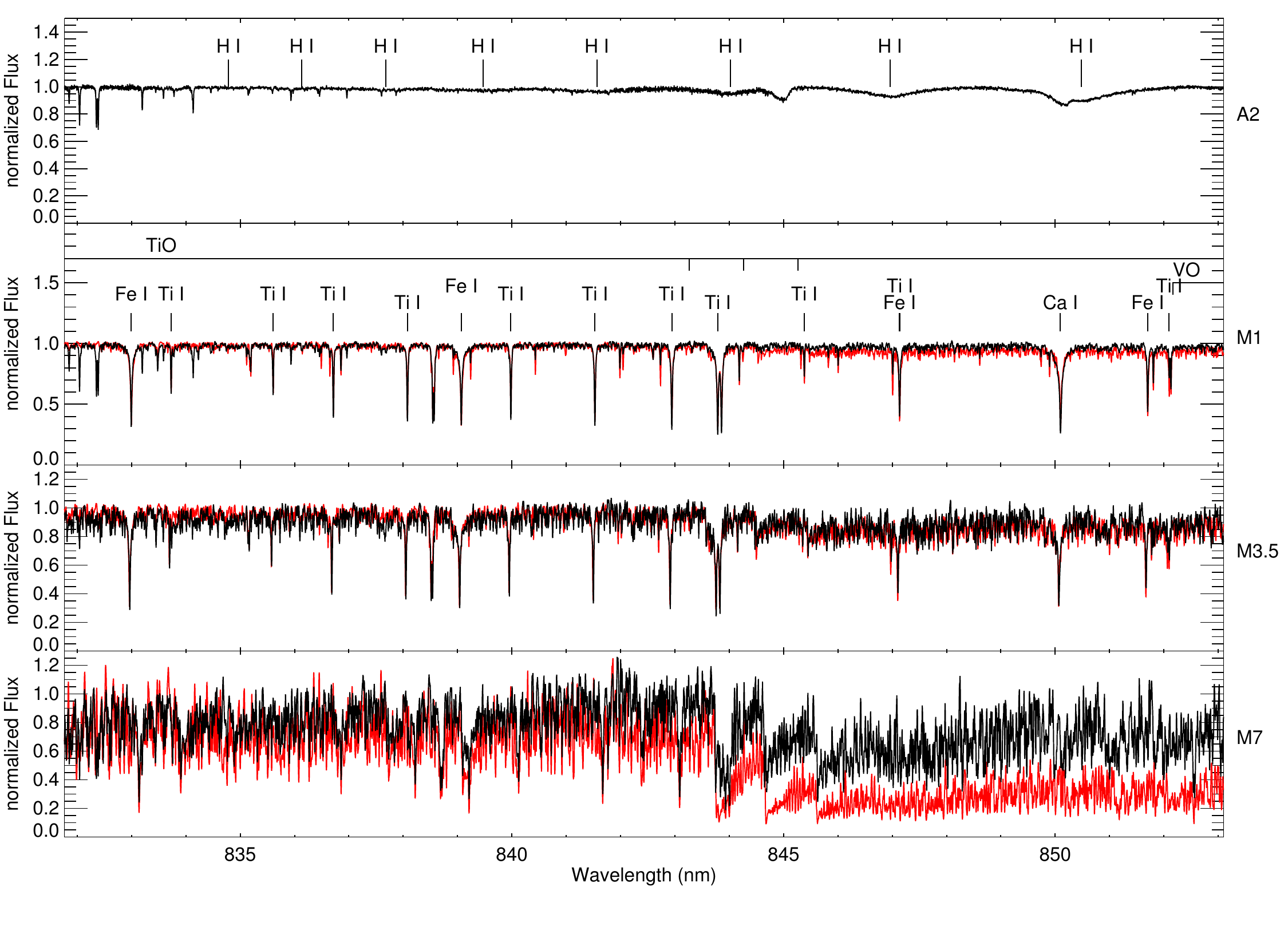}}
  \caption{\label{fig:atlas11}CARMENES spectral atlas.}
\end{figure*}

\begin{figure*}
  \resizebox{.97\hsize}{!}{\includegraphics[angle=90]{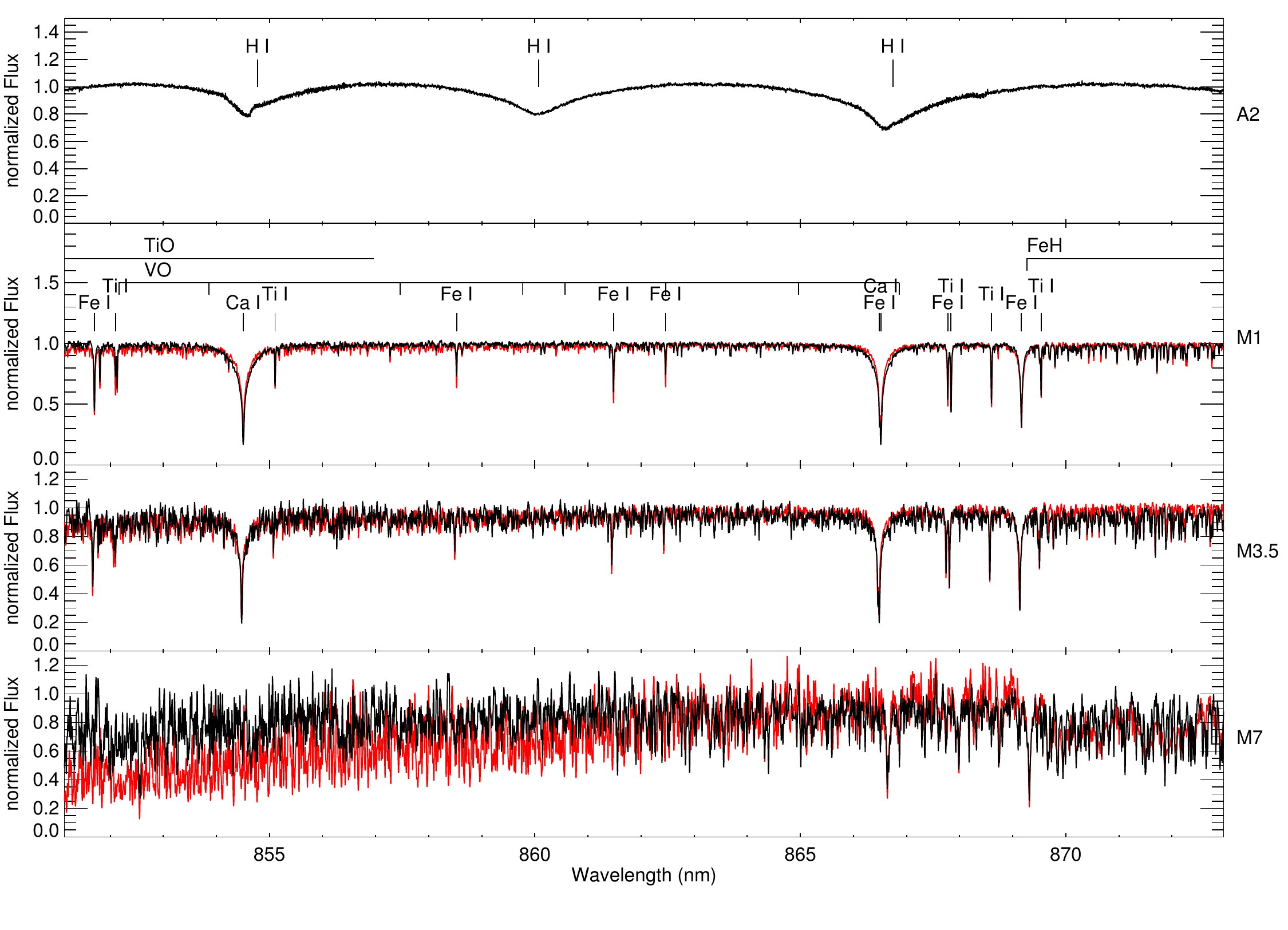}}
  \caption{\label{fig:atlas12}CARMENES spectral atlas.}
\end{figure*}

\begin{figure*}
  \resizebox{.97\hsize}{!}{\includegraphics[angle=90]{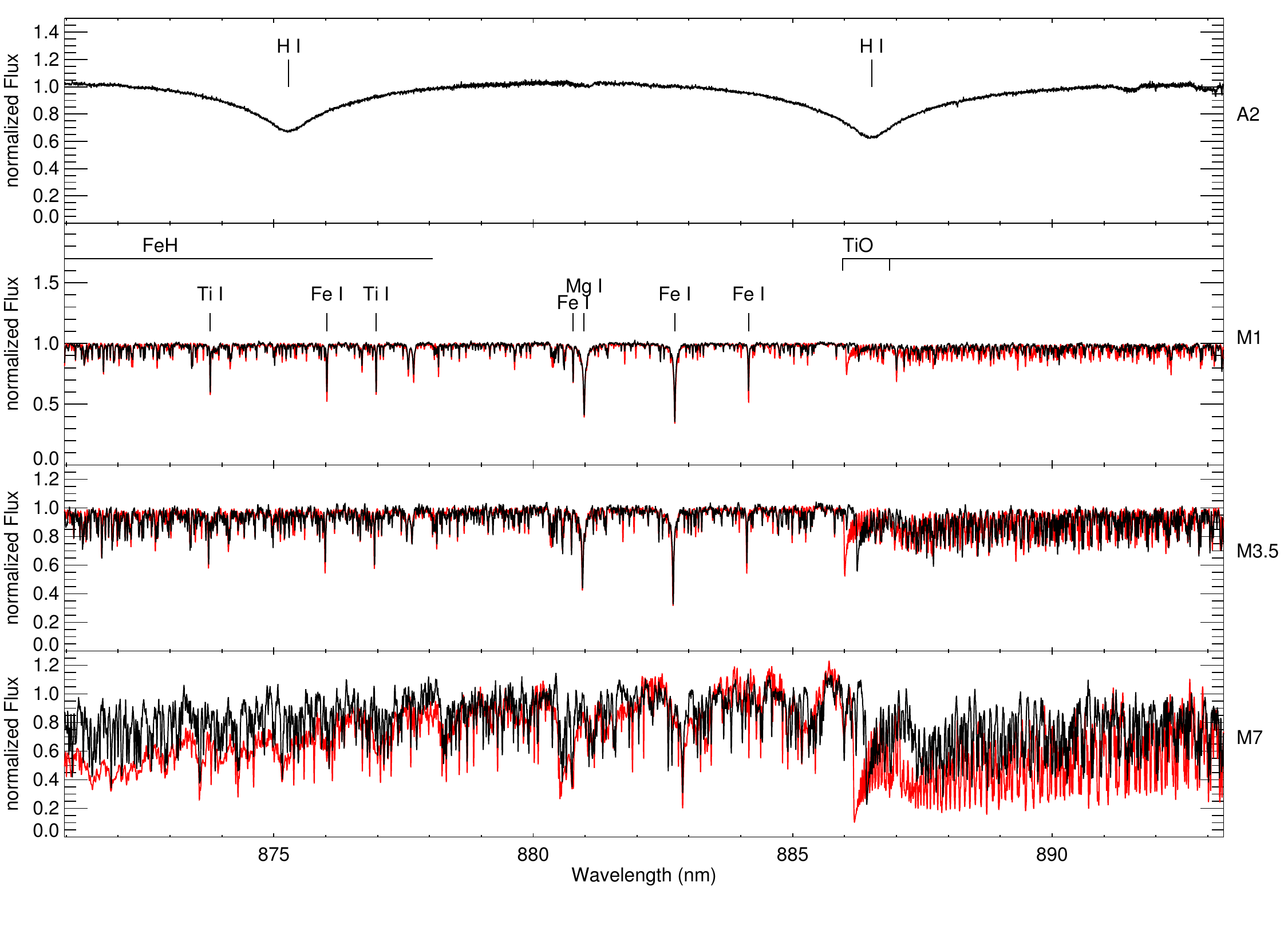}}
  \caption{\label{fig:atlas13}CARMENES spectral atlas.}
\end{figure*}

\begin{figure*}
  \resizebox{.97\hsize}{!}{\includegraphics[angle=90]{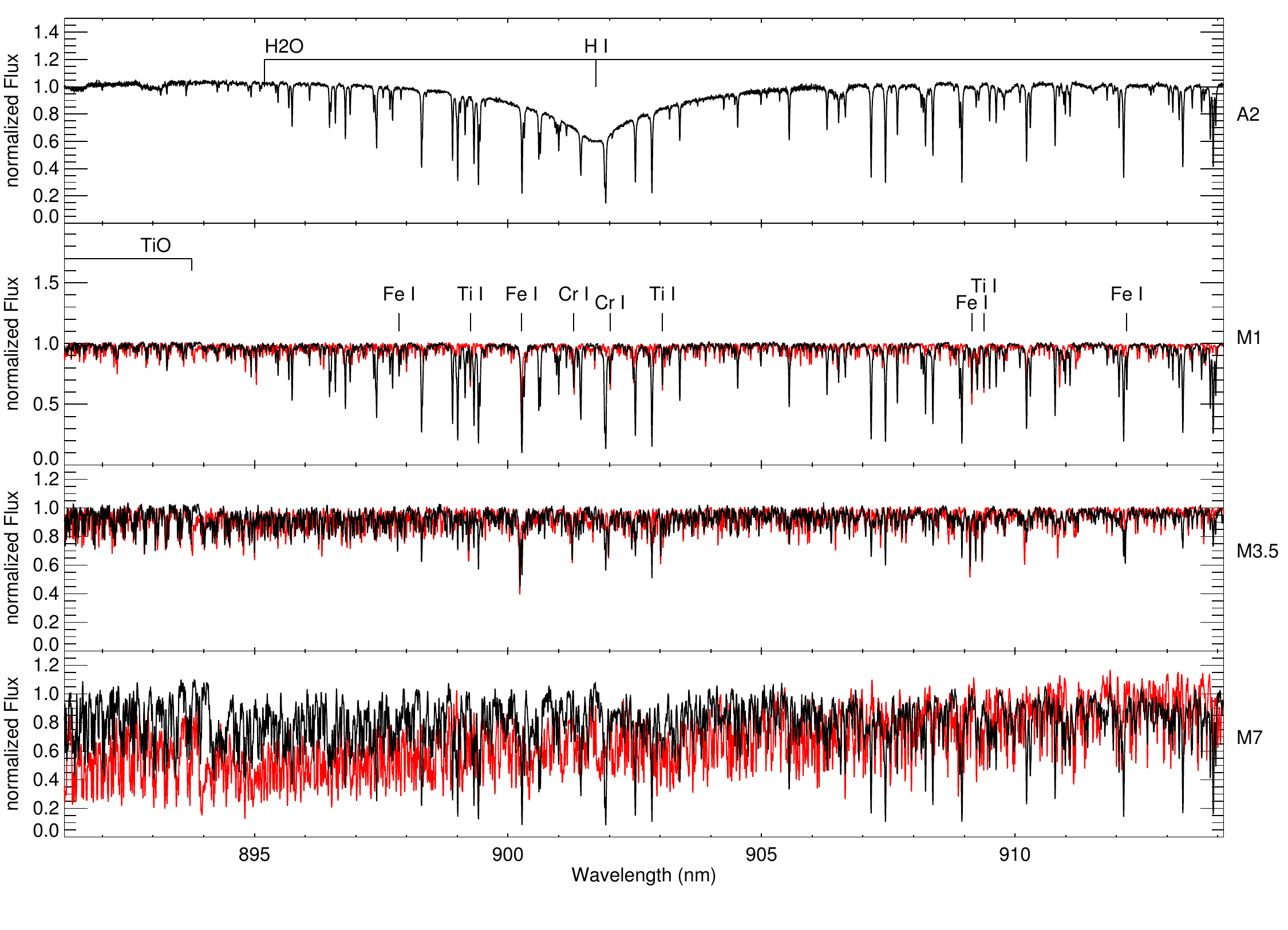}}
  \caption{\label{fig:atlas14}CARMENES spectral atlas.}
\end{figure*}

\begin{figure*}
  \resizebox{.97\hsize}{!}{\includegraphics[angle=90]{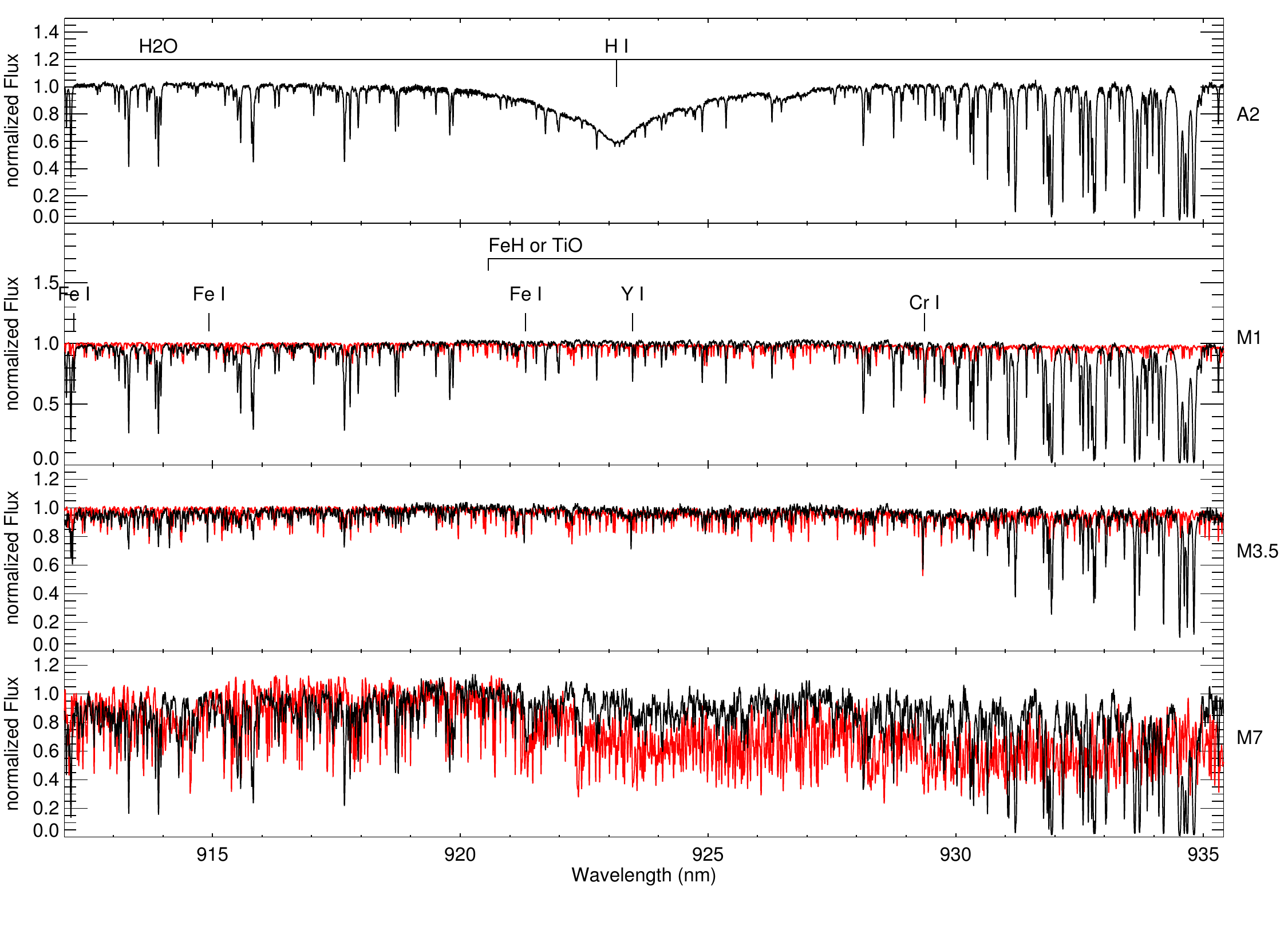}}
  \caption{\label{fig:atlas15}CARMENES spectral atlas.}
\end{figure*}

\begin{figure*}
  \resizebox{.97\hsize}{!}{\includegraphics[angle=90]{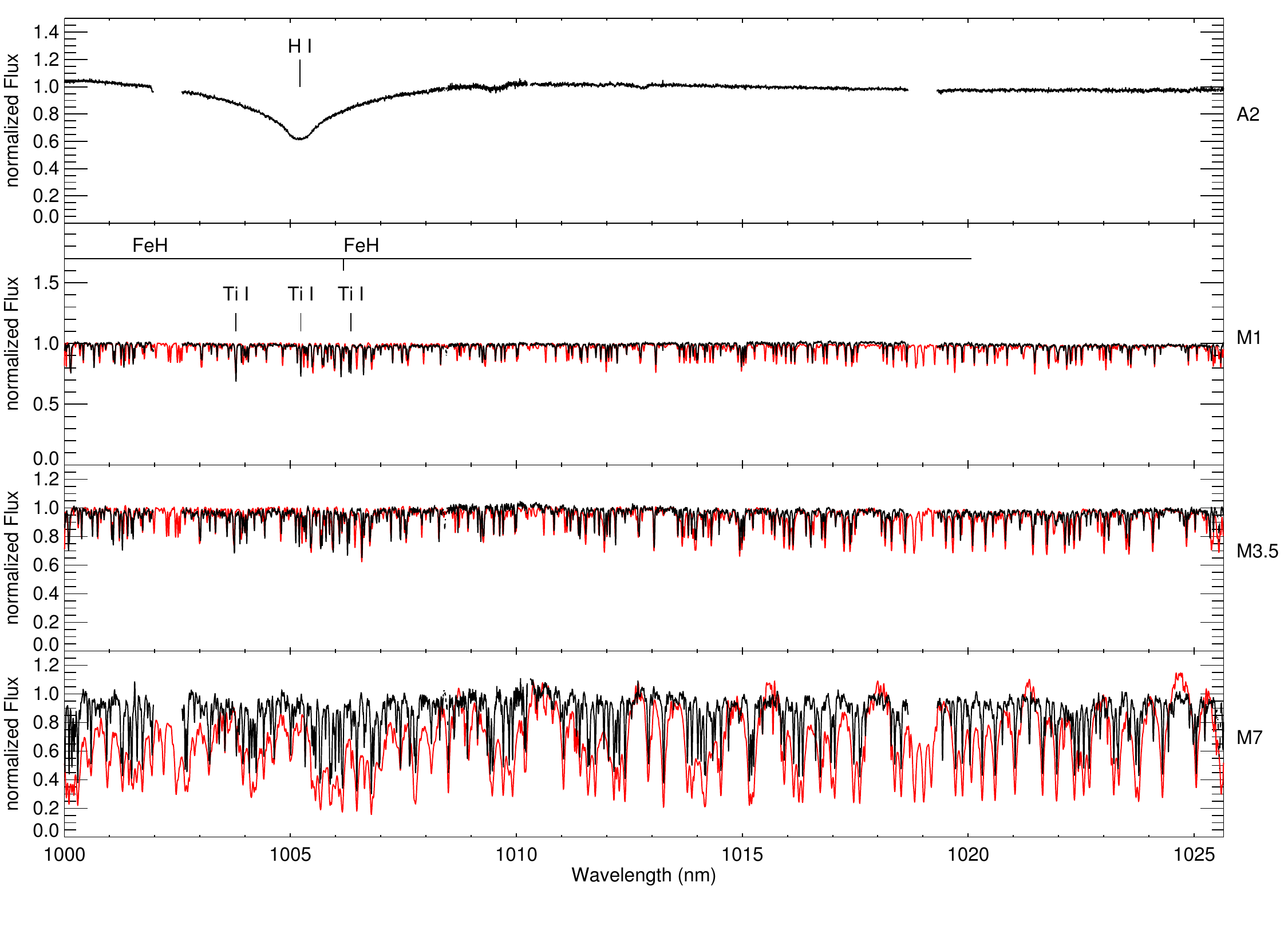}}
  \caption{\label{fig:atlas19}CARMENES spectral atlas.}
\end{figure*}

\begin{figure*}
  \resizebox{.97\hsize}{!}{\includegraphics[angle=90]{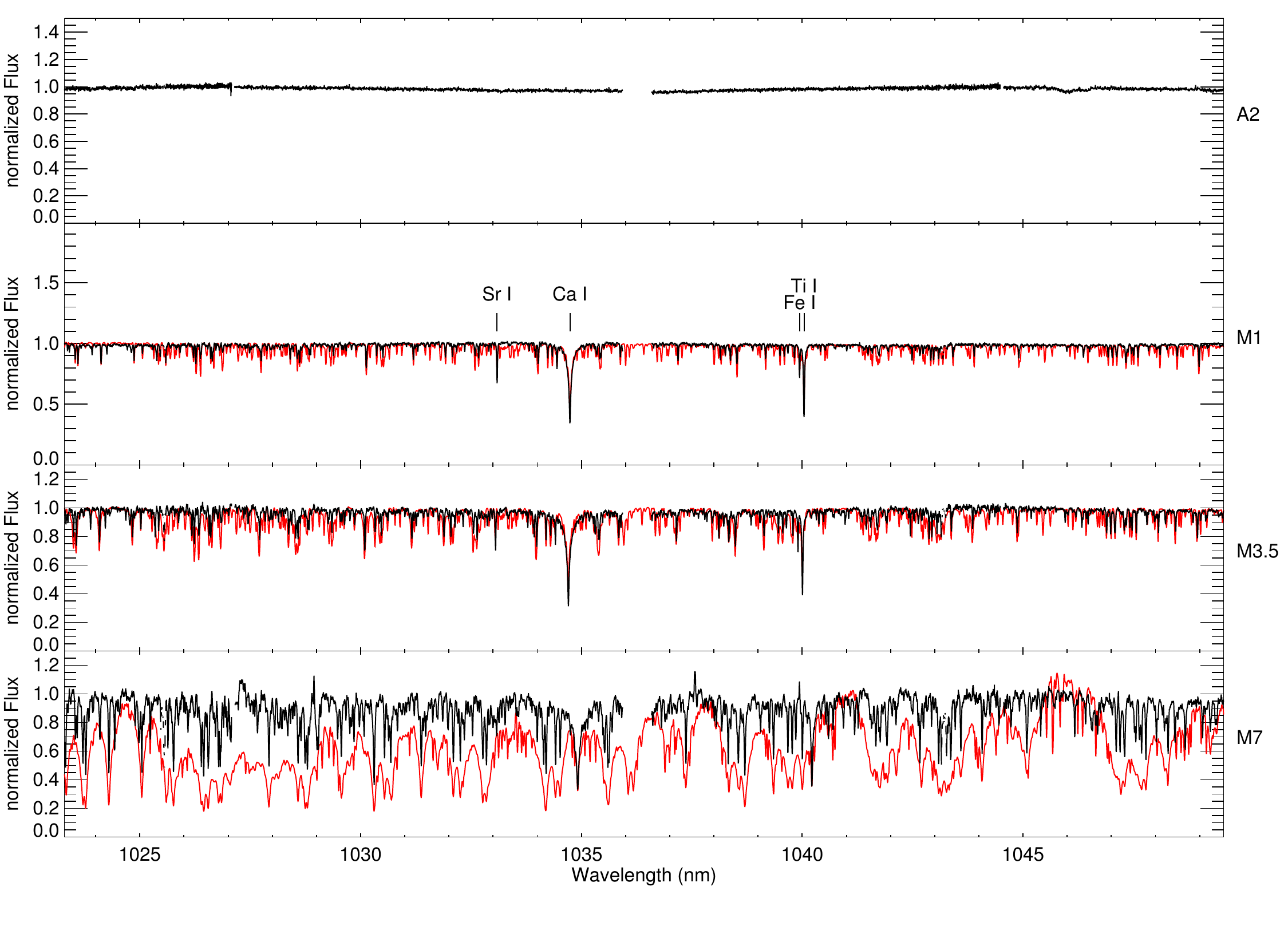}}
  \caption{\label{fig:atlas20}CARMENES spectral atlas.}
\end{figure*}

\begin{figure*}
  \resizebox{.97\hsize}{!}{\includegraphics[angle=90]{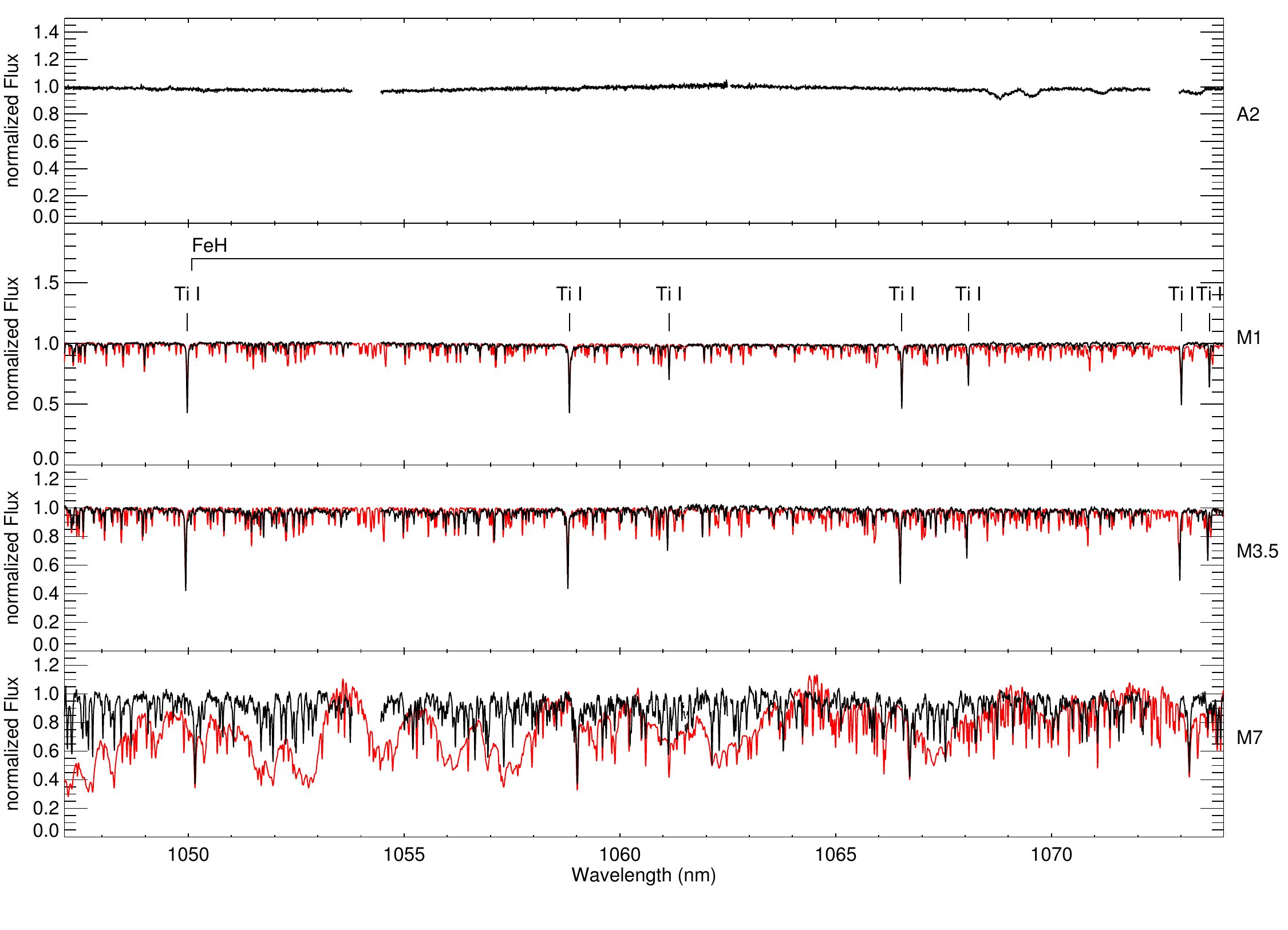}}
  \caption{\label{fig:atlas21}CARMENES spectral atlas.}
\end{figure*}

\begin{figure*}
  \resizebox{.97\hsize}{!}{\includegraphics[angle=90]{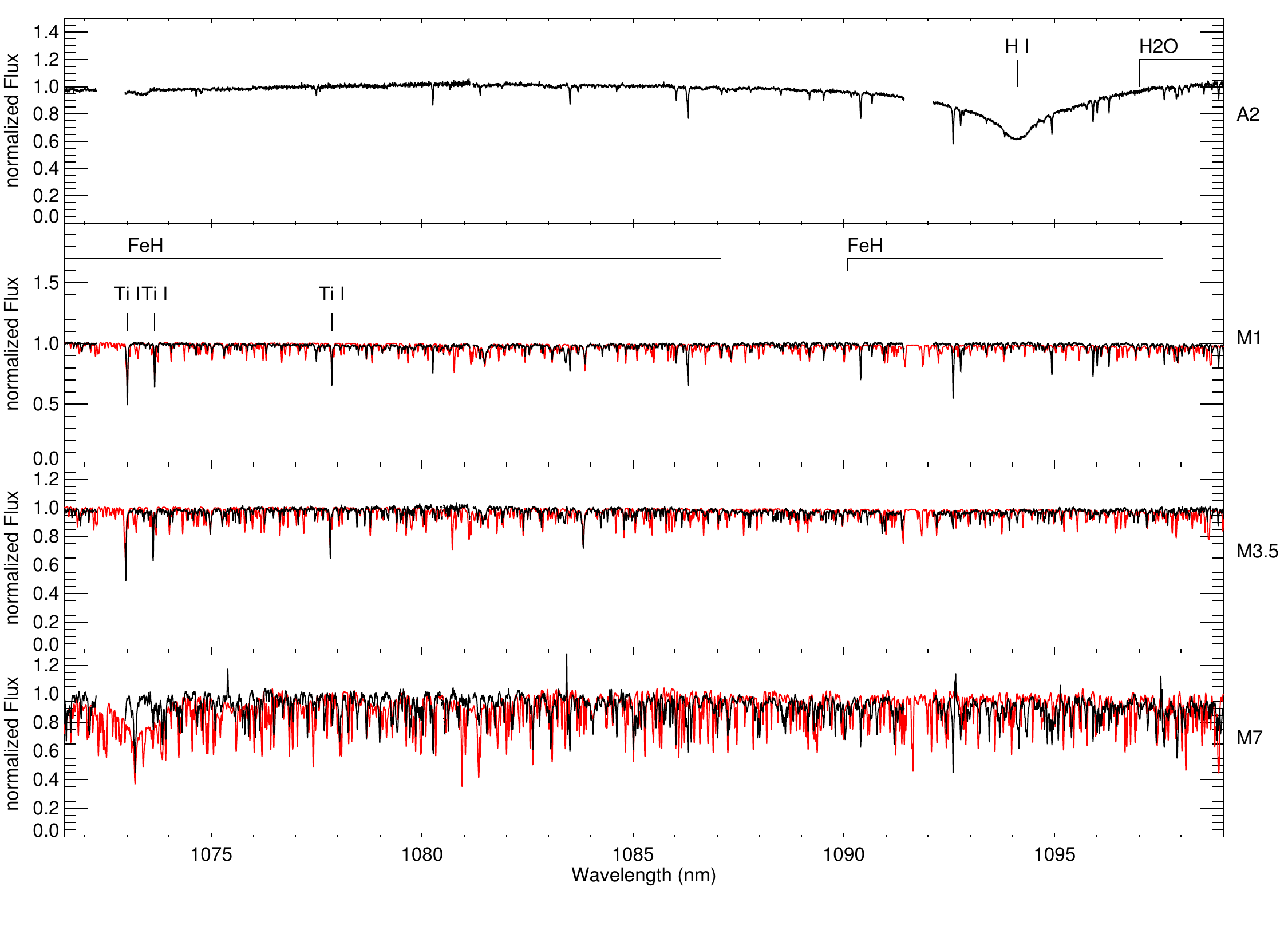}}
  \caption{\label{fig:atlas22}CARMENES spectral atlas.}
\end{figure*}

\begin{figure*}
  \resizebox{.97\hsize}{!}{\includegraphics[angle=90]{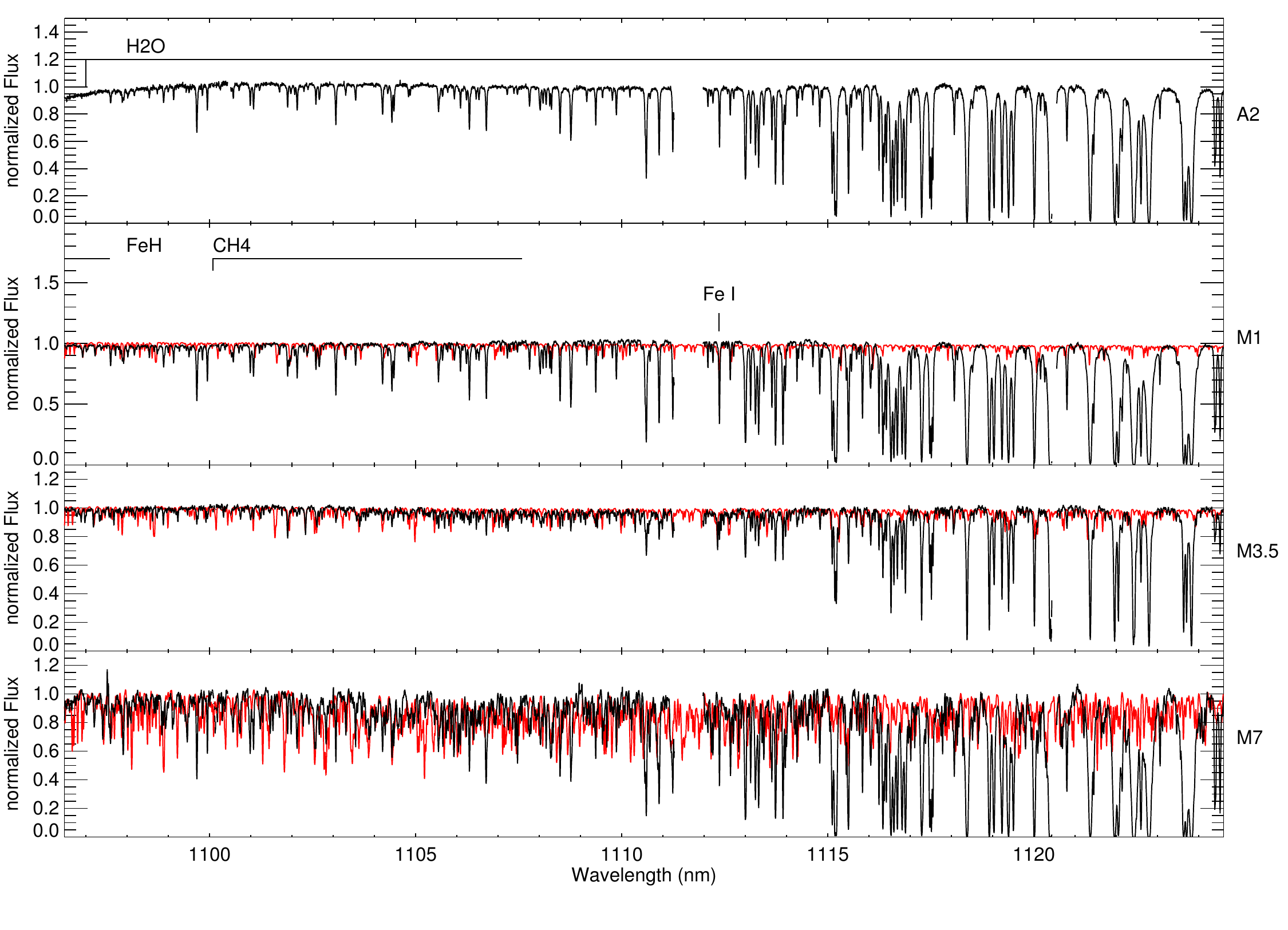}}
  \caption{\label{fig:atlas23}CARMENES spectral atlas.}
\end{figure*}

\begin{figure*}
  \resizebox{.97\hsize}{!}{\includegraphics[angle=90]{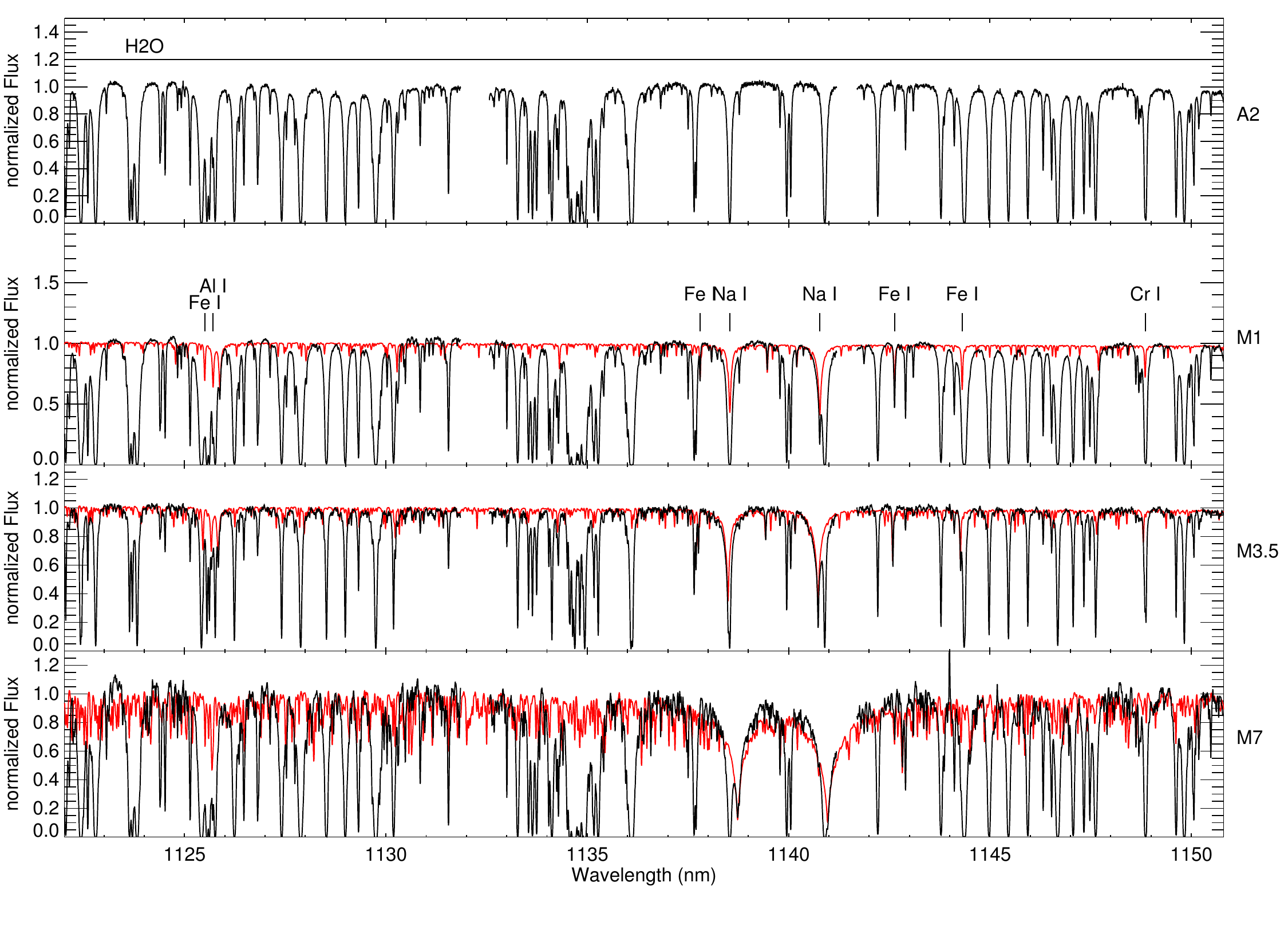}}
  \caption{\label{fig:atlas24}CARMENES spectral atlas.}
\end{figure*}

\begin{figure*}
  \resizebox{.97\hsize}{!}{\includegraphics[angle=90]{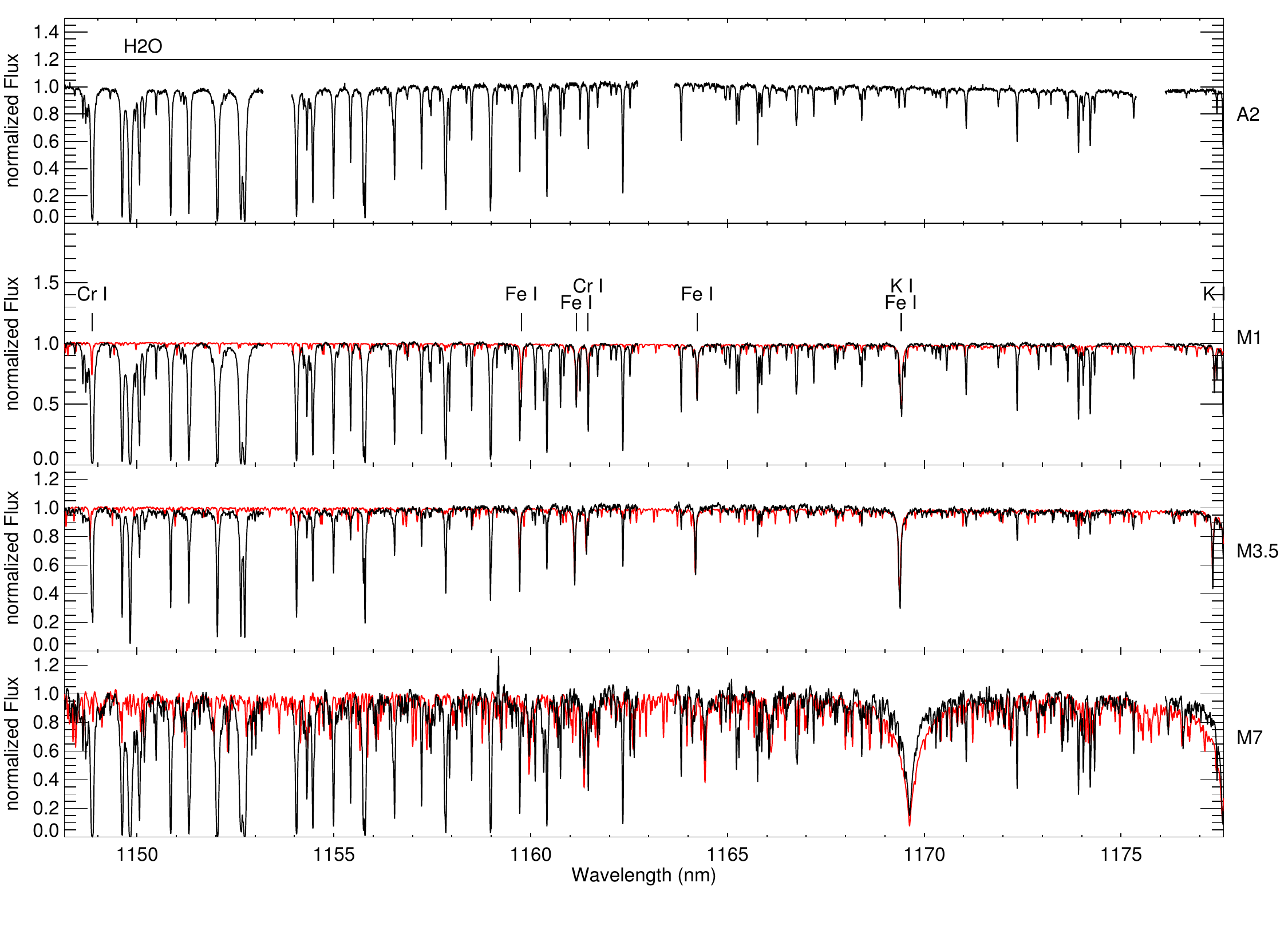}}
  \caption{\label{fig:atlas25}CARMENES spectral atlas.}
\end{figure*}

\begin{figure*}
  \resizebox{.97\hsize}{!}{\includegraphics[angle=90]{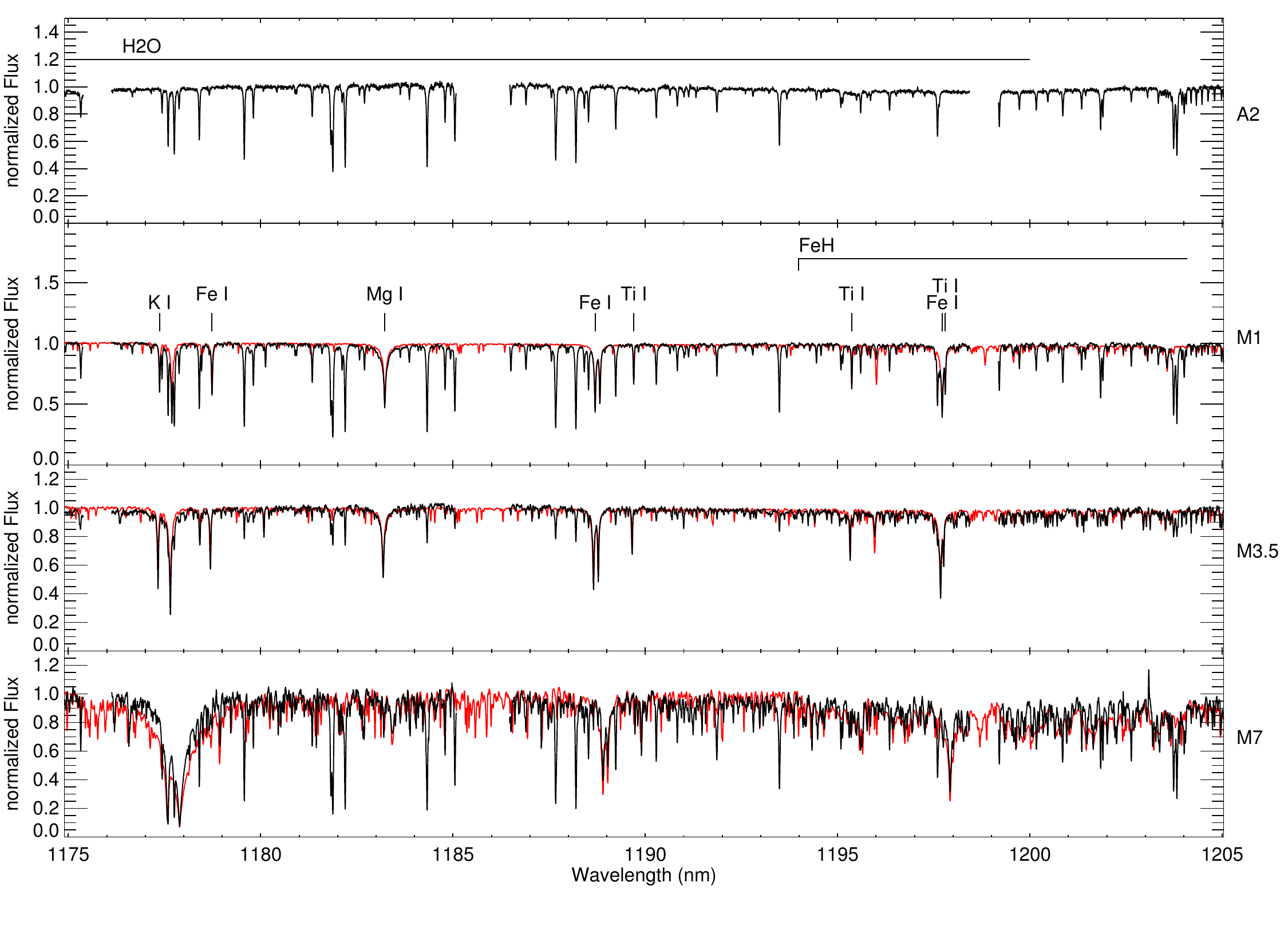}}
  \caption{\label{fig:atlas26}CARMENES spectral atlas.}
\end{figure*}

\begin{figure*}
  \resizebox{.97\hsize}{!}{\includegraphics[angle=90]{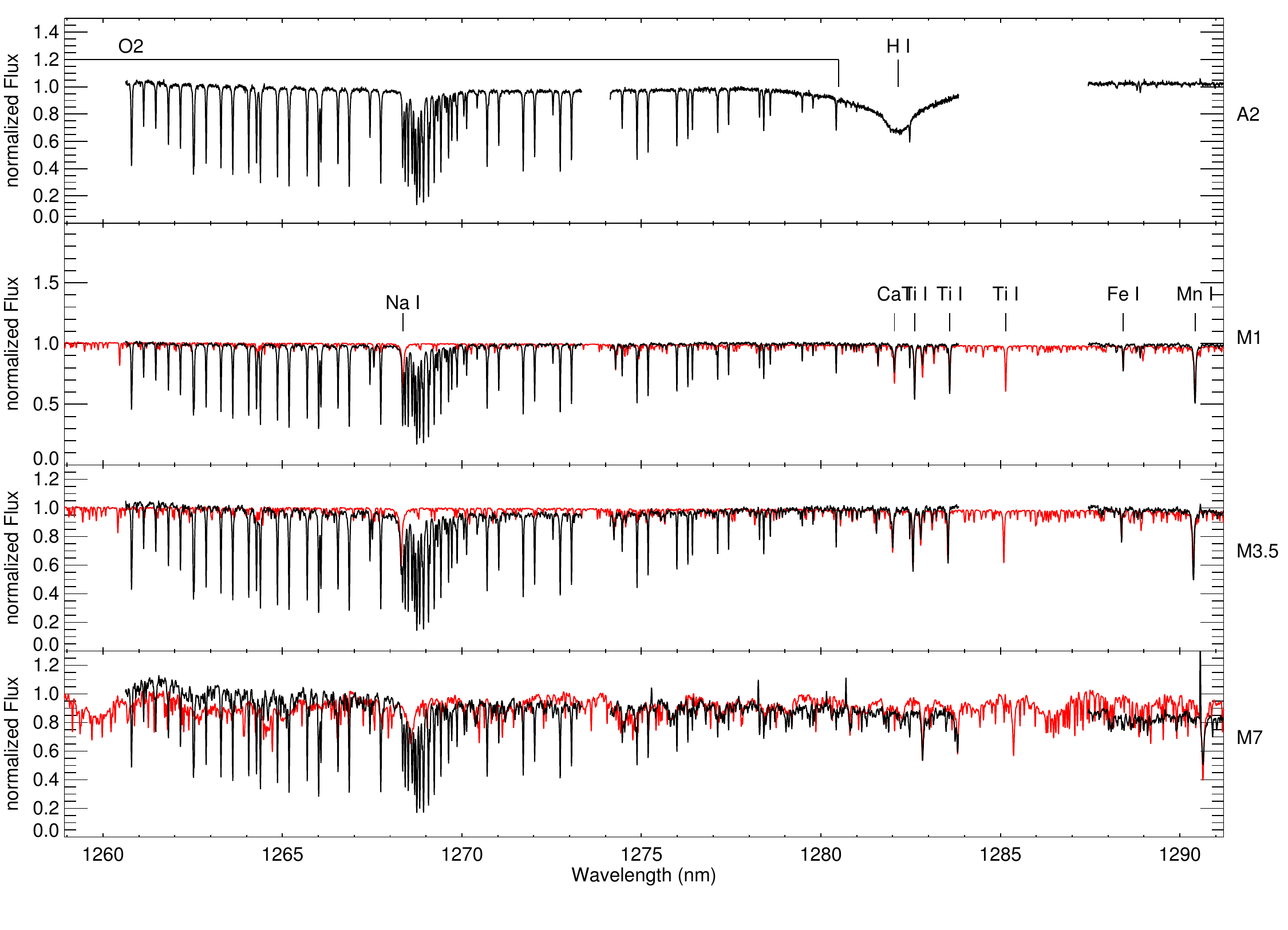}}
  \caption{\label{fig:atlas29}CARMENES spectral atlas.}
\end{figure*}

\begin{figure*}
  \resizebox{.97\hsize}{!}{\includegraphics[angle=90]{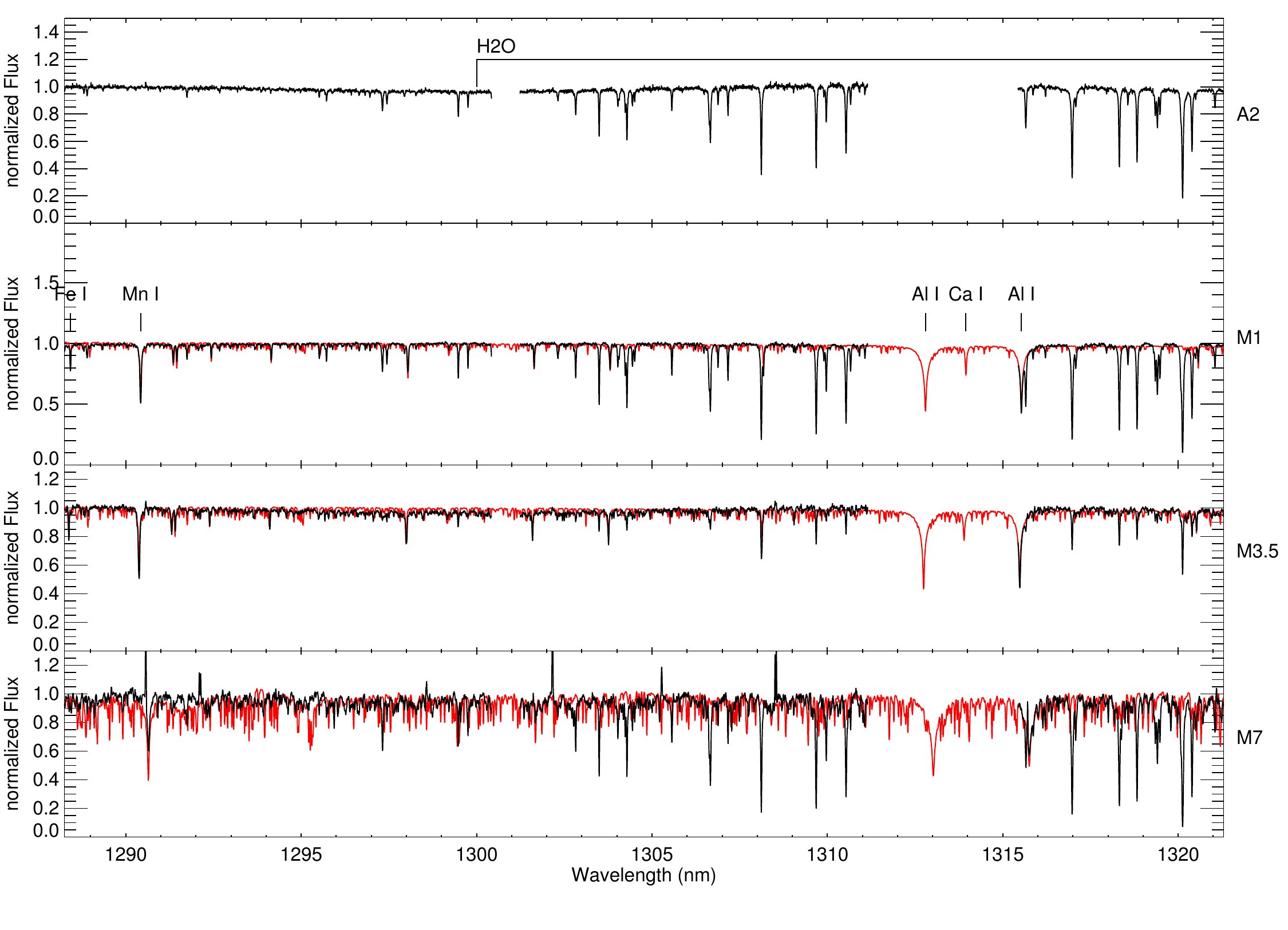}}
  \caption{\label{fig:atlas30}CARMENES spectral atlas.}
\end{figure*}

\begin{figure*}
  \resizebox{.97\hsize}{!}{\includegraphics[angle=90]{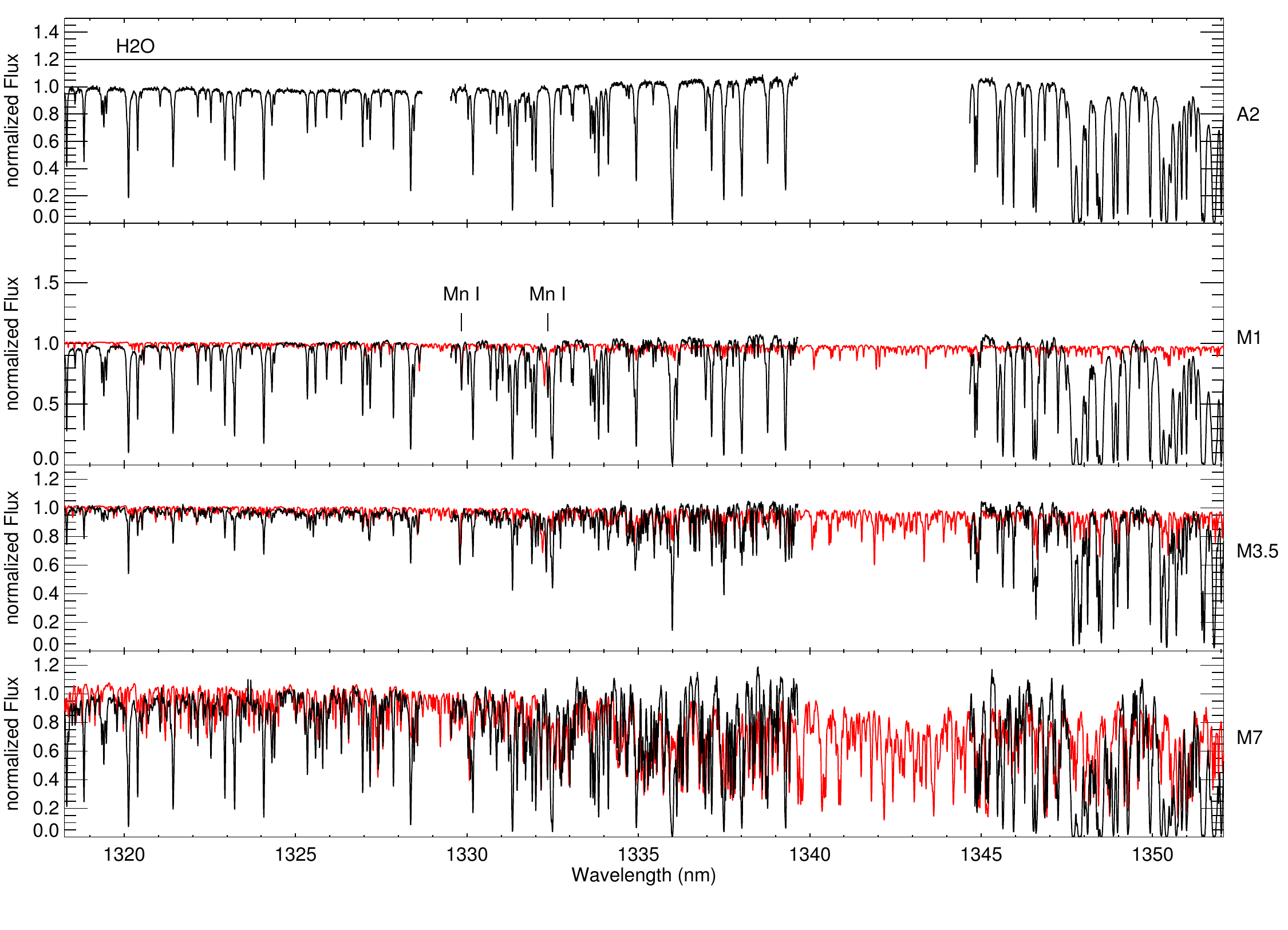}}
  \caption{\label{fig:atlas31}CARMENES spectral atlas.}
\end{figure*}

\begin{figure*}
  \resizebox{.97\hsize}{!}{\includegraphics[angle=90]{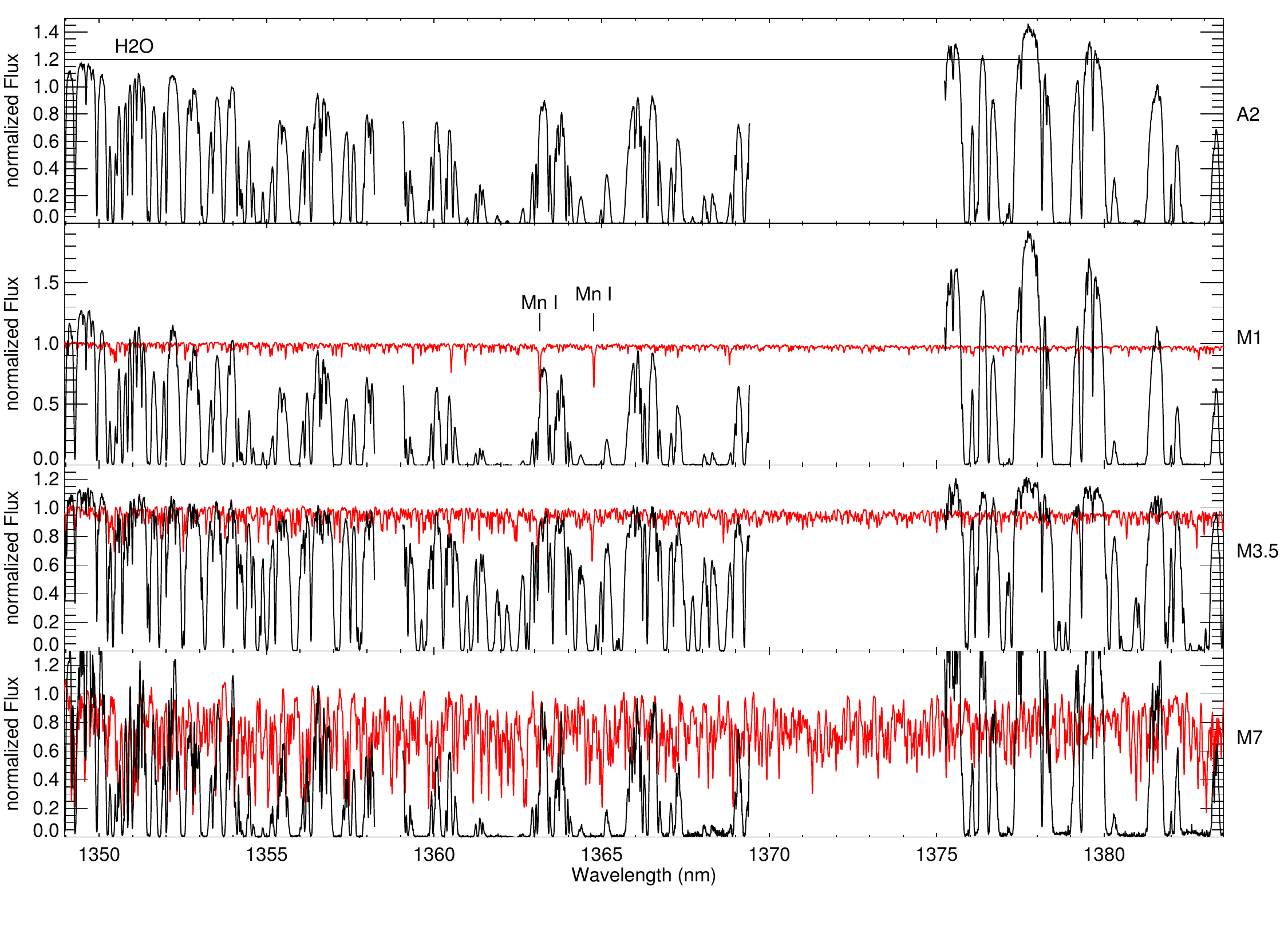}}
  \caption{\label{fig:atlas32}CARMENES spectral atlas.}
\end{figure*}

\begin{figure*}
  \resizebox{.97\hsize}{!}{\includegraphics[angle=90]{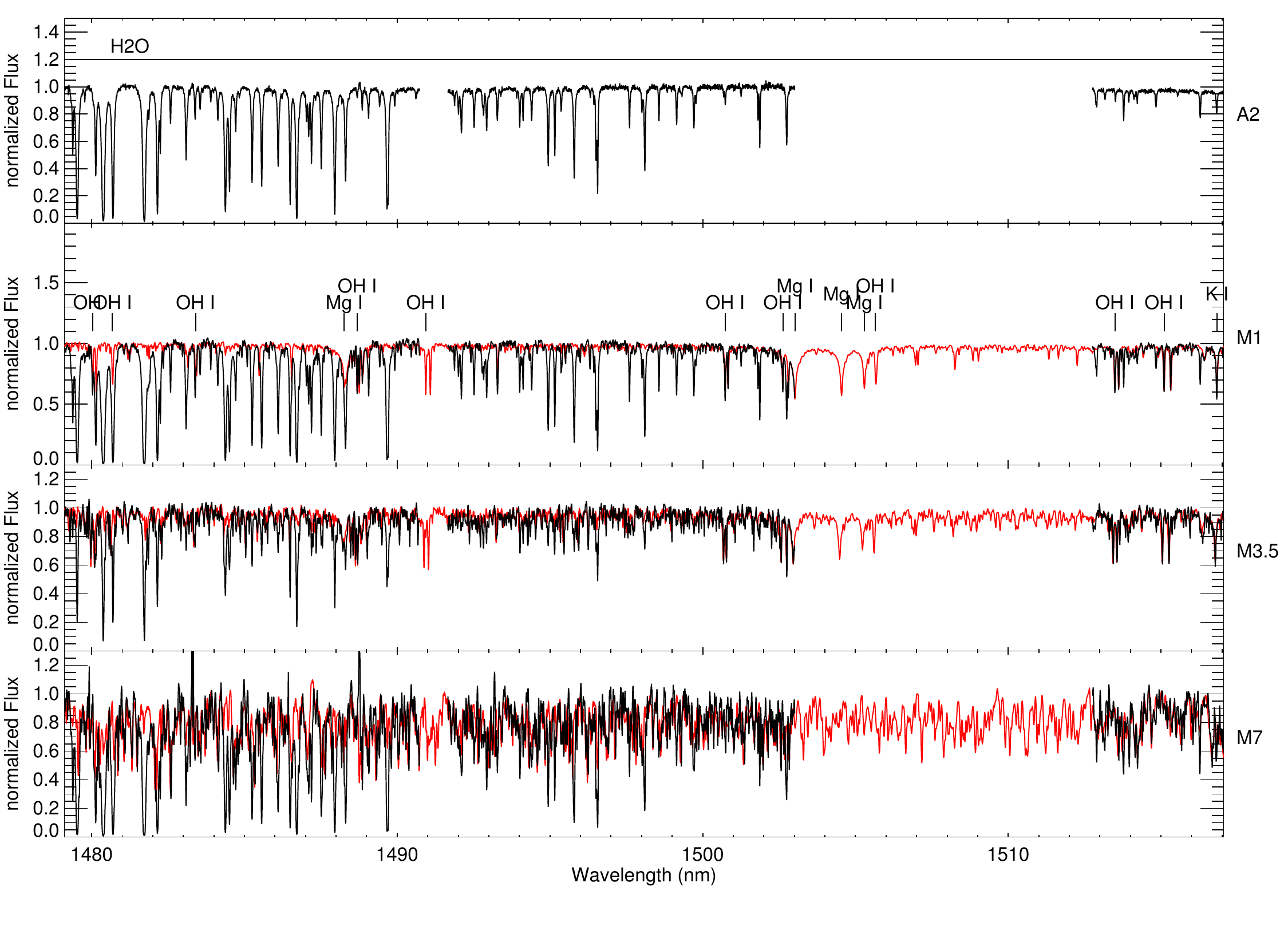}}
  \caption{\label{fig:atlas36}CARMENES spectral atlas.}
\end{figure*}

\begin{figure*}
  \resizebox{.97\hsize}{!}{\includegraphics[angle=90]{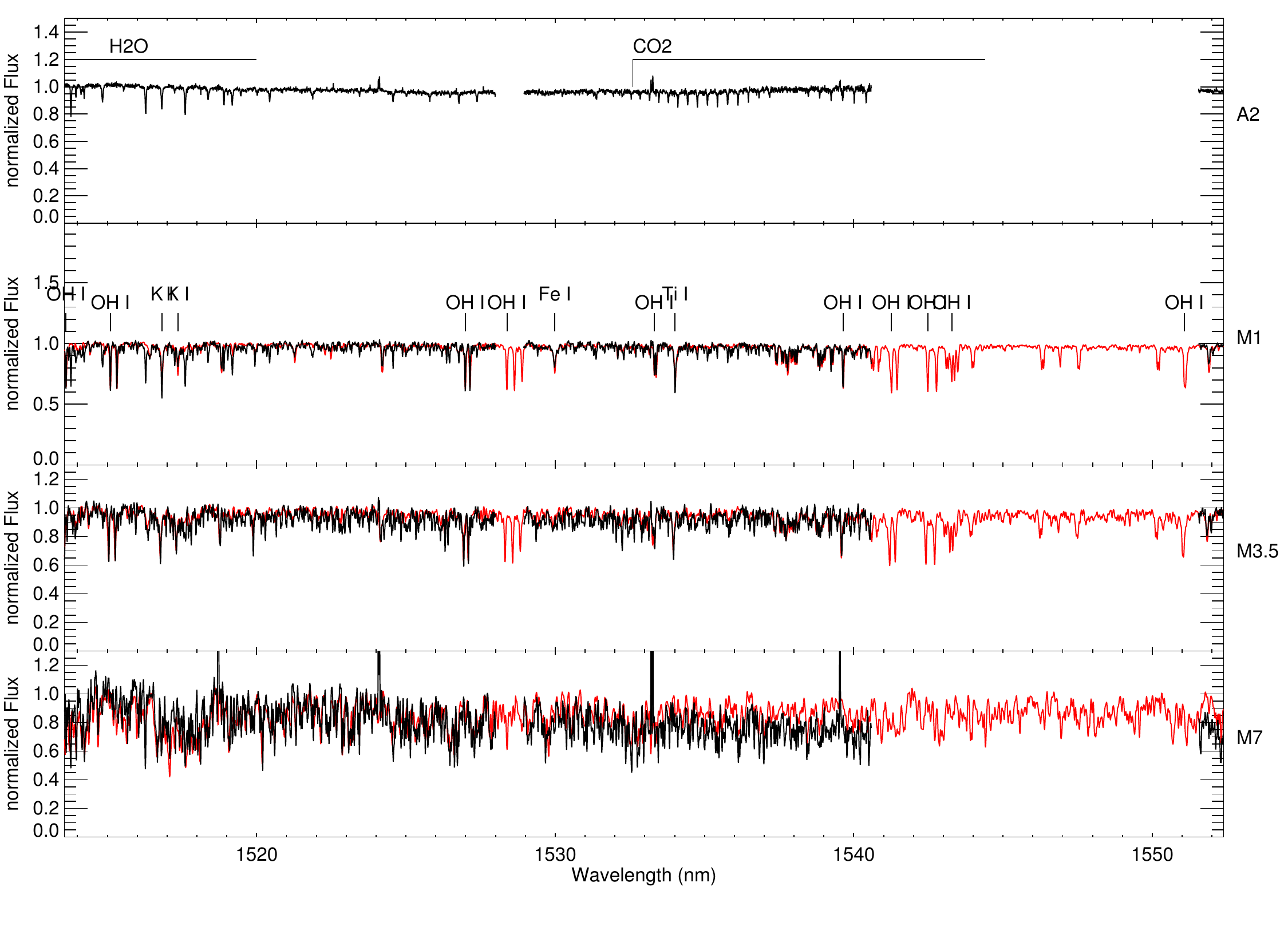}}
  \caption{\label{fig:atlas37}CARMENES spectral atlas.}
\end{figure*}

\begin{figure*}
  \resizebox{.97\hsize}{!}{\includegraphics[angle=90]{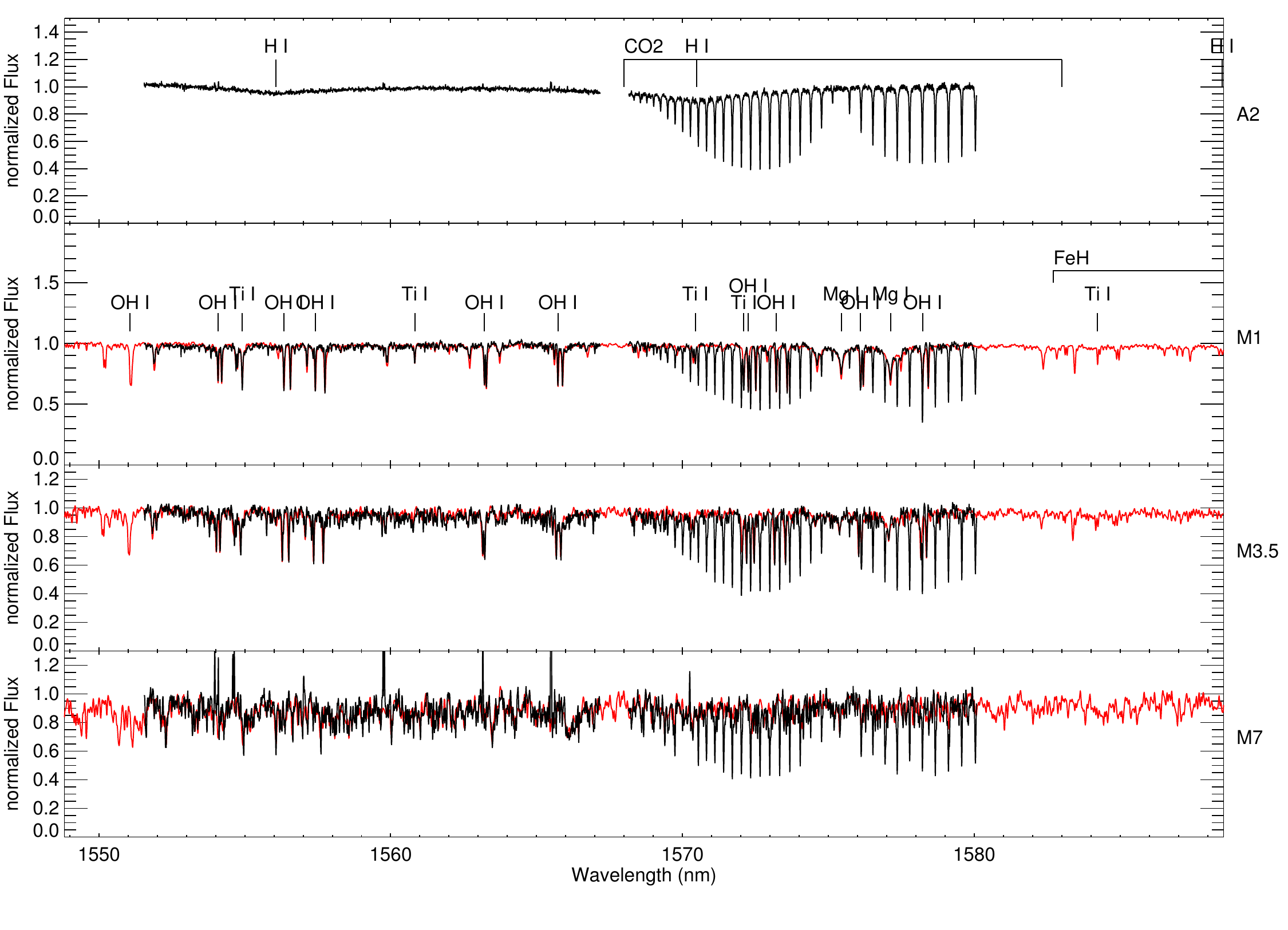}}
  \caption{\label{fig:atlas38}CARMENES spectral atlas.}
\end{figure*}

\begin{figure*}
  \resizebox{.97\hsize}{!}{\includegraphics[angle=90]{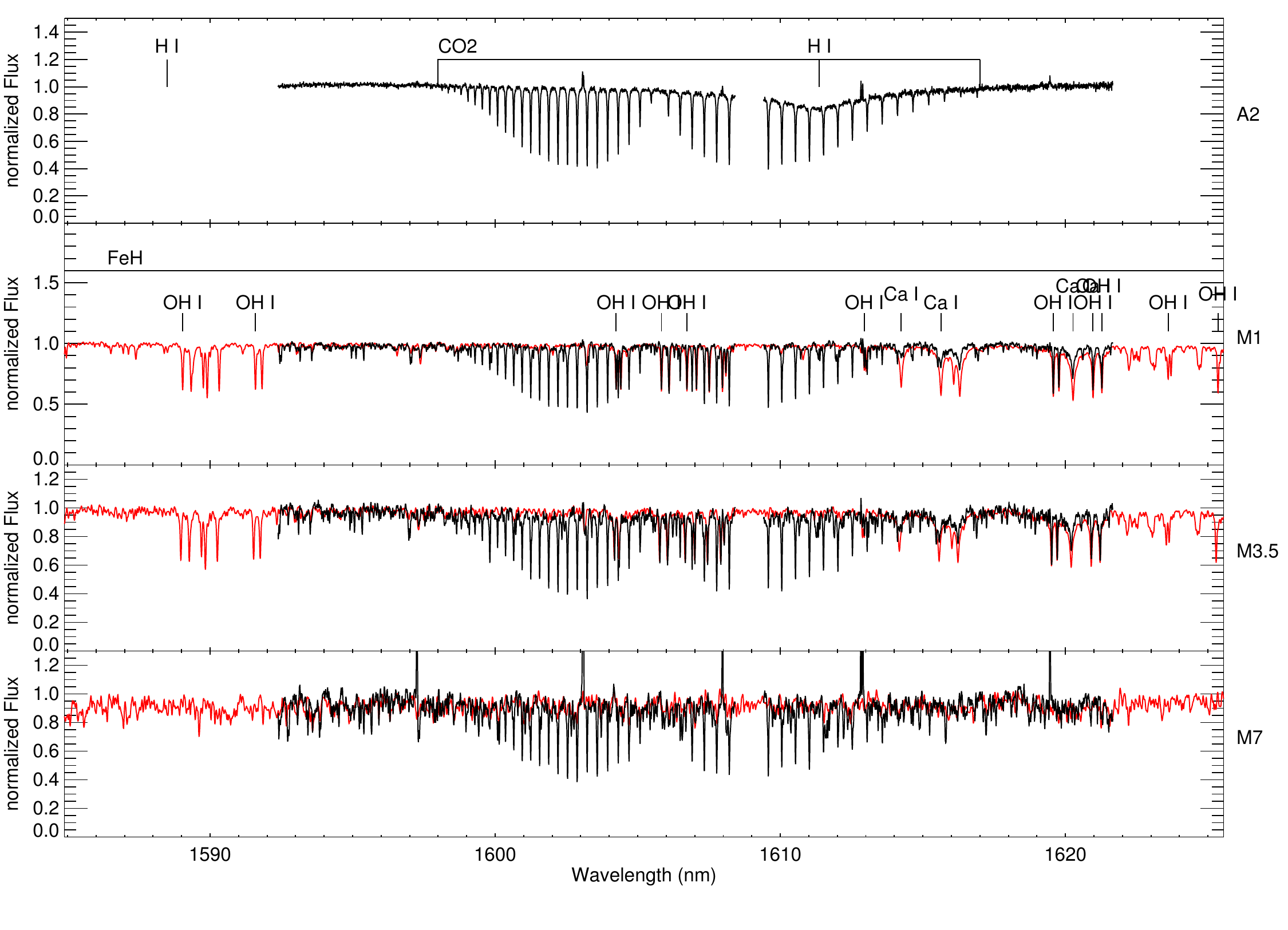}}
  \caption{\label{fig:atlas39}CARMENES spectral atlas.}
\end{figure*}

\begin{figure*}
  \resizebox{.97\hsize}{!}{\includegraphics[angle=90]{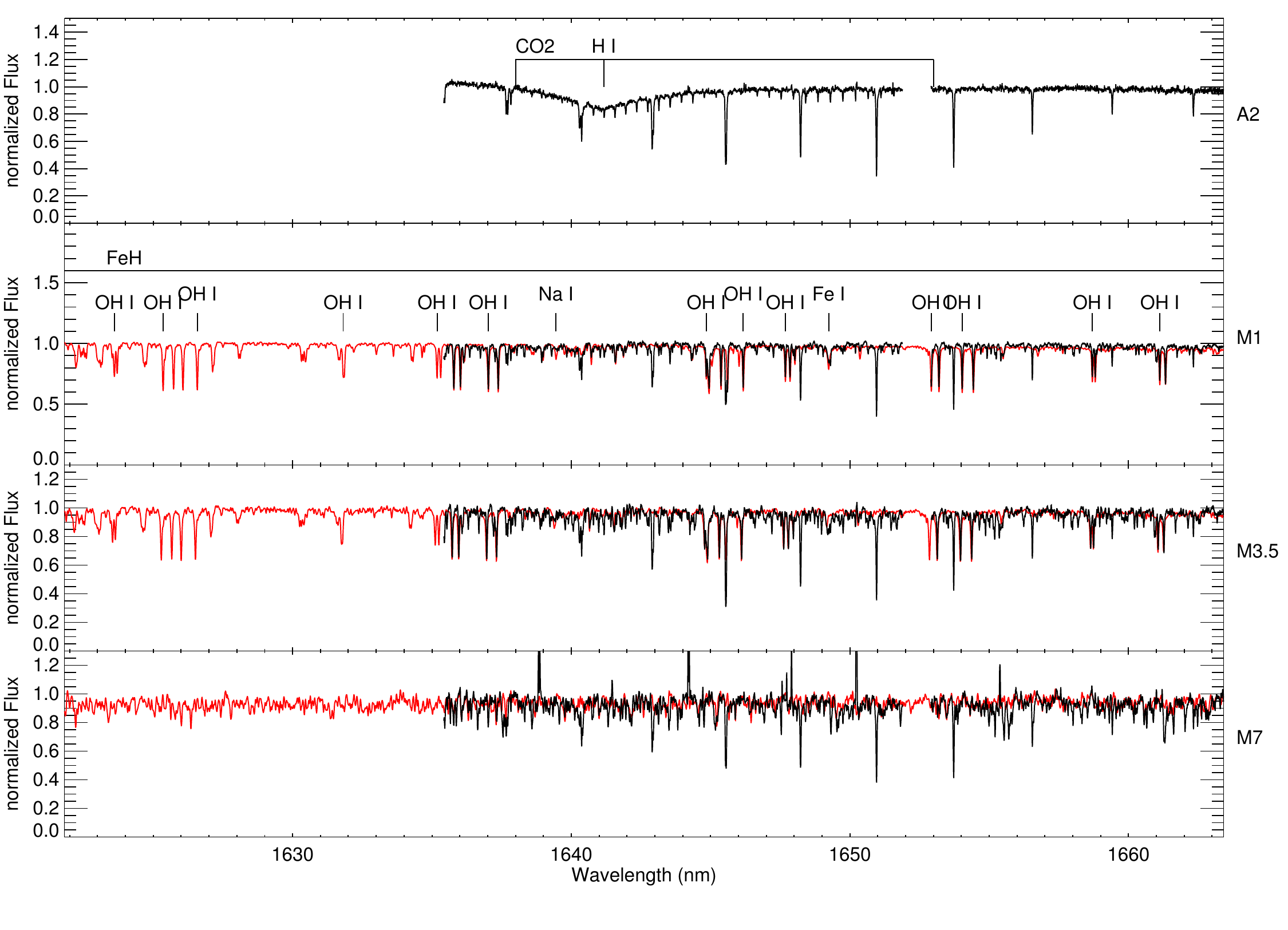}}
  \caption{\label{fig:atlas40}CARMENES spectral atlas.}
\end{figure*}

\begin{figure*}
  \resizebox{.97\hsize}{!}{\includegraphics[angle=90]{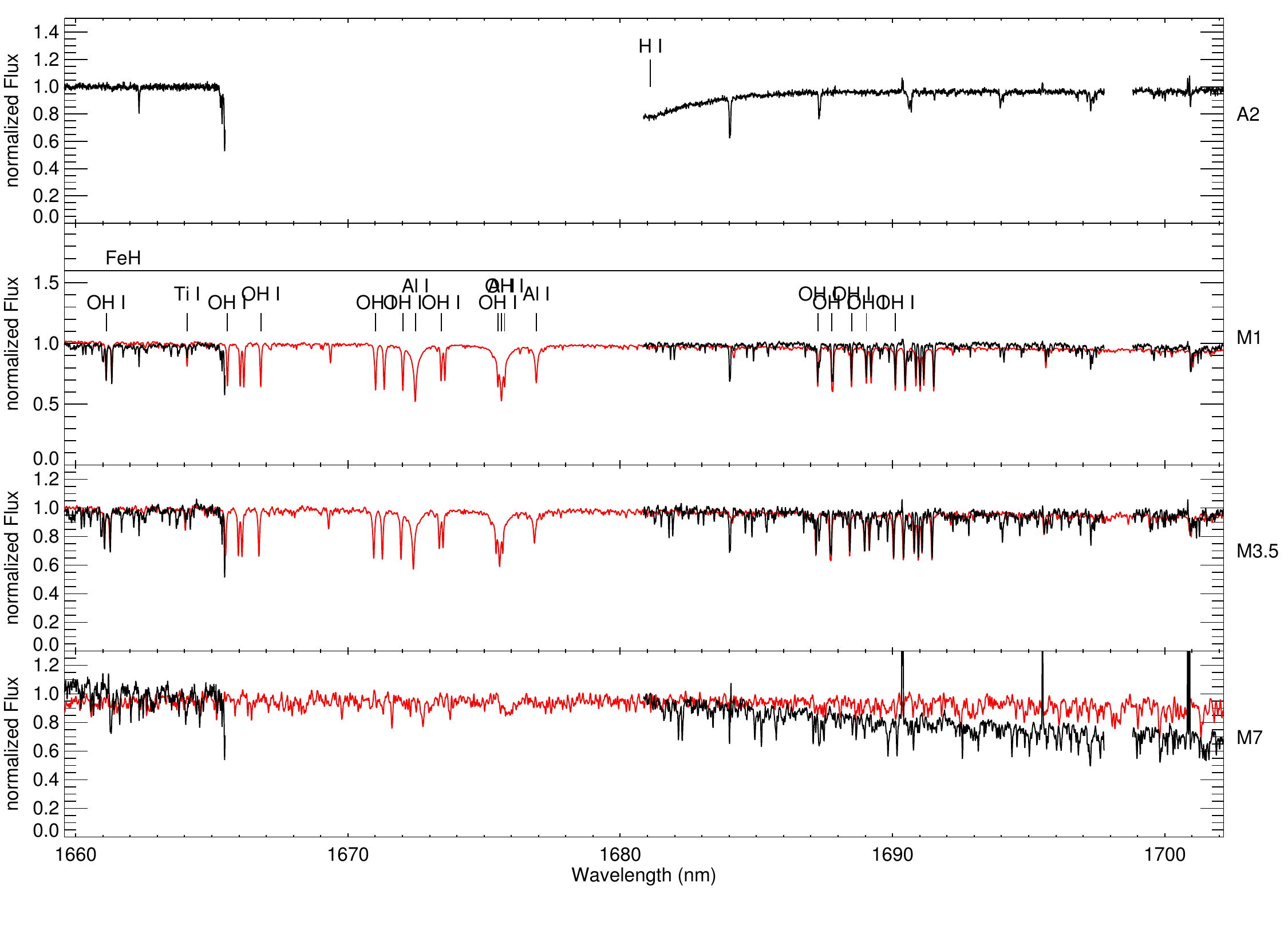}}
  \caption{\label{fig:atlas41}CARMENES spectral atlas.}
\end{figure*}

\begin{figure*}
  \resizebox{.97\hsize}{!}{\includegraphics[angle=90]{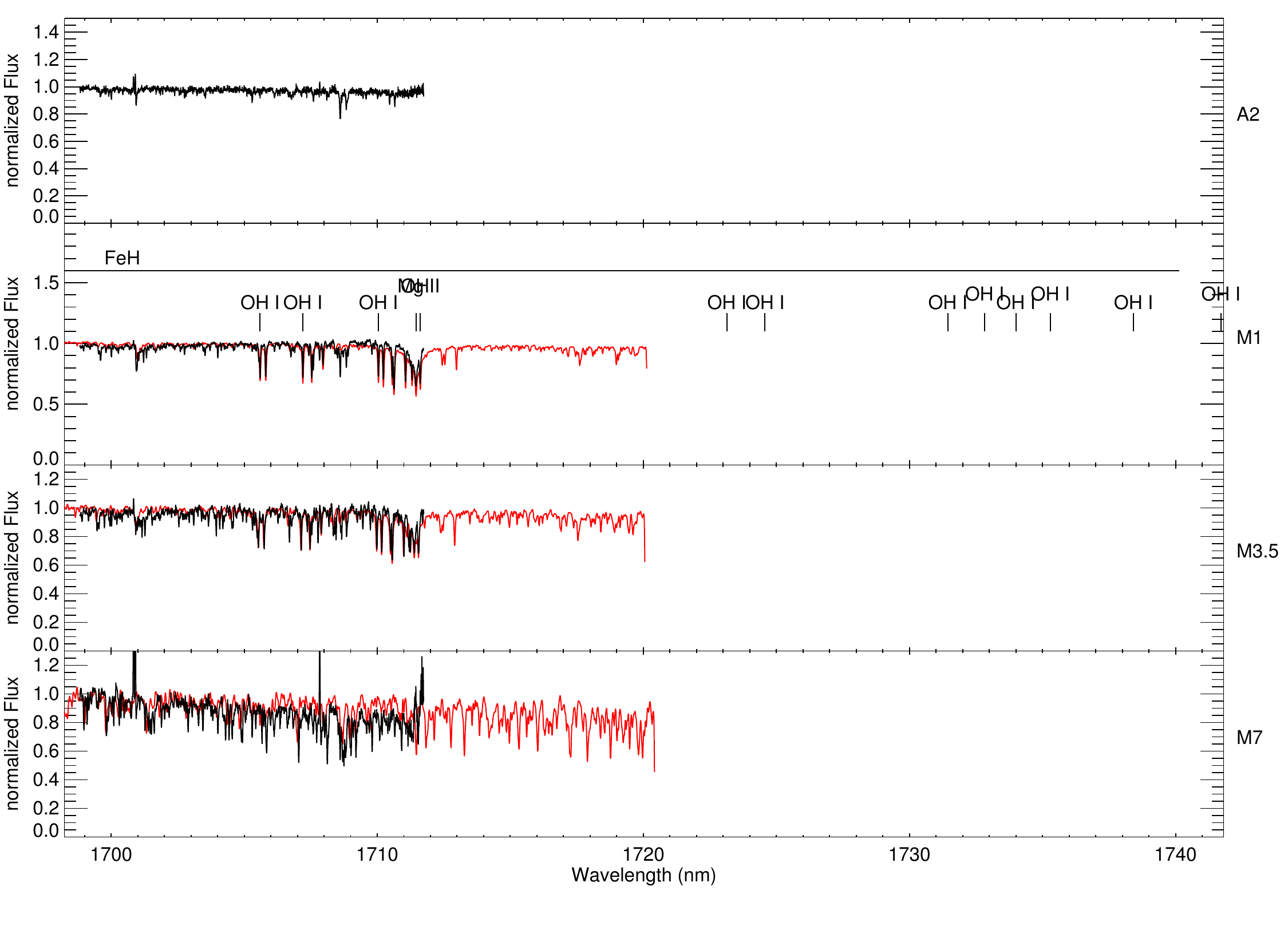}}
  \caption{\label{fig:atlas42}CARMENES spectral atlas.}
\end{figure*}

In Figs.\,\ref{fig:atlas00} -- \ref{fig:atlas42} we plot spectra of a
telluric standard star and three survey targets in black and
the synthetic
model spectra in red. See Section\,\ref{sect:atlas} for details. 

 \clearpage

 \clearpage

 \clearpage

 \clearpage

 \clearpage

 \clearpage

 \clearpage

 \clearpage

 \clearpage

 \clearpage

 \clearpage

 \clearpage

 \clearpage

 \clearpage

 \clearpage

 \clearpage

\begin{figure*}
  \resizebox{.97\hsize}{!}{\includegraphics[angle=90]{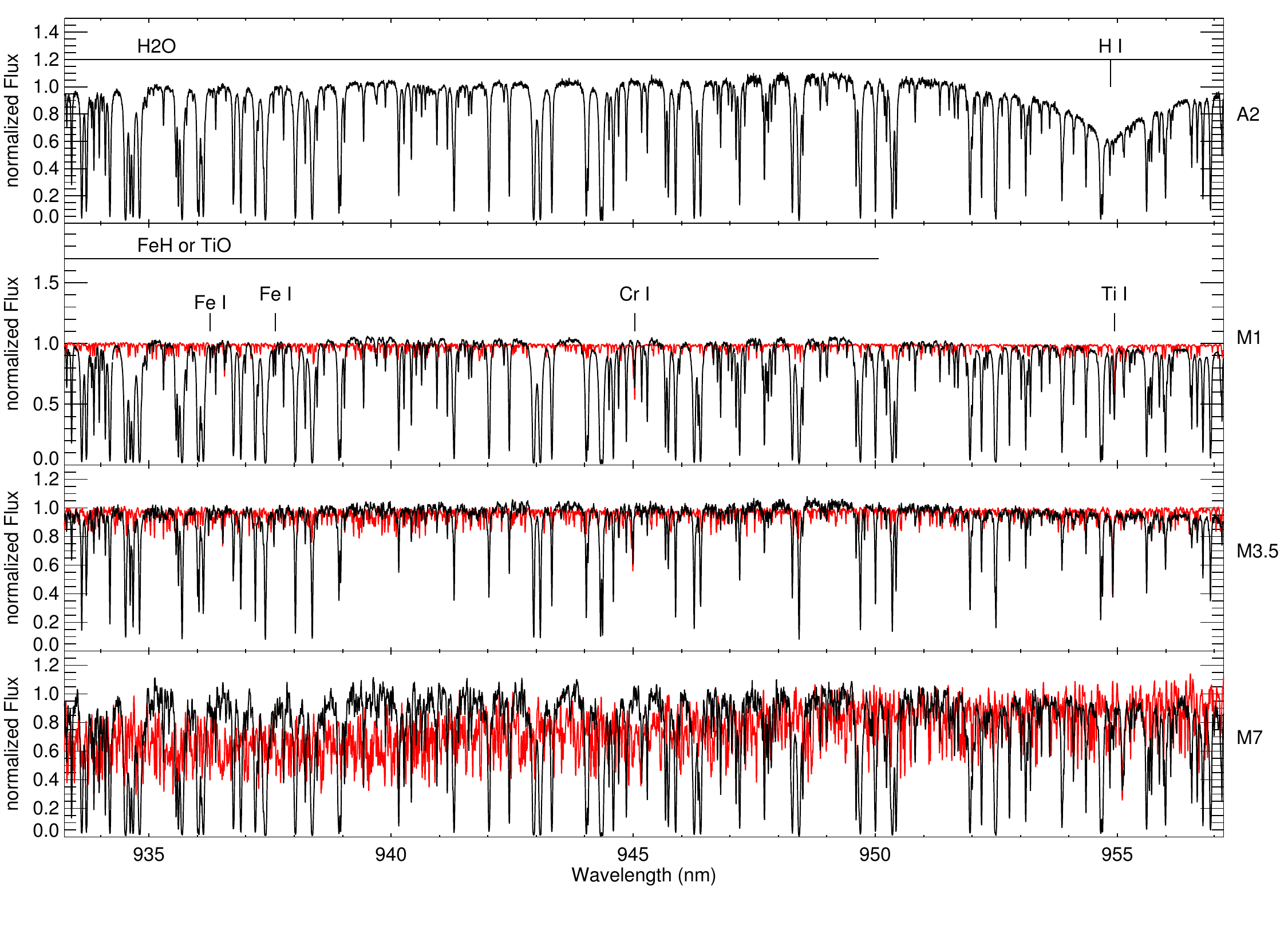}}
  \caption{CARMENES spectral atlas.}
\end{figure*} \clearpage

\begin{figure*}
  \resizebox{.97\hsize}{!}{\includegraphics[angle=90]{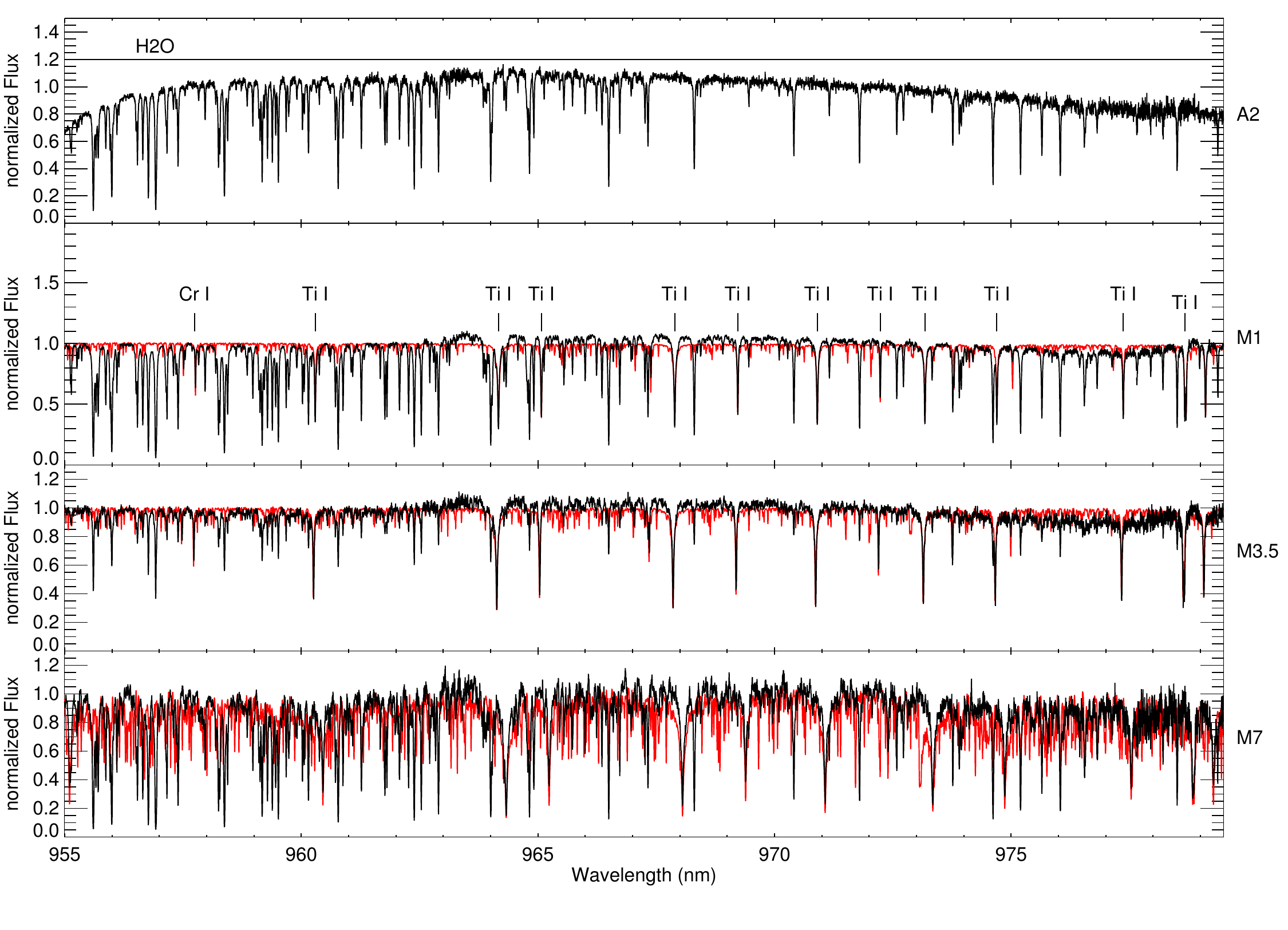}}
  \caption{CARMENES spectral atlas.}
\end{figure*} \clearpage

\begin{figure*}
  \resizebox{.97\hsize}{!}{\includegraphics[angle=90]{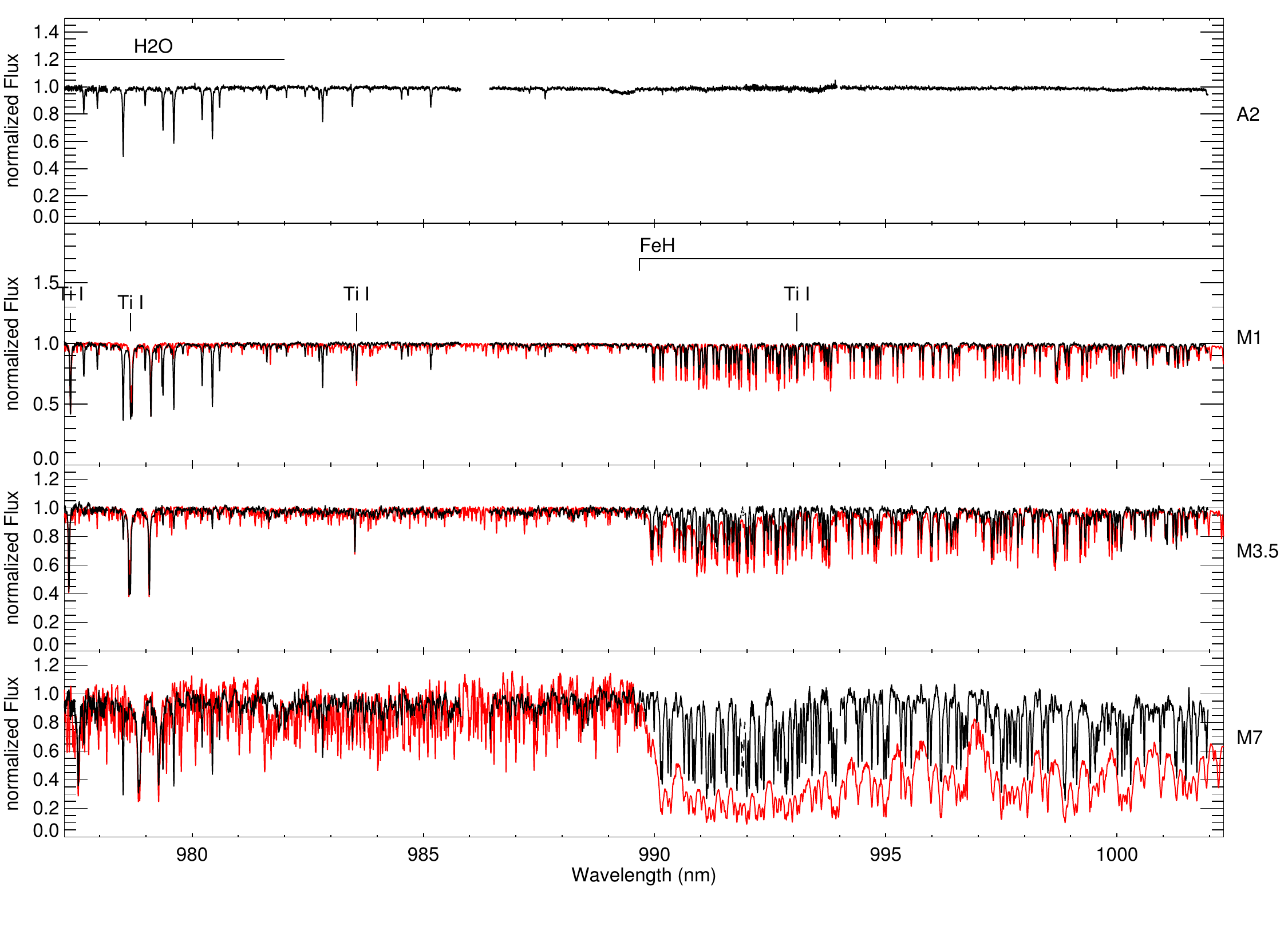}}
  \caption{CARMENES spectral atlas.}
\end{figure*} \clearpage

 \clearpage

 \clearpage

 \clearpage

 \clearpage

 \clearpage

 \clearpage

 \clearpage

 \clearpage

\begin{figure*}
  \resizebox{.97\hsize}{!}{\includegraphics[angle=90]{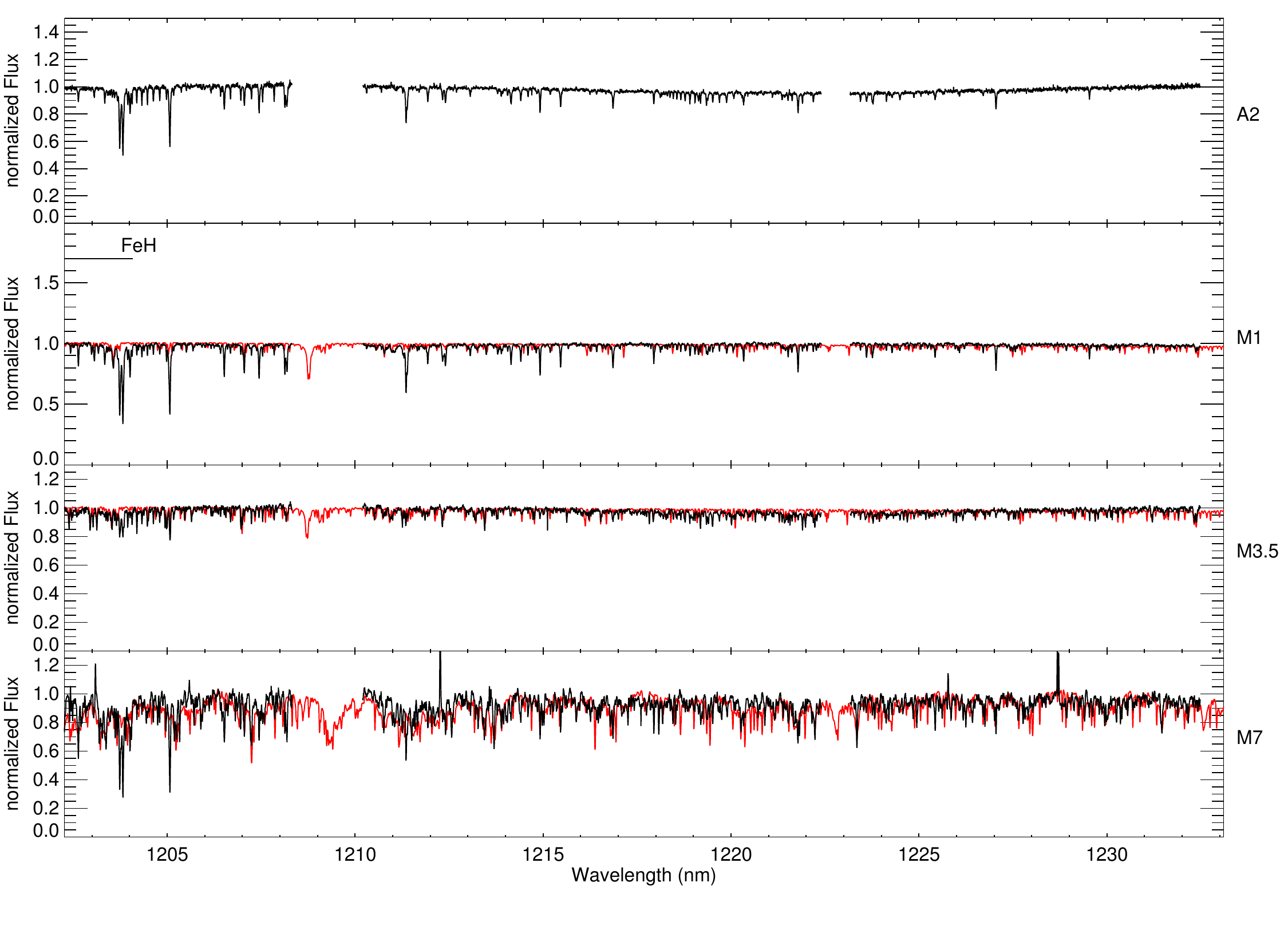}}
  \caption{CARMENES spectral atlas.}
\end{figure*} \clearpage

\begin{figure*}
  \resizebox{.97\hsize}{!}{\includegraphics[angle=90]{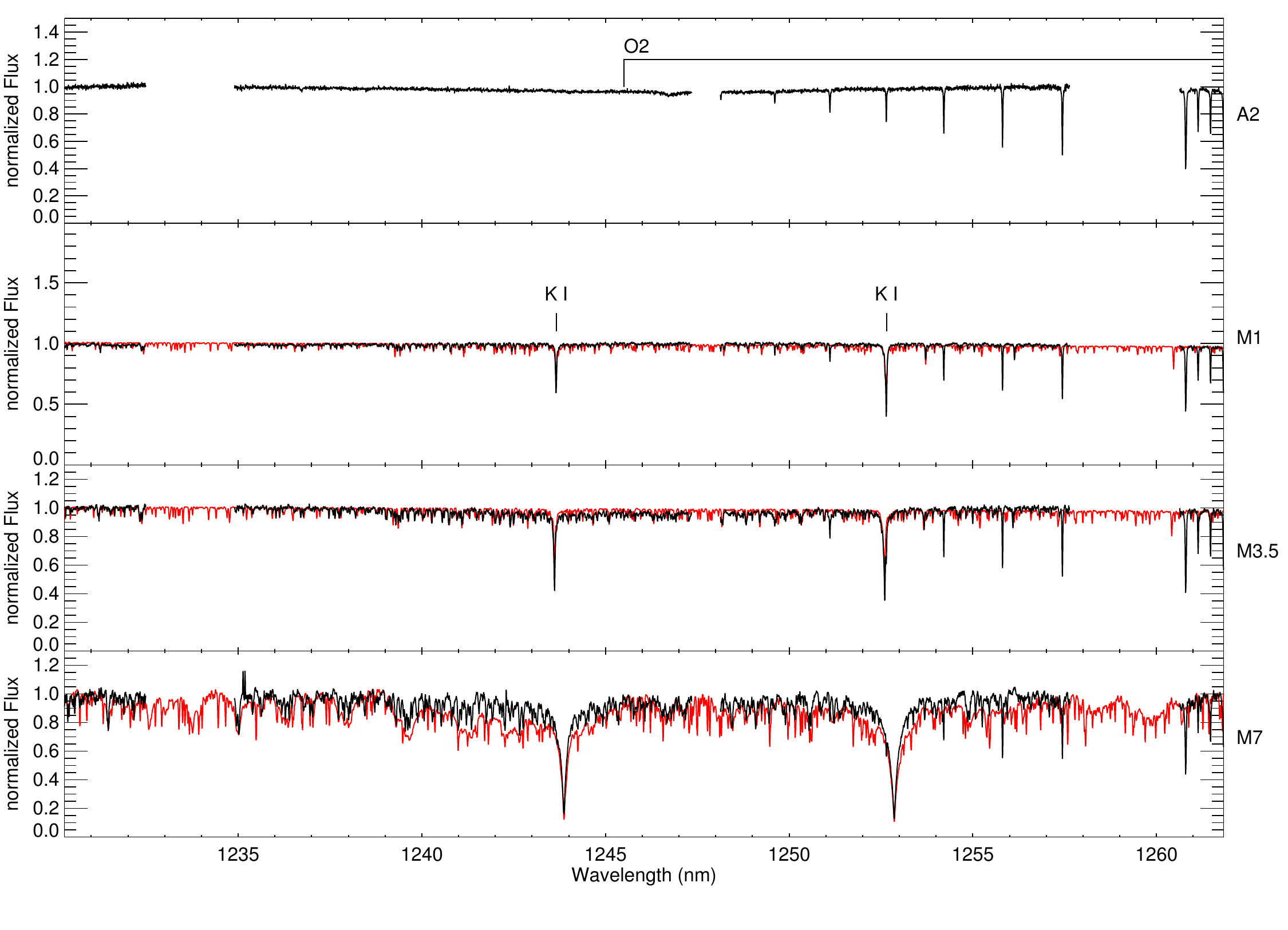}}
  \caption{CARMENES spectral atlas.}
\end{figure*} \clearpage

 \clearpage

 \clearpage

 \clearpage

 \clearpage

 \clearpage

 \clearpage

 \clearpage

 \clearpage

 \clearpage

 \clearpage

 \clearpage

\section{Table of Targets}

Basic data for the CARMENES GTO targets are provided in
Table\,\ref{tab:Table}. Columns are discussed in the main
text. Uncertainties for estimates of the inclination angle $i$ are
computed only from the uncertainties in $\varv\,\sin{i}$.

\longtab[1]{
\begin{landscape}
\begin{longtable}{clrlcrrrcccccc}
\caption{\label{tab:Table} Basic data of the CARMENES GTO sample}\\
\hline\hline
No.& ID & GJ & Name & SpT & $J$ & $M$ & $\varv_{\rm rad}$ & $P$ & Ref. & $\varv\,\sin{i}$ & $P_{\varv\,\sin{i}}/\sin{i}$ & $i$ & $\log{\rm L_{H\alpha}/L_{\rm bol}}$ \\
   &   &  &  &  & (mag) & (M$_{\odot}$) & (km\,s$^{-1}$) & (d) & & (km\,s$^{-1}$) & (d) & deg & \\
\hline
\endfirsthead
\caption{continued.}\\
\hline\hline
No.& ID & GJ & Name & SpT & $J$ & $M$ & $\varv_{\rm rad}$ & $P $ & Ref. & $\varv\,\sin{i}$ & $P_{\varv\,\sin{i}}/\sin{i}$ & $i$ & $\log{\rm L_{H\alpha}/L_{\rm bol}}$ \\
   &   &  &  &  & (mag) & (M$_{\odot}$) & (km\,s$^{-1}$) & (d) & & (km\,s$^{-1}$) & (d) & deg & \\
\hline
\endhead
\hline
\endfoot
001 &  J00051$+$457 &          2 &       GJ 2 &     M1.0 &  6.70 & 0.56 &                 -0.392 &                      &         &               $<2$ &             $>$ 13.3 &                      &         \\
002 &  J00067$-$075 &       1002 &    GJ 1002 &     M5.5 &  8.32 & 0.11 &                -40.058 &                      &         &               $<2$ &             $>$  3.5 &                      & $-4.83$ \\
003 & J00162$+$198E &      1006B & LP 404-062 &     M4.0 &  8.89 & 0.30 &                 -1.457 & $ 104.00 \pm 31.20 $ &   Irw11 &               $<2$ &             $>$  7.4 &           $\le 90  $ &         \\
004 &  J00183$+$440 &        15A &     GX And &     M1.0 &  5.25 & 0.45 &                 11.671 &                      &         &               $<2$ &             $>$ 10.5 &                      &         \\
005 &  J00184$+$440 &        15B &     GQ And &     M3.5 &  6.79 & 0.16 &                 10.876 &                      &         &               $<2$ &             $>$  4.6 &                      &         \\
006 &  J00286$-$066 &       1012 &    GJ 1012 &     M4.0 &  8.04 & 0.39 &                -12.531 &                      &         &               $<2$ &             $>$  9.1 &                      &         \\
007 &  J00389$+$306 &         26 &  Wolf 1056 &     M2.5 &  7.45 & 0.47 &                 -0.499 &                      &         &               $<2$ &             $>$ 11.1 &                      &         \\
008 &  J00570$+$450 &            &  G 172-030 &     M3.0 &  8.10 & 0.39 &                  6.396 &                      &         &               $<2$ &             $>$  9.3 &                      &         \\
\end{longtable}
\begin{flushleft}
\textbf{Period References:}  Irw11: \citet[][]{2011ApJ...727...56I}; New16a: \citet{2016ApJ...821...93N}; SM16a: \citet{2016A&A...595A..12S}; Kira12: \citet{2012AcA....62...67K}; KS07: \citet[][]{2007AcA....57..149K}; HM15:  \citet[][]{2015ApJ...801..106H}; Chu74: \citet{1974IzKry..52....3C}; FH00: \citet{2000AJ....120.3265F}; Har11:  \citet{2011AJ....141..166H}; WF11: \citet{2011MNRAS.418.1822W}; West15: \citet[][]{2015ApJ...812....3W}; SM15: \citet{2015MNRAS.452.2745S}; Nor07: \citet{2007A&A...467..785N}; Ba83: \citet{1983ApJ...275..752B}; Kor10: \citet{2010AN....331..772K}; Tes04: \citet{2004ApJ...617..508T}; KS13: \citet[][]{2013AcA....63...53K}

\end{flushleft}
\end{landscape}
}

\end{appendix}

\end{document}